\documentclass[twocolumn]{aastex62}
\usepackage{tabularx} 
\usepackage{needspace}
\usepackage{appendix}
\usepackage{float}
\usepackage{booktabs}     
\usepackage{tabularx} 
\usepackage{graphicx}
\usepackage{appendix}
\usepackage{url}
\usepackage{array}

\newcommand\feh{$\mathrm{[Fe/H]}$}
\newcommand\cmd{color-magnitude diagram}
\newcommand\EBV{\textit{E(B-V)}}
\newcommand\VminK{\textit{V}--\textit{K}}
\newcommand\VminI{\textit{V}--\textit{I}}
\newcommand\Vmag{\textit{V}}
\newcommand\Bmag{\textit{B}}
\newcommand\Imag{\textit{I}}
\newcommand\Jmag{\textit{J}}
\newcommand\Hmag{\textit{H}}
\newcommand\Kmag{\textit{K}}

\newcommand\Nvar{$N_{\rm m,var}$}
\newcommand\Nprim{$N_{\rm m,prim}$}
\newcommand\Kepmag{$K_{\mathrm{p}}$}
\newcommand\dmag{$\Delta{\rm mag}$}
\newcommand\msun{$\mathrm{M_{\odot}}$}
\newcommand\Pm{${P_{\mathrm{m}}~>~0.5}$}

\newcommand\dance{\texttt{DANCe}}

\newcommand\PMSF{$P_{\mathrm{MSF}}$}
\newcommand\Gdor{$\gamma$~Doradus}
\newcommand\dScuti{$\delta$~Scuti}
\newcommand\Modot{$\rm M_{\odot}$}
\newcommand\Rsun{$\rm R_{\odot}$}
\newcommand\Lsun{{$\rm L_{\odot}$}}
\newcommand\Teff{$T_{\mathrm{eff}}$}

\newcommand\SNpinkeq{{S/N}_{\rm pink}} 
\newcommand\SNpink{${S/N}_{\text{pink}}$} 

\newcommand\RJ{$\rm R_{J}$}
\newcommand\RpA{$0.35~\pm~0.04$}
\newcommand\KpA{15.4}
\newcommand\tdA{2.9}
\newcommand\PeriodA{1.256774}
\newcommand\tdurA{1.12}
\newcommand\RsA{$0.70~\pm~0.07$}
\newcommand\varpiA{$1.401~\pm~0.047$}
\newcommand\distA{$714~\pm~25$}
\newcommand\TeffA{$4896~\pm~216$}

\newcommand\RpAunits{\RpA{}~\RJ{}}
\newcommand\KpAunits{\Kepmag{}~$=$~\KpA{}~mag}
\newcommand\tdAunits{\tdA{}~millimag}
\newcommand\PeriodAunits{\PeriodA{}~days}
\newcommand\tdurAunits{0.05~days (1.12~hours)}
\newcommand\RsAunits{\RsA{}~\Rsun{}}
\newcommand\incAunits{$87~\pm~3 ^{\circ}$}
\newcommand\varpiAunits{$\varpi$~=~\varpiA{}}
\newcommand\distAunits{\distA{}~pc}
\newcommand\TeffAunits{\Teff{}~=~\TeffA~K}

\newcommand\RpB{$0.72~\pm~0.02$}
\newcommand\KpB{15.7}
\newcommand\PeriodB{15.253697}
\newcommand\tdB{2.2}
\newcommand\tdurB{0.77}
\newcommand\RsB{$1.65~\pm~0.04$}
\newcommand\varpiB{$0.46~\pm~ 0.06$}
\newcommand\distB{$2.2~\pm~0.3$}
\newcommand\TeffB{$3856~\pm~82$}

\newcommand\RpBunits{\RpB{}~\RJ{}}
\newcommand\KpBunits{\Kepmag{}$~=$~\KpB{}~mag}
\newcommand\tdBunits{\tdB{}~millimag}
\newcommand\RsBunits{\RsB{}~\Rsun{}}
\newcommand\incBunits{$87~\pm~3 ^{\circ}$}
\newcommand\PeriodBunits{\PeriodB{}~days}
\newcommand\tdurBunits{0.032~days (0.77~hours)}
\newcommand\TeffBunits{\Teff{}~=~\TeffB{}~K}
\newcommand{\totcat}{1,143}

\newcommand{\blendtot}{313}

\newcommand{\nblendtot}{830}

\newcommand{\tottransit}{two}
\newcommand{\tottransitf}{0.2}

\newcommand{\totEB}{44}
\newcommand{\totEBf}{4}
\newcommand{\EBM}{one}
\newcommand{\EBN}{five}

\newcommand{\totvar}{1,097}

\newcommand{\pulsating}{251}

\newcommand{\rotating}{561}

\newcommand{\MRot}{232}
\newcommand{\MPulse}{51}
\newcommand{\NRot}{16}
\newcommand{\NPulse}{13}
\newcommand{\totfield}{740}

\newcommand{\totM}{331}

\newcommand{\Mmvd}{274}

\newcommand{\Mdeblend}{274}
\newcommand{\Mblend}{57}

\newcommand{\VoneM}{10}
\newcommand{\VtwoM}{18}

\newcommand{\totN}{56}

\newcommand{\Nmvd}{34}

\newcommand{\Ndeblend}{34}
\newcommand{\Nblend}{22}

\newcommand{\VoneN}{13}
\newcommand{\VtwoN}{18}

\newcommand{\NGaiaM}{358}

\newcommand{\NBouy}{1,120}

\newcommand{\NKharM}{1,084}
\newcommand{\fKharM}{95}
\newcommand{\NGaiaN}{51}

\newcommand{\NKharN}{752}

\newcommand{\NDiasN}{72}

\newcommand{\fap}{0.4}
\newcommand{\newvar}{523}

\newcommand{\Vmax}{18.7}
\newcommand{\Vmin}{9.7}
\newcommand{\totGaiamatch}{409}

\newcommand{\MDScuti}{18}
\newcommand{\MGDor}{33}
\newcommand{\MMisc}{47}
\newcommand{\NDScuti}{two}
\newcommand{\NGDor}{11}
\newcommand{\NMisc}{22}
\newcommand{\leastPN}{52}
\newcommand{\mostPN}{85}
\newcommand{\MPrimMin}{9.7}
\newcommand{\MPrimMax}{17.2}
\newcommand{\MPrimPulseMax}{14.8}
\newcommand{\MPrimPulsePMax}{3.3}
\newcommand{\MPrimRotPMin}{0.4}
\newcommand{\MPrimRotPMax}{10.6}

\newcommand{\ratioa}{74}
\newcommand{\ratiob}{21}
\newcommand{\totMb}{245}
\newcommand{\totNb}{12}
\newcommand{\Mmvdb}{203}
\newcommand{\Nmvdb}{3}

\shortauthors{Soares-Furtado et al.}
\begin{document}
\title{A Catalog of Periodic Variables in Open Clusters M35 \& NGC~2158}

\correspondingauthor{Melinda Soares-Furtado}
\email{msoares@princeton.edu}

\author[0000-0001-7493-7419]{M. Soares-Furtado}
\altaffiliation{National Science Foundation Graduate Research Fellow}
\affiliation{Department of Astrophysical Sciences, Princeton University, Princeton, NJ 08544, USA}

\author[0000-0001-8732-6166]{J.~D.~Hartman}
\affiliation{Department of Astrophysical Sciences, Princeton University, Princeton, NJ 08544, USA}

\author[0000-0002-0628-0088]{W.~Bhatti}
\affiliation{Department of Astrophysical Sciences, Princeton University, Princeton, NJ 08544, USA}

\author[0000-0002-0514-5538]{L.~G.~Bouma}
\affiliation{Department of Astrophysical Sciences, Princeton University, Princeton, NJ 08544, USA}

\author[0000-0002-4843-345X]{T.~Barna}
\affiliation{Department of Physics \& Astronomy, Rutgers University, 136 Frelinghuysen Rd., Piscataway, NJ 08854}

\author[0000-0001-7204-6727]{G.~\'A.~Bakos}
\altaffiliation{Packard Fellow}
\altaffiliation{Alfred P. Sloan Research Fellow}
\affiliation{Department of Astrophysical Sciences, Princeton University, Princeton, NJ 08544, USA}

\begin{abstract}
We present a catalog of \totcat{}  periodic variables, compiled from our image-subtracted photometric analysis of the K2 Campaign-0 super stamp. This super stamp is centered on the open clusters M35 and NGC~2158. 
Approximately 46\% of our periodic variables were previously unreported. 
Of the catalog variables, we find that \totM{} are members of M35 and \totN{} are members of NGC~2158 (\Pm{}).
Our catalog contains \tottransit{} new transiting exoplanet candidates, both of which orbit field stars.
The smaller planet candidate has a radius of \RpAunits{} and orbits a K dwarf (\KpAunits{}) with a transit depth of \tdAunits{}. 
The larger planet candidate has a radius of \RpBunits{} and orbits a late G type star (\KpBunits{}) with a transit depth of \tdBunits{}. 
The larger planet candidate may be an unresolved binary or a false alarm. 
Our catalog includes \totEB{} eclipsing binaries, including ten new detections. 
Of the eclipsing binaries, one is an M35 member and \EBN{} are NGC~2158 members.  
Our catalog contains a total of \totvar{} non-transiting variable stars, including a field $\delta$~Cepheid exhibiting double mode pulsations, \rotating{} rotational variables, and \pulsating{} pulsating variables (primarily \Gdor{} and \dScuti{} types).
The periods of our catalog sources range between 43 minutes to 24 days. 
The known ages of our reported cluster variables will facilitate investigations of a variety of stellar evolutionary processes.
\end{abstract}

\section{Introduction}
\label{sec:introduction}

In 2017, we released 3,960 light curves \citep{SoaresFurtado2017} for stars of \Vmin{}~mag~$<$~\Vmag{}~$<$~\Vmax{}~mag in the K2 Campaign-0 (hereafter K2C0) super stamp.
The super stamp is centered on the open clusters Messier 35 (NGC~2168, hereafter M35) and NGC~2158. 
We obtained high-precision photometry with reduced blending by developing an image subtraction pipeline tailored to address the K2-specific mission systematics. 
With only two operating gyroscopic reaction wheels, the K2 systematics are dominated by the low-frequency motion that is induced by solar pressure and the subsequent thruster firings that are intended to correct for the drift of the spacecraft. Sources drift across two to three pixels throughout each cycle.
Our K2 image subtraction pipeline methodology is discussed in-depth by \cite{Huang2015}.

Image subtraction facilitates the analysis of time-series data in densely populated regions, such as crowded stellar cluster fields, particularly in regions near the cluster cores.
When compared with other methods used to examine this same dataset, our TFA-corrected light curves were found to have marginally higher photometric precision across all the timescales investigated. 
The K2C0 light curves are hosted at \url{http://k2.hatsurveys.org/archive/}. 
In addition to the K2C0 analysis, this method has been extended to analyze other K2 regions. 
\cite{Huang2015} performed an analysis of nearly 23,000 sources in the K2 Campaign-1 field.  
It has been used to extract high-precision microlensing signals near the Galactic bulge from the K2 Campaign-9 dataset \citep{Zhu2017,Zhu2017b}.  
The method has also been applied to the globular cluster M4, resulting in the detection of a new variable class known as the millimagnitude RR Lyrae \citep{Wallace2019,Wallace2019b}.

Despite the challenges due to crowding, variability searches in cluster fields have led to some important discoveries.
Much of the recent success was made possible by the cluster time-series data provided by the Kepler and K2 transit surveys.
Among the clusters observed are the Hyades, the Pleiades, M4, M18, M21, M25, M35, M44 (Praesepe), M45, M67, M80, NGC~1647, NGC~2158, NGC~6717, NGC~6774 (Ruprecht~147), NGC~6791, NGC~6811, NGC~6819, and NGC~6866. 
To date, the transit search technique has revealed ${\sim}10$ confirmed exoplanets orbiting stars in open clusters \citep[and references therein]{David2018}. 
The first exoplanet detected in an open cluster by the transit method was found using Kepler data less than six years ago \citep{Meibom2013}.
The majority of cluster exoplanet candidates have been revealed by the K2 survey \citep[and references therein]{Ciardi2018}.
It is likely that a census analysis of exoplanets found in stellar clusters would shed light on the underlying physical processes driving planetary formation and evolution, particularly since these systems span a wide age and metallicity range. 
Determining the planet occurrence rates and planet architectures within clusters would have far-reaching significance, as all stars, and therefore all planetary systems, likely form in clustered environments \citep[e.g.][]{Portegies2010}.
To paint a complete picture of the pertinent physical mechanisms at play, more cluster planet detections are required, as the majority have been found among solar-like stars in the highly-irradiated regime. 

There is also value in the detection and characterization of eclipsing binaries (EBs) in stellar clusters.
The precise radii measurements of evolved EBs provides a useful age estimate for stellar clusters, offering tight constraints on both the distance and reddening of cluster members. 
\cite{Sandquist2013} demonstrated this technique by using Kepler observations of the EB system WOCS~23009 to constrain the age of NGC~6819. 
This resulted in an age estimate of $2.62~\pm~0.25$~Gyr with stellar model physics responsible for systematic uncertainties at the level of $10\%$. 
This age estimate was further constrained to unprecedented precision when the number of detected EBs in the system substantially increased \citep{Brewer2016}.

M35 is a compelling region to search for periodic variability, as the stellar constituents provide a significant contribution to the high-mass end of the initial-final mass relation (IFMR)
\citep{Williams2009}. These stars are important probes for our models of stellar evolution. They represent the lower mass limit for the transition between the class of stars that will evolve to produce white dwarfs and those that will end their lives as violent supernovae.
Detecting EBs in M35 not only has the potential to enhance the precision of the high-mass end of the initial-final mass relation for white dwarfs, it may also reduce uncertainties in the masses of progenitor stars within the white dwarf-neutron star transition. 
Section~\ref{sec:EBs} describes the cluster-affiliated EBs in our dataset.

Detections of Cepheids and RR Lyrae stars are also important, as these sources map the substructures and stellar populations of the Milky Way.  
Over the course of its mission, Kepler observations significantly increased the number of Cepheid and RR Lyrae detections, bringing the counts into the hundreds and thousands, respectively \citep{Molnar2018}. 
While Cepheid variables are less common than RR Lyrae stars, their greater luminosity makes them easier to detect.
This often permits their detection in regions where blending has not been mitigated.
RR Lyrae stars, in contrast, often require blending mitigation techniques in order to be detected. 

Stellar rotation rates can be obtained from the time series analysis of photometric variations produced by rotating starspots \citep[e.g.][]{Irwin2009}. 
Rotational variables found in stellar clusters are particularly important probes to test stellar evolutionary models. 
The field of gyrochronology was built upon the work of \cite{Skumanich1972} whereby the projected stellar rotational velocities ($v \sin{i}$) among stars within the Hyades and Pleiades clusters were shown to decrease as the square root of the star's age.  
For some main sequence (MS) stars, gyrochronology is the most precise chronometer available \citep[e.g.][]{Epstein2014}. 
The rotation periods of late MS stars are also dependent upon mass \citep[e.g.][]{Barnes2007,Irwin2009,Meibom2009,Meibom2011,Meibom2015} and much work has been done to model the empirical relations between these stellar parameters \citep[e.g.][]{Kawaler1989,Barnes2003,Barnes2007,Barnes2010,Mamajek2008,Angus2015}, as well as to probe the underlying physical processes driving these relations \citep[e.g.][]{Reiners2012,Gallet2013}. 
Young coeval MS stellar populations have been shown to display rotation rates that range by two orders of magnitude \citep[e.g.][]{Irwin2008,Irwin2009,Hartman2010}. 
As a cluster ages, the stellar rotation rates quickly converge \citep[e.g.][]{Hartman2009,Meibom2011}.
It has been proposed that the mechanism driving this convergence is the loss of angular momentum via stellar winds \citep{Weber1967}, however, there are still many associated uncertainties with the corresponding models.
To effectively probe theoretical models of angular momentum evolution stellar rotation rates must be observed across a wide range of stellar ages and masses. 

To contribute to these efforts, we used data from the K2 transit survey to search open cluster systems for the presence of planets, eclipsing binaries, rotational variables, and pulsational variables.
In this paper, we present the results of a periodic variability search performed on the sources in the K2C0 super stamp. We assign cluster membership probabilities to our identified variables, compare our sources to the results from prior searches conducted in this field, discuss the newly identified variables, and make our results publicly available.  
Section~\ref{sec:clusters} provides some background information for the K2C0 open clusters. 
The K2C0 photometric observations are described in Section~\ref{subsec:data}.
Section~\ref{sec:cull} describes the technical steps involved in our assimilation of the periodic variable catalog. 
Section~\ref{sec:census} summarizes the contents of our variable catalog. 
This section includes relevant catalog census information, such as the distributions of stellar type, variable class, and cluster membership. 
We also discuss the period--magnitude relation obtained for M35 rotational variables.
Section~\ref{sec:variabilitysearch} compares our variable catalog results to the results generated by prior searches conducted within this field. 
Section~\ref{sec:newvar} describes the new variables that were found in our search.
In Section~\ref{sec:exo} we review our candidate transiting exoplanet detections.
We discuss our EB detections in Section~\ref{sec:EBs}.
In Section~\ref{subsec:Cepheid} we discuss our analysis of the double-mode $\delta$~Cepheid variable, V0371 Gem.
In Section~\ref{sec:RotVarPulse}, we provide example light curves for a subset of rotational variables, \Gdor{} variables, and \dScuti{} variables. 
In Section~\ref{sec:summary} we summarize our findings.

\section{The K2C0 Open Clusters \& Data} 
\label{sec:clusters}
\subsection{M35}
\label{subsec:M35}
With an age of ${\sim}150$~Myr \citep{Meibom2009}, M35 is a critical system to probe for exoplanets, as planets undergo rapid evolutionary changes during the first few hundred million years after formation \citep{Adams2006}. 
Such a cluster is a useful system to test theories of planetary formation and planet migration timescales. 
The angular diameter of the cluster is $0.5^{\circ}$ and the stars are well dispersed relative to other open clusters.
This reduces source confusion and permits multi-object spectroscopy follow-up.
Crowding is a concern near the core of the cluster.
\cite{Cantat2018} (hereafter CG18) used Gaia DR2 data to infer a distance of $861$~pc to the cluster.
The inferred reddening for this system is estimated to be $\EBV{}=0.255~\pm~0.024$~mag \citep{Sung1999}. 

The stars comprising the M35 MS span ${\sim}14$ magnitudes in brightness in the \Vmag{} vs.\ \textit{B-V} plane.
The hottest M35 members are classified as B3 in spectral type \citep{Kalirai2003}. 
Using the overshooting models of \cite{Girardi2000}, the cluster main sequence turn-off (MSTO) mass has been estimated as ${\sim}3.75$~\Modot{}, which indicates that the age of M35 is similar to  that of the Pleiades (M35 is ${\sim}150$~Myr, while the Pleiades is ${\sim}125$~Myr) \citep{Vidal1973}.
In contrast to the Pleiades, however, M35 is rather metal-poor with a  metallicity of \feh{}~$=-0.21~\pm~0.10$ \citep{Barrado2001}. 

The median proper motion of the cluster is measured as $(\mu_{\alpha} \cos \delta,\mu_{\delta})=(2.308,-2.905)\pm(0.239,0.235)$~mas/yr \citep{Cantat2018}.
Due to this low proper motion, as well as a low Galactic latitude, there has been much discrepancy regarding the number of cluster members.
Early estimates determined ${\sim}500$ stellar members \citep{Cudworth1971}. 
A counting analysis of MS members performed 30 years later found ${\sim}1,700$ cluster members \citep{Barrado2001}.  
The most recent estimate, determined by CG18, found 1,325 stellar members (\Pm{}). 
Unfortunately, this study did not include the most radially distant members within the cluster (as measured by the 2D projected distance from the cluster core).
In an effort to ensure completeness in our M35 membership analysis, we employed results from multiple membership catalogs. These catalogs are described in Section~\ref{sec:membership}.

\subsection{NGC~2158}
\label{subsec:N2158}
Separated from M35 by $25{\arcmin}$, the angular diameter of NGC~2158 is $5{\arcmin}$.
Severe blending is a concern for this densely packed cluster, which lies at a distance of $4.5$~kpc \citep{Cantat2018}.
The system was actually considered to be a globular cluster until \cite{Shapley1930} carefully investigated individual sources. 
With an age estimate of $2$~Gyr, NGC~2158 is too young to be classified as a globular cluster \citep{Carraro2002} and has been deemed a member of the old thin disk population \citep{Jacobson2009}. 
The age is old enough for the open cluster to have undergone mass segregation.

The light from the cluster is dominated by an underlying population of yellow stars with the hottest members classified as F0 in spectral type. The cluster MSTO mass is estimated at ${\sim}1.3$~\msun{}, which was determined using the cluster age and a Z~=~0.0048 isochrone \citep{Carraro2002}.
Despite the fact that NGC~2158 is much older than M35, it is richer in metals (\feh{}$~=-0.03~\pm~0.14$). 
The inferred reddening for the cluster is estimated at $\EBV{}~=~0.55~\pm~0.10$~mag \citep{Carraro2002}.

\citep{Cantat2018} estimated a cluster proper motion of $(\mu_{\alpha} \cos \delta,\mu_{\delta})=(-0.177,-2.002)\pm(0.185,0.173)$~mas/yr. 
While NGC~2158 is one of the most populous open clusters, its low proper motion poses a formidable challenge to cluster membership analysis.
In an effort to ensure completeness in our cluster membership analysis, we employed multiple membership catalogs. 
These catalogs were produced using proper motions as well as multi-wavelength photometry (optical and near-infrared). 
Given the spheroidal shape of this cluster, we incorporated a King model fit to weight the associated membership probabilities. 
Our cluster membership analysis is described further in Section~\ref{sec:membership}.

\subsection{The K2C0 Dataset \& Processing}
\label{subsec:data}
The full K2 field of view (FOV) is $115$ square degrees and comprises an angular diameter of six degrees on the sky. 
The FOV is produced by the combination of 21 CCD modules, each of which is composed of two $2200\times 1024$ pixel CCDs.  
The entirety of the K2C0 super stamp lies on a single module. 
The K2C0 super stamp is an aggregate of 154 postage stamp images that are toggled across a field centered on the open clusters M35 and NGC~2158.
The super stamp is $0.5 ^{\circ}$ in angular diameter (about the size of M35). 
Each individual postage stamp is $50~\times~50$ pixels in size. 
Both the presence of the open clusters and the proximity to the dense Galactic anti-center results in significant crowding. 

The K2C0 field was observed from March--May 2014.
The telescope was mistakenly placed in coarse-pointing tracking mode during the first half of the campaign.
Our analysis of the K2C0 super stamp solely employed cadences captured when the instrument tracking was set to the fine-pointing mode, which reduced our total number of cadences to 1,551 (beginning with long cadence number 89347).
Each individual cadence observation consists of a 29-minute integrated exposure and these data span an observation window of 31~days. 
The K2 survey never returned to this same field in subsequent campaigns.
We examined stars in the K2C0 field as faint as \Vmag{}~=~18.7~mag, corresponding to a Kepler magnitude of \Kepmag{}~$=16$~mag. 
To perform source extraction we utilized \textit{The Fourth U.S. Naval Observatory CCD Astrograph Catalog} (UCAC4) and \textit{The Ecliptic Plane Input Catalog} (EPIC).
UCAC4 is supplemented by the \textit{Two Micron All-Sky Survey} (2MASS), which provides photometric data for 110 million stars and the AAVSO Photometric All-Sky Survey (APASS), which provides photometric data for over 50 million stars. 
We applied an image subtraction process to the data, using the technique outlined by \cite{Alard1998,Alard2000}.

This process is summarized by the following sequence of steps: 
\begin{itemize}
\item[1.] For each cadence, the target pixel frames (TPFs) from our module of interest are assembled into a super stamp. 
\vspace{-1.0mm}
\item[2.] For each super stamp cadence, we perform source extraction using the UCAC4 star catalog. We determine the astrometric transformation between UCAC4 sources and the extracted K2 sources.
\vspace{-1.0mm}
\item[3.] A sharp super stamp image with median directional pointing is selected. We call this the ``astrometric reference frame."
\vspace{-1.0mm}
\item[4.] All cadences are spatially transformed to a common coordinate frame. This step minimizes spacecraft drift and allows for a more accurate model of the instrument's motion in our detrending procedure.
\vspace{-1.0mm}
\item[5.] A stacked median average of all the translated K2C0 frames are generated. We call this the ``master photometric reference frame."
\vspace{-1.0mm}
\item[6.] Each cadence is subtracted from the master photometric reference frame, resulting in a variable field with reduced blending.
\vspace{-1.0mm}
\item[7.] Photometry is performed on the sources in each image-subtracted cadence.
\vspace{-1.0mm}
\item[8.] The light curves are assembled for all sources.
\vspace{-1.0mm}
\item[9.] A high-pass filter with one-day binning is applied to the data. The decorrelation procedure modeled by \cite{Vanderburg2014} is applied.
\vspace{-1.0mm}
\item[10.] The TFA-correction as outlined by \cite{Kovacs2005} is applied to remove further systematics.
\end{itemize}
Our data reduction procedure is described in more detail in \cite{Huang2015} and the detrending procedure is described in more detail in \cite{SoaresFurtado2017}. 

\section{From K2C0 Light Curves to Catalog Variables} \label{sec:cull}
From our initial set of 3,960 image-subtracted K2C0 sources, \totcat{} were identified as periodic variables and have been added to our catalog.
A digital version of the catalog has been made publicly available and is hosted online in a flexible format at \url{https://k2.hatsurveys.org/archive/}.
The sources excluded from our catalog were either found not to vary periodically within our investigated period bounds ($0.03$~days to $31$~days) or they were identified as secondary blends. 
We describe our procedure for identifying the primary variable source within a blended group in Section~\ref{subsec:deblend}.

\subsection{Periodogram Searches}
\label{subssec:periodogram}
To detect periodic variables within our nonuniformly-sampled, image-subtracted photometry, we employed three distinct periodogram searches. 
In our search for transit candidates, which includes detached stellar EB systems and transiting exoplanets, we used the Box-fitting Least Squares transit search routine \citep[BLS;][]{Kovacs2002}.
To detect and classify variability that is more sinusoidal in nature, we used the generalized Lomb-Scargle (LS) algorithm \citep{Zechmeister2009,Scargle1982} and the phase dispersion minimization (PDM) algorithm \citep{Stellingwerf1978}. 

We ran period searches on our 3,960 light curves (LCs) using the Python module \texttt{Astrobase v0.2.2} \citep{Bhatti2018} and the command line utility \texttt{VARTOOLS} \citep{Hartman2016}. 
The search results were reviewed both by eye and by implementing automated cut-off metrics, which are described in Sections~\ref{bls}--\ref{pdm}.
Two observers looked through the entire set, generating a culled subset of variables. 
When the variable classification or period designation was unclear, the LCs were flagged and reviewed by two additional observers. 
We include comments in the digital catalog to address ambiguous sources.

\subsubsection{BLS Periodogram Parameters}
\label{bls}
Using \texttt{VARTOOLS}, we employed the BLS algorithm search on our time-series data to reveal LC sources with periodic `box'-shaped features. 
The algorithm relied on two variables: (1) the span of the observation window for a given source and (2) the mean density estimate of the target star, which is directly related to observable parameters. 
The span of the observation window was generally $T_{\mathrm{span}}~=31$~days, however, a subset of sources near the edges of the super stamp were sampled over shorter intervals.
The stellar mean density was determined using the star's \VminK{} color with the assumption that the source was a zero-age main sequence star with a metallicity of \feh{}~=~0.0.
Using density estimates, we performed a bi-linear interpolation of the tabulated stellar isochrone models generated by \cite{Yi2001}. 

The minimum period threshold was set to $a/R_{\star}>1$, where $a/R_{\star}$ is the ratio of the orbital separation to the size of the stellar radius.  
Recalling the relation
\begin{equation}
\frac{a}{R_{\star}}=\bigg(\frac{G T^2 \rho}{3 \pi}\bigg)^{\frac{1}{3}} > 1,
\end{equation}
we set a minimum search period of 
\begin{equation}
T_{\rm min}=\sqrt{3\pi/(\rho G)},
\end{equation}
where $G$ is the gravitational constant and $\rho$ is the mean density of the host star. 
This relation relies on the assumption that the mass of the transiting object is negligible when compared to the mass of the host star.

To determine the number of phase bins needed, we computed $q$, the transit duration in units of phase, for the case of a transiting object on a circular, edge-on orbit.
The stellar density is used in this implementation of the BLS search to determine the range of transit duration values to search at a given period.
The minimum transit duration in phase at the longest period, $q_{\mathrm{min}}$, is equal to $1/2q$. 
This can be represented as
\begin{equation}
q_{\mathrm{min}}=0.5q=0.5\bigg(\pi \frac{a}{R_{\star}}\bigg)^{-1}.
\end{equation}
The number of bins, $n_{\rm bin}$, is calculated as
\begin{equation}
n_{\rm bin}=\frac{2}{q_{\mathrm{min}}}=\bigg(\frac{64\pi^2 G \rho T_{\mathrm{span}}}{3}\bigg)^{1/3}
\end{equation}
and the frequency step-size is given by the relation
\begin{equation}
\Delta f=\frac{0.25}{ (T_{\mathrm{span}}) (q_{\rm min})}.
\end{equation}
For reference, some typical numbers for these parameters are $n_{\rm bin}=850$ and $\Delta f=2\times~10^{-10}$~s$^{-1}$.
As a careful secondary check, we repeated our BLS search using the \texttt{Astrobase} platform.
For this BLS search, we imposed a fixed frequency range of $3.85\times 10^{-7}$ to $2.3\times 10^{-4}$~s$^{-1}$, a uniform frequency step-size of $\Delta f=5.0\times 10^{-4}$ s$^{-1}$, and fixed minimum and maximum in-phase transit durations of $q_{\rm min}=0.01$ and $q_{\rm max}=0.8$.

All BLS results were reviewed by eye for characteristic box-shaped dimming (constant in depth and flat-bottomed). The period, phase, and transit durations were inspected. 
We also checked LCs for the presence of a secondary eclipse that is indicative of an EB. 
To find contact binaries, the corresponding LS and PDM periodograms were checked for the characteristic arc-shaped LC signature that is indicative of W~Ursae Majoris variables (low mass contact binaries). 
We employed a signal-to-noise metric, \SNpink{}, measured using \texttt{VARTOOLS} and defined by \cite{Pont2006} as
\begin{equation}
\SNpinkeq{}=\sqrt{\frac{\delta_{m}^2}{(\sigma_{w}^2/n_{t})+(\sigma_{r}^2/N_{t})}},
\label{eqC}
\end{equation}
where $n_{t}$ is the number of data points present in the transit and $N_{t}$ is the number transits sampled. 
The transit depth is given by $\delta{m}$, which provides a proxy of the signal. The red noise of the light curve is represented by $\sigma_{r}$. This is calculated after subtracting the transit model, binning residuals in time with bins equal in size to the transit duration, and then taking the standard deviation. 
The variable $\sigma_{w}$ represents the root mean square (rms) scatter of the un-binned LC after subtracting the transit model (generally referred to as the `white noise').
Instead of imposing a firm cut-off for the \SNpink{} metric, which would eliminate real transiting sources with highly correlated noise and outliers, we use the threshold \SNpink{$<17$} as a flag for a more in-depth investigation. 
The catalog variables with low \SNpink{} metrics were deemed as compelling variable sources despite this low metric after an inspection of the LC by eye.
For reference, \SNpink{$=36$} is the median value among transiting sources in our catalog.

\subsubsection{LS Periodogram Parameters} 
\label{ls}

To unveil rotating and pulsating variable stars in the K2C0 LC sample, we executed a generalized Lomb-Scargle (LS) periodogram search \citep{Zechmeister2009} with \texttt{VARTOOLS}. 
This search identified the five highest amplitude peaks in the power spectrum. 
We applied a fixed frequency range of $3.85\times~10^{-7}$~s$^{-1}$ to $3.85\times~10^{-4}$~s$^{-1}$ and a uniform frequency step-size of 
\begin{equation}
\Delta f=\frac{sb}{ T_{\mathrm{span}}}=3.73 \times 10^{-9} \rm \ s^{-1}.
\end{equation}
The variable $sb$ represents the subsample, which was set to $sb=0.01$. 
The observation time window was set to $T_{\mathrm{span}} =31$~days. 
Also calculated was the VARTOOLS-derived formal false alarm probability, log($FAP$), and the signal-to-noise ratio for each of the identified peaks, $S/N$. 
Our threshold was set to $S/N>7$, where
$S/N$ is given by the equation
\begin{equation}
S/N = \frac{LS - \overline{LS}}{\sigma_{LS}},
\end{equation}
where 
\begin{equation}
LS = \frac{\chi_{0}^2 - \chi(f)^2}{\chi_{0}^2}.
\end{equation}
Here, $\chi_{0}^2$ is the value of $\chi^2$ about the weighted mean and $\chi(f)^2$ is the value of $\chi^2$ about the best-fit sinusoidal signal with frequency $f$. 
All the identified peak periods are whitened and the periodograms are then recomputed before searching for subsequent peak periods. 
Sources with high log($FAP$) values were flagged as suspicious and carefully investigated by eye. 
Only four catalog sources have a log($FAP$) $>1$, making up less than \fap{$\%$} of our catalog. 
For comparison, we performed a generalized Lomb-Scargle periodogram search using the \texttt{Astrobase} on all LCs using a fixed frequency range of $3.73\times~10^{-7}$~s$^{-1}$ to $3.85\times~10^{-4}$~s$^{-1}$ and the \texttt{Astrobase} default uniform frequency step-size of $\Delta f=1.0\times 10^{-4} \rm \ s^{-1}$. 

\subsubsection{PDM Periodogram Parameters} 
\label{pdm}
Using \texttt{Astrobase}, we performed a PDM periodogram search with a frequency range of $3.73\times~10^{-7}$~s$^{-1}$ to $3.8\times~10^{-4}$~s$^{-1}$ and a uniform frequency step-size of $\Delta f=~1.0\times 10^{-4}$~s$^{-1}$.
The results of the LS and PDM periodograms were carefully compared by eye to discern the most accurate period for each target. 
The best selected period is listed in \textit{Column 6~(P)} of our variable catalog. 

\subsection{Identifying Primary Variables in Blended Groups} 
\label{subsec:deblend}
After the identification of periodic sources in the K2C0 super stamp, we sought to mitigate any degeneracies in variability ---~these are commonly referred to as \textit{blends}. 
We identify if a given target source is blended by comparing the top five periods identified by the power spectrum, as well as the associated harmonics, to those corresponding to the neighboring sources within a 10-pixel radius (equivalent to $40\arcsec$). 
Some stars contained as many as $50$ blended neighbors. 
If no blends were found at this initial step, the target was identified as a conclusive primary. 
The optimal method to distinguish a primary source within a blended group is dependent upon the variable classification of a given source. 
As a first step, we distinguished by eye whether the target LC either (a) displayed the characteristic box-shaped dimming indicative of a transit event or (b) displayed sinusoidal variations indicative of a pulsating, rotating, or erupting variable. 

To distinguish a primary source among the candidate EBs and transiting exoplanets, we compared the transit depth, transit duration, and the \SNpink{} measurement among the members of a blended group. 
Primary sources were designated as those with the following characteristics: (a) the deepest transit depth among sources in the blended group, (b) a sensible transit duration (given the corresponding color information), and (c) the strongest \SNpink{} among sources in the blended group. 
While the third criterion was generally true for our designated blended primaries, this was the least significant criterion in our selection process.  

To identify the primary source in a set of pulsating and rotating variables, we fit an order-3 Fourier series to the differential flux time series for each of the five periods identified in the peak power spectrum. 
This fit was performed on all sources within the blended group.  
We then compared the corresponding Fourier fit amplitudes. 
Primary sources were designated as those with the largest Fourier fit amplitude. 
If the amplitude difference between the target and any member of the blended group was $< 0.4$ millimag, the target was labeled as an ambiguous blend.
We list the blend status for identified variables in \textit{Column 10~(Blend)} of our catalog.
The distribution of identified primaries and ambiguous blends is discussed in Section~\ref{sec:deblenddist}.

\subsection{Periodic Variable Source Classification} 
The \totcat{} periodic variables in our culled catalog were initially separated into two distinct groups: (1) those displaying box-like dips in their LCs (indicative of an EB or exoplanet transit candidate) and (2) those with sinusoidal LC signatures indicative of rotating or pulsating variables.

Sources classified as transiting exoplanet candidates are those with transit depths, transit durations, and orbital separations that are consistent with a star-planet system.
To mitigate false positives among the transiting exoplanet candidates, we incorporated spectroscopic measurements, Gaia parallaxes, stellar density estimates, and relevant cluster parameters (such as age and metallicity). 
Sources classified as candidate transiting exoplanets were labeled with the class identifier \textit{Transit}, also found in \textit{Column~8~(Class)} in our catalog. We discuss our exoplanet candidate sources in Section~\ref{sec:exo}.

To distinguish an EB from an exoplanet transit candidate, we searched for signatures of a secondary eclipse among the phase-folded LCs.  
This secondary eclipse is reminiscent of an EB system, as transiting exoplanet candidates do not exhibit two distinct minima. 
Also added to the EB class were sources with characteristic arc-shaped LC signatures indicative of W~Ursae Majoris variables (low mass contact binaries). 
Sources classified as EBs were labeled with the class identifier \textit{EB} in \textit{Column~8~(Class)} of our variable catalog. 
We discuss our eclipsing binary sources in Section~\ref{sec:EBs}.

Non-transiting variable sources were examined for characteristic LC signatures produced from stellar pulsations, rotation, and/or eruptions. 
Sources were classified using the following parameters: (a) the source period, (b) the shape of the LC, (c) the photometric variability amplitude as measured from the Fourier series fit to the differential magnitude (as described in Section~\ref{subsec:deblend}, except, in this case, we used a Fourier fit on the magnitude rather than the flux), and (e) the Gaia DR2 derived parameters of temperature, radius, and luminosity. 
The reddened-corrected $J-K$ difference values were not used in our classification given the high associated uncertainties.
All the sources designated as rotational variables were checked to ensure that (a) the corresponding rotational velocities did not exceed the stellar break-up speed and (b) the effective temperature does not exceed the expected temperature ceiling of $T_{\rm eff}=6250$~K, above which surface magnetic activity is unexpected among dwarf stars. Sources that violate these conditions are inspected for reclassification and designated as \textit{Misc} if a distinct class cannot be identified.
In \textit{Column 6~(P)} of our variable catalog, we list the source periods used for variable classification.
Most often, this period corresponds to the frequency responsible for the peak maximum in the LS power spectrum. 
There are cases where the PDM identified period was selected or lower amplitude peaks were determined to be responsible for the photometric variations.
The \textit{Column 71~(Comment)} of our variable catalog indicates when a period other than the LS power spectrum peak maximum was selected.

We created a variable classification pipeline using the relevant classes listed in \textit{The AAVSO International Variable Star Index} (VSX) \citep{Watson2006,Watson2017} and in \cite{Karttunen2017}. 
Given our sample period and \dmag{} bounds, our classification pipeline searched for nondescript rotational variables and pulsators of the following types: \dScuti{} (also known as dwarf Cepheids), RR Lyrae, \Gdor{}, Cepheids, $\beta$~Cepheids, and slowly pulsating B-type variables. 
Some of these classes were omitted entirely, as they were outside the period bounds of our sample, such as rapidly oscillating Ap stars and Mira variables.
The pipeline procedure is outlined below.
\begin{itemize}
\item[1.] We begin with the assumption that all the listed periodic variable classes within our period and \dmag{} amplitude bounds are possible for a given target star. 
These periodic variable classes are taken from the VSX, as they were listed in 2015, and from the classes listed in Table~14.1 of \cite{Karttunen2017}. 
\vspace{-1.0mm}
\item[2.] We rule out variable classes in a sequence of steps. 
In our first elimination round, we compare the amplitude calculated from the Fourier series fit to the differential magnitude to the \dmag{} bounds corresponding to each of the potential variable classes.
At this step, classes that do not encompass the target amplitude are eliminated. 
The rotating class cannot be eliminated at this step given its wide range of brightness variations.
\vspace{-1.0mm}
\item[3.] We then determine if the period of the target is compatible with the period bounds for each of the remaining variable class categories.
Classes that do not include the designated source period are eliminated at this step.
\vspace{-1.0mm}
\item[4.] We employ the associated Gaia-derived luminosity and temperature (and the corresponding errors) to determine if further classes may be ruled out. 
Source temperatures are used to see if the target is outside the instability strip ($\pm \log_{10}T_{\rm eff} = 0.5$~K), which would rule out pulsational variability.
\vspace{-1.0mm}
\item[5.] The source is labeled as \textit{Misc} if either all the classification categories have been eliminated or if multiple, indistinguishable classifications subsist.
\vspace{-1.0mm}
\item[6.] We then compare our classification designations to those of preexisting variable catalogs. 
\end{itemize}

More massive stars ($M_{\star}>2$~\Modot{}) evolving off the MS will cross the instability strip at higher luminosities in the HR~Diagram. 
The high mass and luminosity of these stars result in longer periods than the average \dScuti{} star, which can reach up to 1~day \citep[e.g.][]{Buchler2007,Buchler2008, Smolec2010}. 
This is taken into consideration when we review the classification designation of a source. 

Two blended sources that share a common periodic signature do not necessarily share the same variable classification. 
This is because there may be differences in their corresponding stellar classification.
The results of our variable classification procedure are discussed in Section~\ref{sec:cdist}.

\subsection{Determining Cluster Membership}\label{sec:membership}
\subsubsection{M35 Membership}
In order to determine M35 cluster membership probabilities for as many of our catalog variables as possible, it was necessary to employ multiple cluster membership catalogs. 
The cluster membership catalogs and the matching procedures are outlined in Sections~\ref{subsubsec:cat1}--\ref{subsubsec:cat4}.
In order of precedence, the catalogs employed included (1) the CG18 Gaia DR2 membership catalog, (2) the \cite{Bouy2015} multi-epoch \dance{} catalog (hereafter \dance{}), and (3) the \cite{Kharchenko2013} membership catalog (hereafter K13). 
If an M35 membership probability could not be found in the CG18 Gaia DR2 catalog, we searched for a match in the \dance{} catalog. 
If the source remained unmatched, we performed a final search through the K13 catalog.
The CG18 membership probability catalog concentrated on more central regions of the cluster and excluded many sources in the \dance{} catalog, which provided the bulk of our listed M35 membership probabilities.
The K13 catalog provided M35 membership probabilities for four sources that were excluded from both the CG18 and \dance{} catalogs. 
A total of 16 catalog sources remained unmatched after searching each of the three catalogs.
In Table~\ref{Table:membcountA}, we provide the size of each of the cluster membership catalogs, as well as the number of source matches found.
The membership probabilities are listed in \textit{Column 64~(PM35)} of the variable catalog. 

\begin{table}
\centering
\begin{tabularx}{0.23\textwidth}{lrrr}
\toprule
\multicolumn{3}{c}{Cluster Membership} \\
\multicolumn{3}{c}{Probability Catalogs} \\
\toprule
\multicolumn{3}{c}{M35} \\
\toprule
Catalog &  $N_{\mathrm{cat}}$ & $N_{\mathrm{m,LC}}$  \\
\midrule
CG18 & 1,705 & \NGaiaM{} \\
\dance{} & 338,392 & \NBouy{} \\
K13 & 100,801 & \NKharM{} \\
\toprule
\multicolumn{3}{c}{NGC~2158} \\
\toprule
Catalog & $N_{\mathrm{cat}}$ & $N_{\mathrm{m,LC}}$\\
\midrule
CG18 & 1,633 & \NGaiaN{} \\
K13 & 20,640 & \NKharN{} \\
D06 & 330 & \NDiasN{} \\
\bottomrule
\end{tabularx}
\caption{The cluster membership probability catalogs used in our membership analysis of M35 \& NGC~2158.
The total number of sources in each catalog is given by $N_{\mathrm{cat}}$. 
The total number of matches found when comparing our variables to those listed in the catalogs is given by $N_{\mathrm{m,LC}}$.
The catalogs are listed in descending order of precedence. 
}
\label{Table:membcountA} 
\end{table} 

\subsubsection{NGC~2158 Membership}

\begin{figure}
\centering
\includegraphics[width=0.47\textwidth]{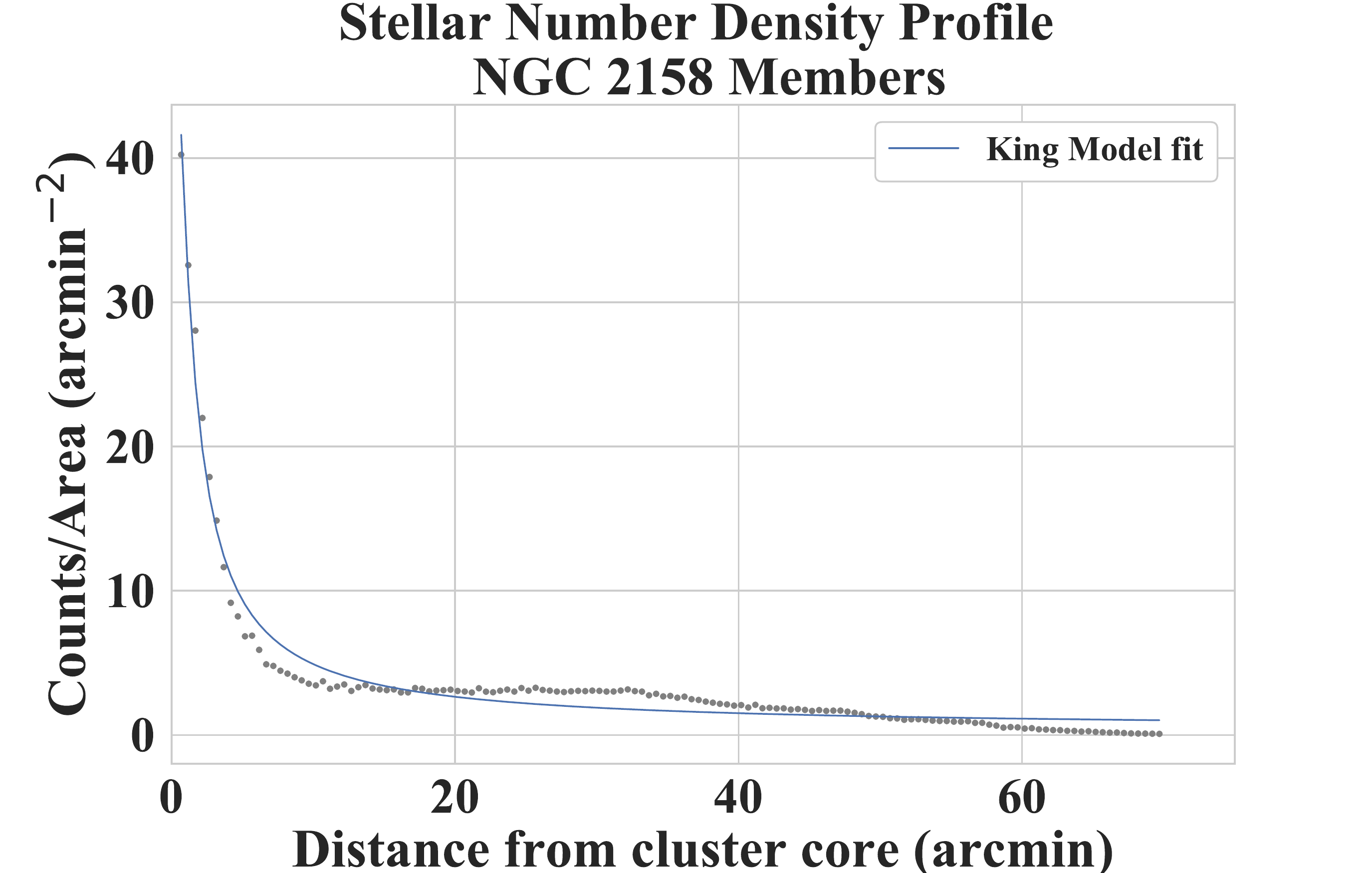}
\caption{King model fit to ${\sim}$97,000 sources in the CG18 Gaia DR2 dataset, positioned about the core of NGC~2158. 
Our fit is shown in \textit{blue}, which determined a core radius of $r_{\mathrm{c}}=0.9{\arcmin}$ (${\sim}1$~pc) and a tidal radius of $r_{\mathrm{t}}=16{\arcmin}$ (${\sim}21$~pc).}
\label{fig:KingFit}
\end{figure}

We followed a similar approach to ascribe NGC~2158 membership probabilities.
In order of precedence, the catalogs employed included (1) the CG18 Gaia DR2 derived catalog, which focused primarily on the more central regions of the cluster, (2) the aforementioned K13 catalog, and (3) the \cite{Dias2006} catalog (hereafter D06).\footnote{The \dance{} catalog did not include NGC~2158 membership probabilities and is omitted from the procedure.}
The K13 catalog supplied the vast majority of membership probabilities. 
We found 51 matches in the CG18 catalog.
The D06 catalog provided membership probabilities for four sources excluded by the combined efforts of CG18 and K13. 
One of these four sources, the rotating variable UCAC4-571-024298, is an NGC~2158 member.
A total of 388 catalog sources remained unmatched after searching the three catalogs.
In Table~\ref{Table:membcountA} we provide the size of each of the cluster membership catalogs, as well as the number of source matches found. 
The cluster membership catalogs and the matching procedures are outlined in Sections~\ref{subsubsec:cat1}--\ref{subsubsec:cat4}.

We found that many sources possessed high membership probabilities despite being positioned well outside the cluster tidal radius. 
Given the cluster's low proper motion, similar to that of field stars, this was unsurprising.
The spheroidal shape of the cluster permits a 2D King model fit to the spatial distribution of these sources \citep{King1962}. 
To ensure a conservative membership likelihood for our sources, we weighted the catalog-derived membership probabilities by a King model fit.
In the cases where a catalog match was not found, the King model weight was used as a membership proxy. 
The membership probabilities are listed in \textit{Column 65~(PN2158)} of the variable catalog. 

To calculate the membership probability derived by the King model ($P_{\rm K}$) for a given source, we used the method outlined in \cite{Li2018}. 
The probability is given by $P_{K}=(f_{K}-b)/(f_{K})$, where $f_{K}$ is the King model fit to the stellar density profile and $b$ is the background estimate.
Our fit was performed on ${\sim}$97,000 Gaia DR2 sources out to a distance of $70{\arcmin}$ from the cluster core.
We employed the central coordinates measured by \cite{Cantat2018}. 
Our fit is shown in Figure~\ref{fig:KingFit}. 
From this fit, we estimate a core radius of $r_{\mathrm{c}}=0.9{\arcmin}$ (${\sim}1$~pc) and a tidal radius of $r_{\mathrm{t}}=16{\arcmin}$ (${\sim}21$~pc). 

\subsubsection{Gaia DR2 Cluster Membership Catalog (CG18)}
\label{subsubsec:cat1}
The CG18 membership catalogs employed Gaia DR2 data to calculate cluster membership probabilities for stars in 1,212 Milky Way stellar clusters \citep{Cantat2018}. 
Cluster membership probabilities were determined using the unsupervised membership assignment code, \texttt{UPMASK}, which performs a k-means clustering analysis on the proper motions and parallaxes for a population of sources. 
The CG18 M35 membership catalog contains a total of 1,705 sources and the CG18 NGC~2158 membership catalog contains 1,633 sources.
To find matches for our variable sources, we searched the membership catalog for sources with a separation $<5\arcsec$ and a Gaia magnitude difference of $<0.4$~mag. The Gaia magnitude difference is the difference between the reported DR2 Gaia magnitude and our calculated source Gaia magnitude using the \textit{B~-~V} Johnson-Cousins algorithm described in \cite{Jordi2010}.  
Among the two CG18 cluster membership probability catalogs, we found a total of \totGaiamatch{} source matches with \NGaiaM{} matches in the M35 catalog and \NGaiaN{} in the NGC~2158 catalog. 
 
\subsubsection{The \dance{} Cluster Membership Catalog}
\label{subsubsec:cat2}

The \dance{} cluster membership catalog was assembled using proper motion measurements and multi-wavelength (optical and near-infrared) photometry from some of the best ground-based archival datasets collected over the past 18 years \citep{Bouy2015}. 
The team applied this method to two clusters, the Pleiades \citep{Sarro2014} and M35 \citep{Bouy2015}.
The M35 \dance{} analysis resulted in significantly reduced membership probabilities for many sources that were considered cluster members in prior catalogs. 
Membership probabilities have not been provided for NGC~2158 by this group.

To identify cluster members, the team employed a training set, which was supplied by \cite{Barrado2001} for M35. 
While \cite{Bouy2015} found that contamination from nearby NGC~2158 members did not reach more than ${\sim}2\%$, they noted that it is a challenging system to analyze due to the cluster's low proper motion and the overlap observed in the \cmd{} (CMD) among cluster members and field sources.
The \dance{} catalog contains a total of 338,892 sources, 4,349 of which have cluster membership probabilities of \Pm{}.
Searching this catalog, we found a total of \NBouy{} source matches, representing nearly our entire sample of K2C0 variables.
This catalog provided membership probabilities for 772 sources unmatched in the CG18 catalog. 
Similar to our approach of source matching in the CG18 catalog, we searched for matches with a $<5\arcsec$ separation as well as a magnitude difference of $<0.4$~mag (\Jmag{} or \Hmag{}, depending on the data that was available for the target).
We used this same source matching criteria for the K13 and D06 catalogs. 
These catalogs are described in Sections~\ref{subsubsec:cat3} and \ref{subsubsec:cat4}.

\subsubsection{The K13 Cluster Membership Catalog}
\label{subsubsec:cat3}
The K13 cluster membership catalog was part of the \textit{Milky Way Star Clusters} project \citep{Kharchenko2013}, which resulted in the assignment of cluster membership probabilities for sources in 3,006 Milky Way clusters.
The results were obtained by employing kinematic and near-infrared photometric data from the all-sky PPMXL catalog \citep{Roeser2010} and 2MASS \citep{Curtis2013}. 
The cluster membership probabilities were determined using the stellar source location with respect to the reference sequences, which include isochrones in photometric diagrams and/or the average cluster proper motion in kinematic diagrams.
We found \NKharM{} sources with corresponding M35 membership probabilities, representing \fKharM{$\%$} of our variable catalog. 
This catalog provided M35 membership probabilities for four sources that were excluded from both the CG18 and \dance{} catalogs.
We found \NKharN{} source matches with corresponding NGC~2158 membership probabilities. 
This catalog provided NGC~2158 membership probabilities for 708 sources that were excluded from the CG18 catalog.

\subsubsection{The D06 Cluster Membership Catalogs}
\label{subsubsec:cat4}

The D06 cluster membership probability catalog is part of a cluster membership survey performed on 112 Milky Way open clusters, including M35 \citep{Dias2001} and NGC~2158 \citep{Dias2006}. We used only the NGC~2158 cluster membership catalog, which was produced using positions and proper motions from the UCAC2 Catalogue.
To calculate the probability of cluster membership, the team employed a method proposed in \cite{Sanders1971}, which fits the relative proper motions to a maximum likelihood statistical model. We found a total of \NDiasN{} source matches in the D06 catalog for NGC~2158.
This catalog provided NGC~2158 membership probabilities for four sources that were excluded from both the CG18 catalog and the K13 catalog.

\section{Variable Catalog Census} 
\label{sec:census}

\subsection{Ambiguous Blend Distribution} 
\label{sec:deblenddist}
\begin{figure}
\centering
\includegraphics[width=0.45\textwidth]{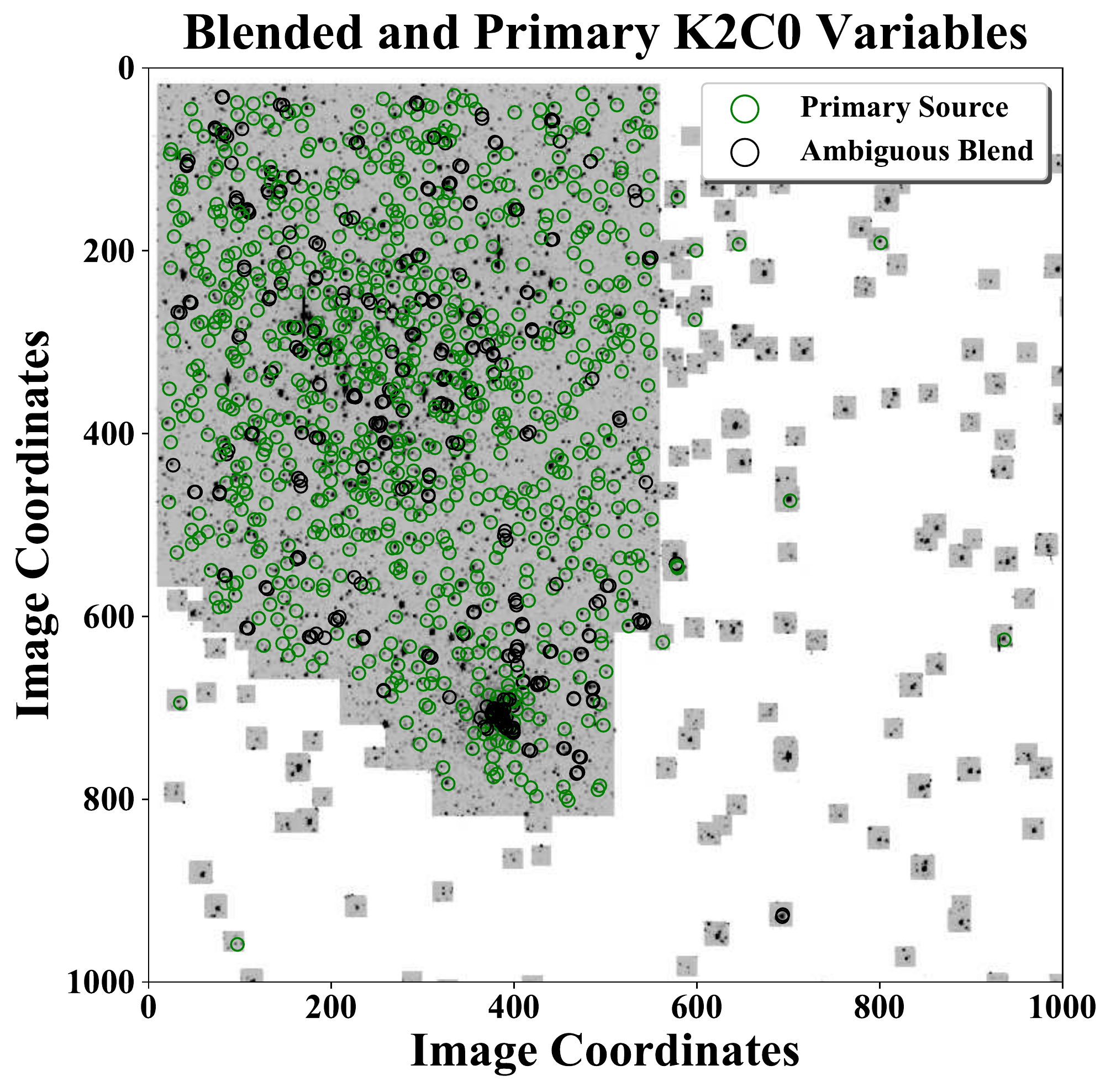}
\caption{2D projected spatial location of our identified primary variables (\textit{green circles}) and ambiguously blended variables (\textit{gray circles}).
A collection of ambiguously blended variables is observed near the core of NGC~2158.}
\label{fig:blendloc}
\end{figure}

Of our \totcat{} catalog variables, \nblendtot{} sources were identified as the primary variable source, while an unambiguous primary could not be determined for the remaining \blendtot{} variables within blended groups. 
In Figure~\ref{fig:blendloc}, we illustrate the 2D projected spatial distribution of our primary variables (\textit{green circles}) and our ambiguously blended variables (\textit{gray circles}). 
A collection of ambiguously blended variables is observed near the core of NGC~2158.
This is expected given the low angular separation of stars in this region, which results in enhanced blending. 

\subsection{Classification Distribution}
\label{sec:cdist}

\begin{figure}
\centering
\includegraphics[width=0.45\textwidth]{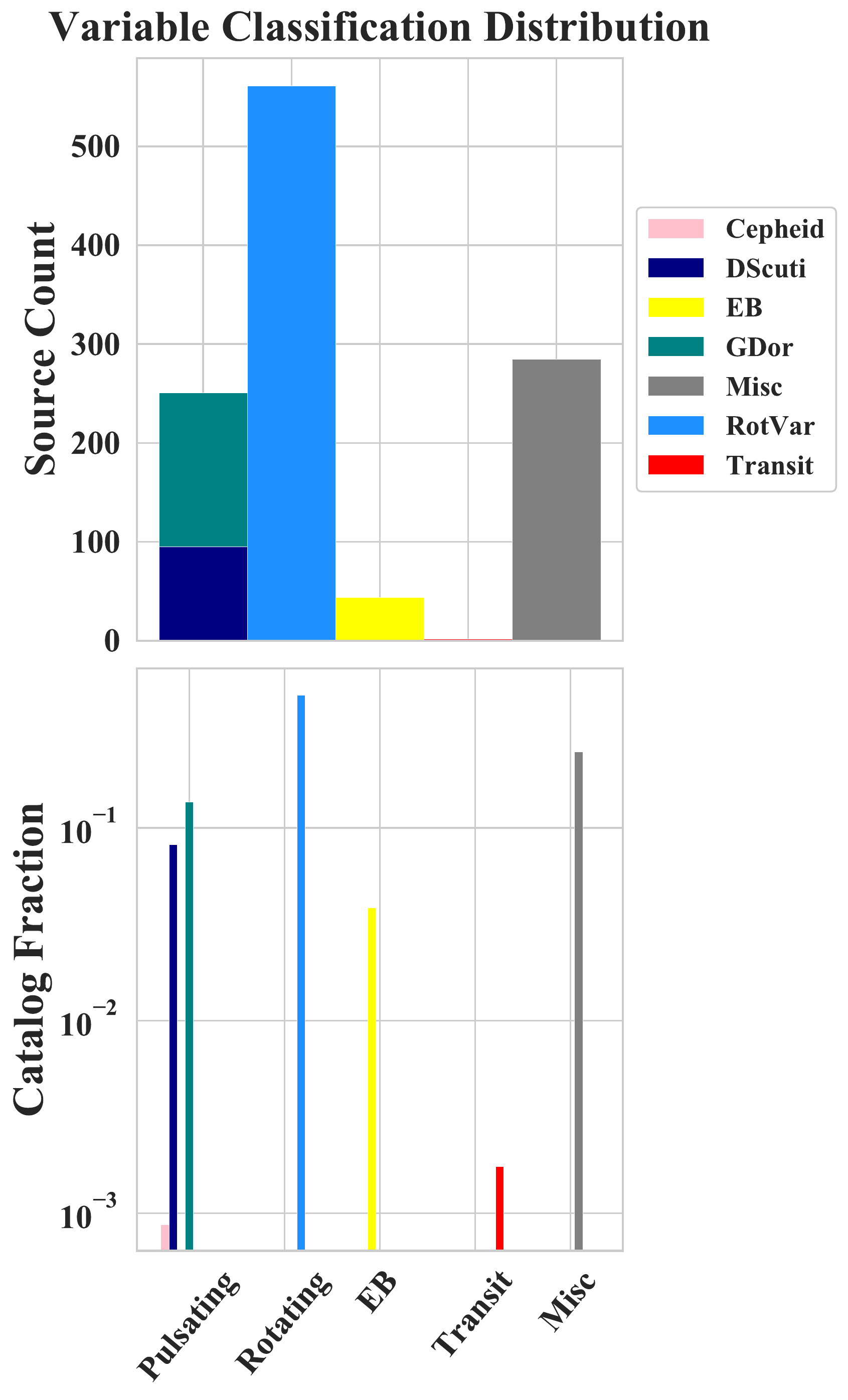}
\caption{\textit{Top:} The source count for our five variable classes. 
\textit{Bottom:}.
The catalog fraction for variable classes with subclassification groups shown separately. 
Rotational variables: \rotating{}, pulsating variables: \pulsating{} (primarily \Gdor{} and \dScuti{} variables), EBs: \totEB{}, transiting exoplanet candidates: \tottransit{}.}
\label{fig:catcount}
\end{figure}

The variable classification distribution among our catalog variables is shown in the top and bottom panels of Figure~\ref{fig:catcount}.
Our identified variable classes are listed as follows: \textit{Pulsating, Rotating, Misc, EB,} and \textit{Transit}.
Subclassification identifiers were used to more precisely distinguish variable type when such a distinction was possible. 
For example, an identified  \Gdor{} star is listed with the class identifier \textit{Pulsating} and the subclassification identifier \textit{GDor}. 
Rotating variables make up the most highly populated variable class group, containing \rotating{} objects and comprising nearly half of all catalog variables. 
Pulsating variables make up the second most populated variable classification group, containing \pulsating{} sources.  
Nearly all our pulsating variables are \Gdor{} and \dScuti{} pulsators. 
The \totEB{} detected EBs make up \totEBf{$\%$} of our variable catalog and our \tottransit{} exoplanet transit candidates make up \tottransitf{$\%$}. 
We review the variable class distribution of cluster-affiliated sources in Section~\ref{sec:memberdist}.

\begin{figure}
\centering
\includegraphics[width=0.45\textwidth]{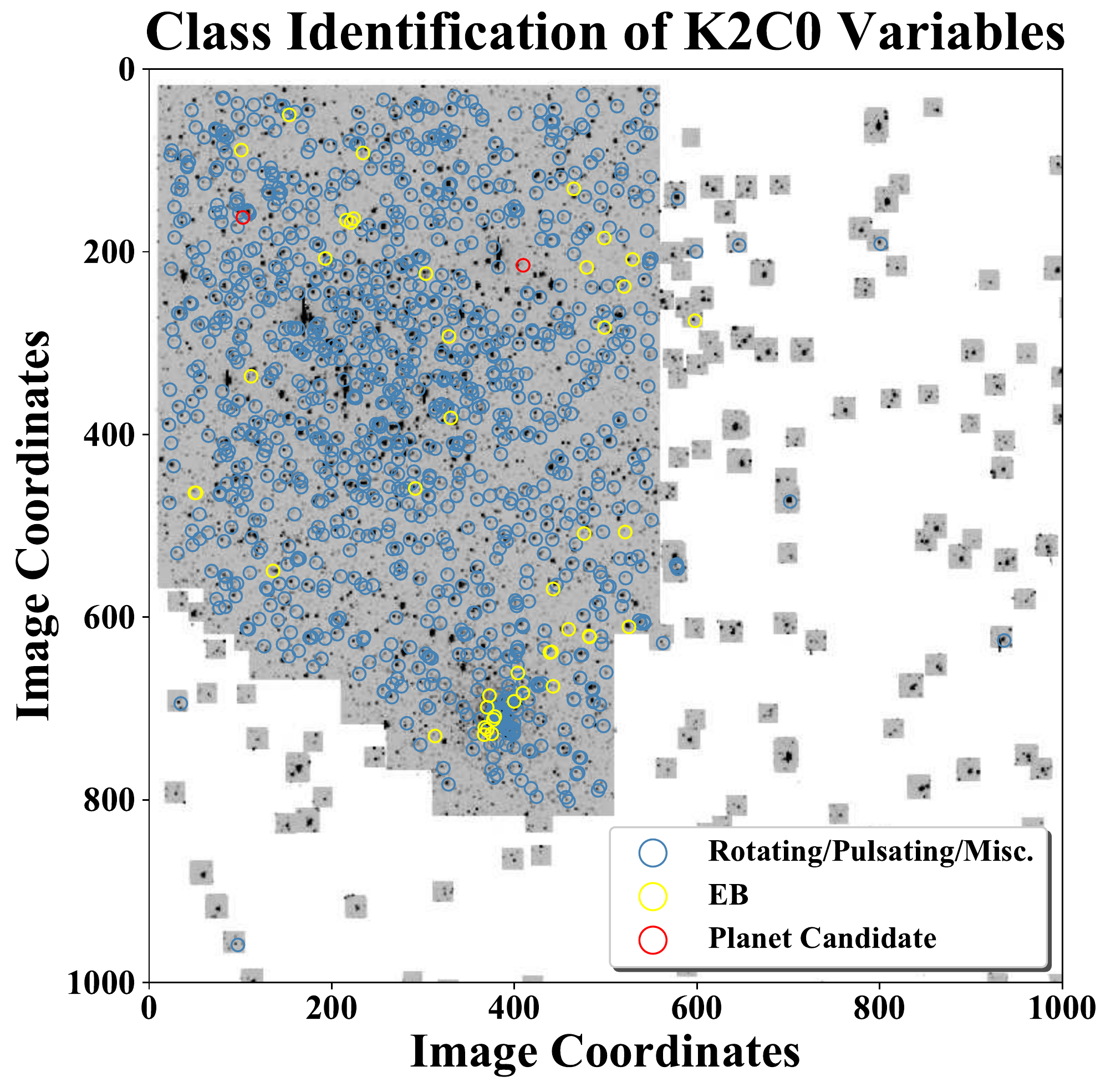}
\caption{The 2D projected spatial distribution of our catalog variables. 
Sources are color-coded to reflect variable classification.}
\label{fig:variablelocation}
\end{figure}

The 2D projected spatial distribution of our catalog variables is shown in Figure~\ref{fig:variablelocation}. Candidate transiting exoplanets are encircled in \textit{red}, candidate EB sources are encircled in \textit{yellow}, and rotating/pulsating variable sources (as well as indeterminate or \textit{Misc.} sources) are encircled in \textit{blue}.

\subsection{Cluster Membership Distribution} 
\label{sec:memberdist}
\begin{figure}
\includegraphics[width=0.4\textwidth]{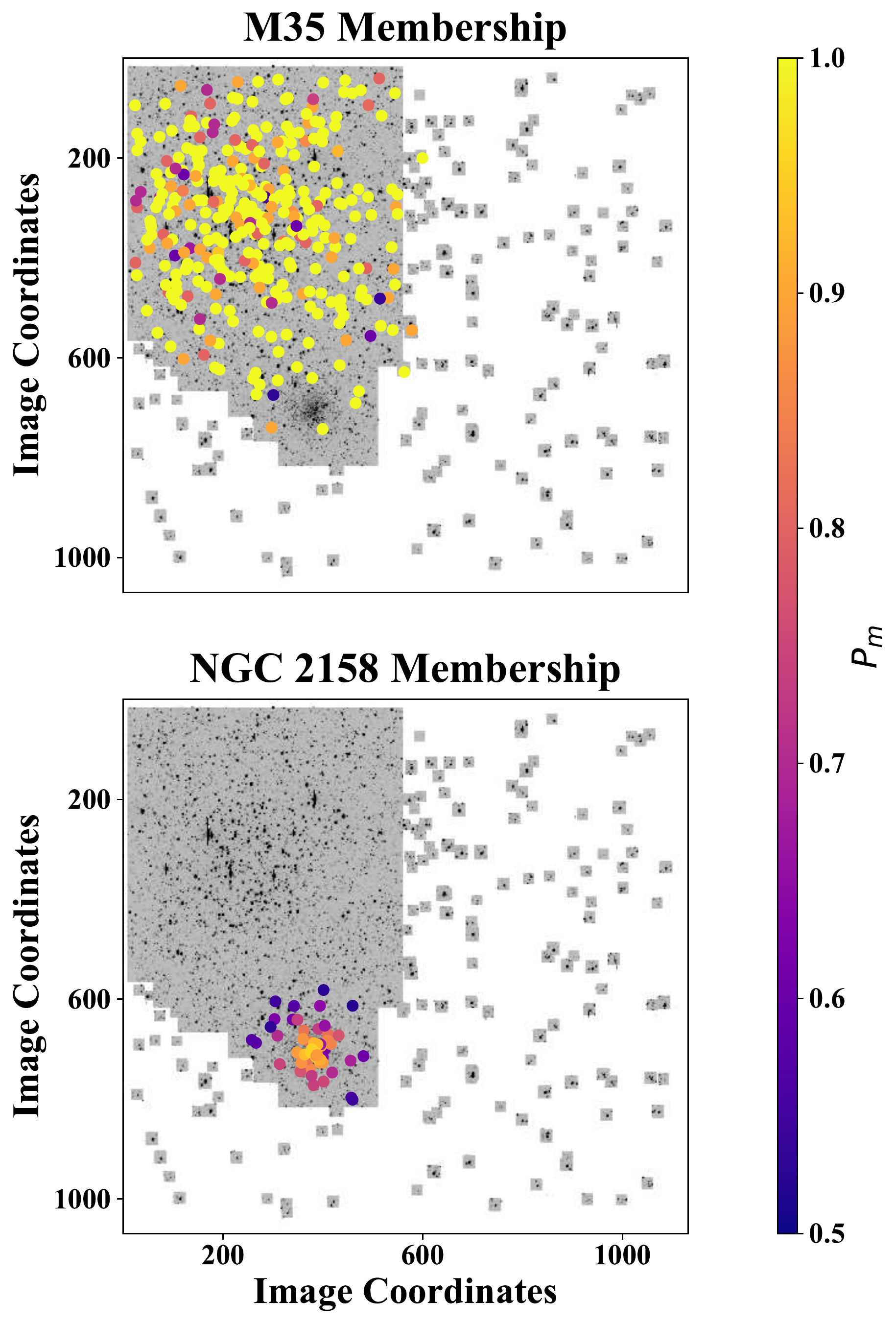}
\caption{The spatial distribution of cluster members (\Pm{}), color-coded to reflect membership probability. M35 members are shown in the \textit{top panel} and NGC~2158 members are shown in the \textit{bottom}. 
High-probability M35 members ($P_{\mathrm{m}}~>~0.9$) are found across the K2C0 super stamp, while high-probability NGC~2158 members are positioned closer to the cluster core.}
\label{fig:membershipprob}
\end{figure}

For the cluster membership distribution analysis, we consider ``high-probability members" as sources with $P_{\rm m}>0.9$ and ``probable members" as sources with \Pm{}.
A total of \totfield{} variables were designated as field sources ($P_{\rm m}~<0.5$ for both clusters).
The 2D projected spatial distribution for probable cluster members is shown in Figure~\ref{fig:membershipprob}. 
The color corresponds to membership probability, ranging between 0.5~--~1. 
M35 members are well dispersed across the super stamp, while the NGC~2158 variables are positioned close to the cluster core. 

\begin{table}
\centering
\begin{tabularx}{0.28\textwidth}{lrr}
\toprule
 &  M35 & NGC~2158 \\
\midrule
\multicolumn{1}{c}{}   & \multicolumn{2}{c}{$P_{\rm m}>0.5$}     \\
\Nvar{} &  \totM{} & \totN{} \\
\Nprim{} &  \Mmvd{} & \Nmvd{} \\
\midrule
\multicolumn{1}{c}{}   & \multicolumn{2}{c}{$P_{\rm m}>0.9$}     \\
\Nvar{} & \totMb{} & \totNb{} \\
\Nprim{} & \Mmvdb{} & \Nmvdb{} \\
\bottomrule
\end{tabularx}
\caption{\label{Table:membresults} 
Census data for our cluster variables. 
Results are listed for sources with \Pm{} (\textit{top}) and for high-probability members with $P_{\rm m}>0.9$ (\textit{bottom}).
The number of cluster members found in our K2C0 variable catalog is given in the column \Nvar{}.
The number of primary members (not blends) in the variable catalog is given by \Nprim{}. 
}
\end{table} 

Table~\ref{Table:membresults} summarizes the number of cluster members found among our \totcat{} catalog variables for both probable members and high-probability members. 
Within the catalog are \totM{} M35 cluster members (\ratioa{\%} are high-probability members) and \totN{} NGC~2158 cluster members (\ratiob{\%} are high-probability members).
Of the M35 variables, \Mmvd{} were designated as primary sources (not ambiguously blended).
Of the NGC~2158 catalog variables, \Nmvd{} were designated as primary sources.

\begin{figure}
\centering
\includegraphics[width=0.4\textwidth]{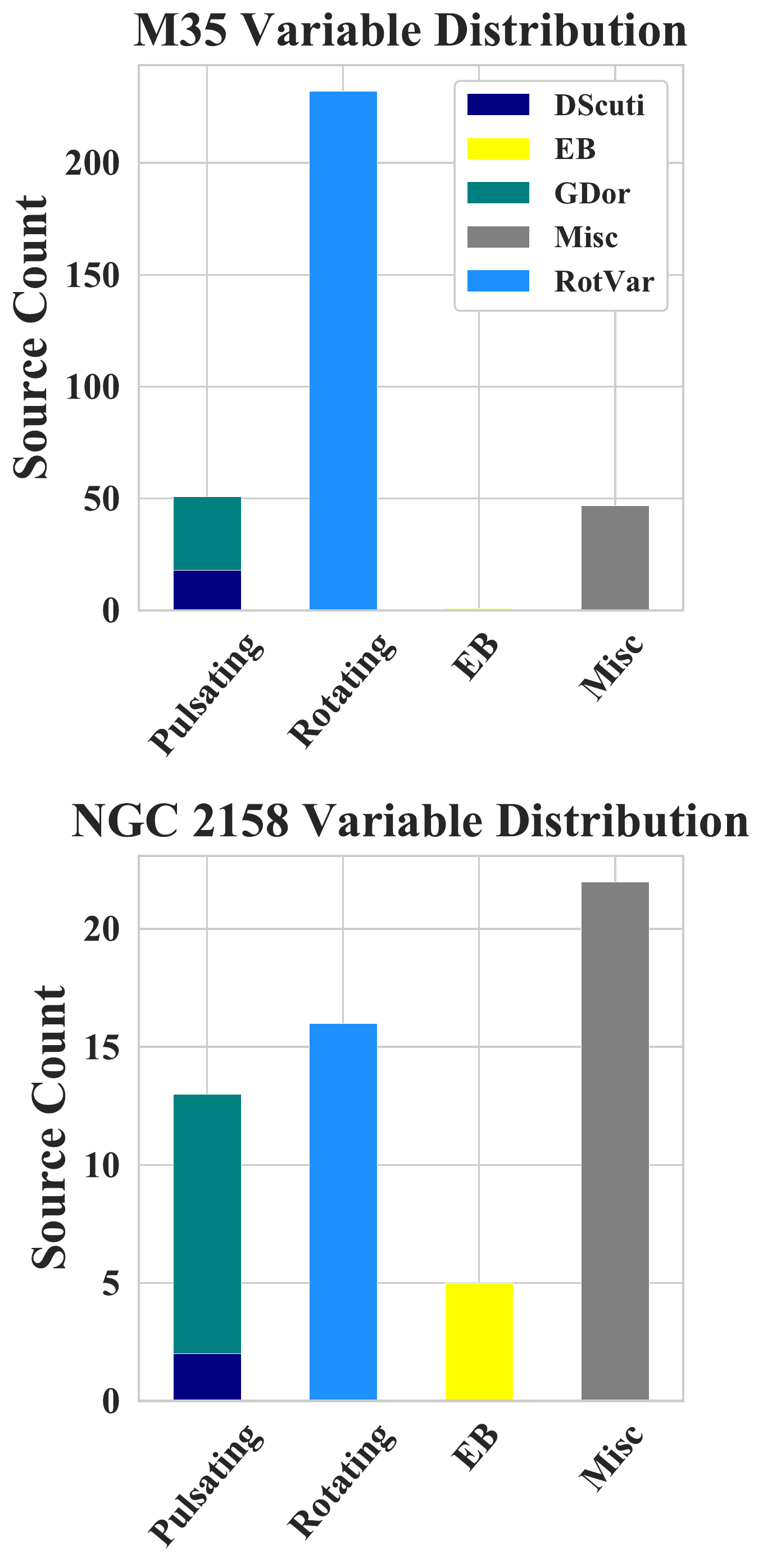}
\caption{Variable classification distribution for probable M35 members (\textit{top panel}) and probable NGC~2158 members (\textit{bottom panel}).
Note the differences in y-axis bounds to account for the fact that our catalog contains ${6\times}$ more M35 members than NGC~2158 members.
M35: \EBM{} EB, \MPulse{} pulsating variables, and \MRot{} rotating variables. NGC~2158: \EBN{} EBs, \NPulse{} pulsating variables, and \NRot{} rotating variables.}
\label{fig:clusterbreakdown}
\end{figure}

In Figure~\ref{fig:clusterbreakdown}, we illustrate the variable classification distribution for members of M35 (top) and members of NGC~2158 (bottom).
The cluster variables consist of EBs, pulsating variables, rotating variables, and those of indeterminate type.
No cluster-associated transiting exoplanet candidates were found.
In the case of M35, we found \EBM{} EB candidate, \MRot{} rotating variables, \MPulse{} pulsating variables (\MDScuti{} \dScuti{} variables and \MGDor{} \Gdor{} variables), and \MMisc{} variables of indeterminate type. 
In the case of NGC~2158, we found \EBN{} EBs, \NRot{} rotating variables, \NPulse{} pulsating variables (\NDScuti{} \dScuti{} variables and \NGDor{} \Gdor{} variables), and \NMisc{} variables of indeterminate type. 
The proximity of M35 results in a greater sensitivity to low amplitude variations, which accounts for the larger number of total variables. 
NGC~2158 is richer in the total stellar count and this accounts for the greater number of EBs, which generally display large amplitude variations. 
\vspace{4.0mm}

\subsection{Color-Magnitude Distribution} 
\label{sec:cmddist}

\begin{figure*}
\centering
\includegraphics[width=0.75\textwidth]{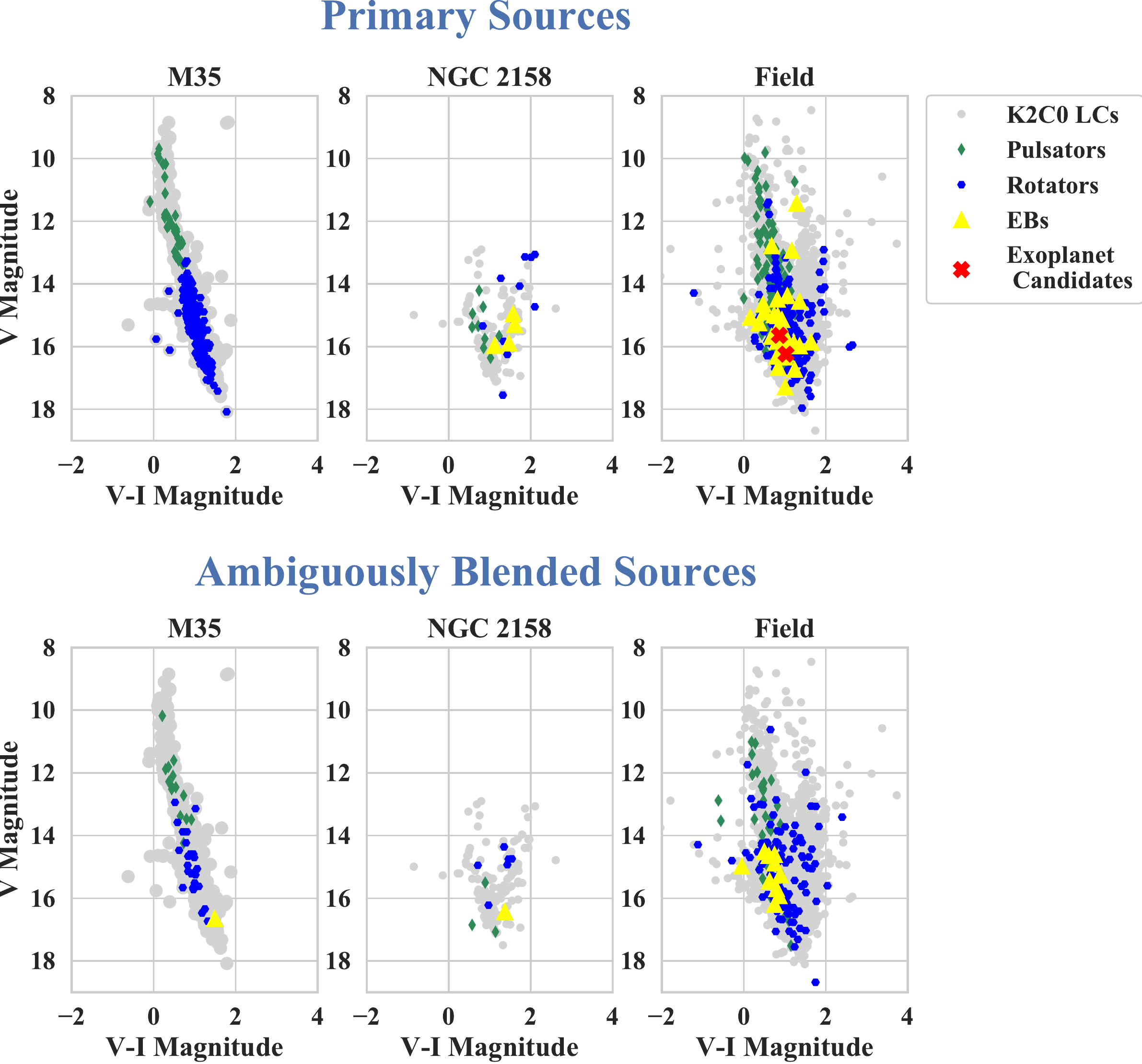}
\caption{\VminI{} vs.\ \Vmag{} CMDs for primary sources (top row) and blended sources  (bottom row) in M35 (first column), NGC~2158  (second column), and the field  (third column). The \Vmag{} and \Imag{} measurements were obtained using UCAC4. 
The \textit{gray points} in each panel depict sources from our full set of K2C0 LCs.}
\label{fig:CMD}
\end{figure*}

In Figure~\ref{fig:CMD}, we illustrate the color-magnitude distribution of our catalog variables in \VminI{} vs.\ \Vmag{} format.
Our original 3,960 K2C0 LCs are plotted as \textit{gray} points in the background of each of the CMD panels. 
Our \totM{} probable M35 variables are shown in the first column, our \totN{} NGC~2158 probable variables are shown in the second column, and our \totfield{} field sources are shown in the third column. 
Sources identified as primary variables are shown in the top row and ambiguous blends are shown in the bottom row. 
We illustrate the class of rotating variables as \textit{blue points}, pulsating variables as \textit{green diamonds}, EB candidates as \textit{yellow triangles}, and transiting exoplanet candidates as \textit{red crosses}. 
Note that there is some overlap between the open clusters and the field sources. 
These data are in good agreement with the cluster CMD distributions produced using photometry data from \cite{Carraro2002} (for M35) and \cite{Barrado2001} (for NGC~2158). 
We obtained the comparison photometry data from the open cluster database, WEBDA.\footnote{The Open Cluster database WEBDA is available at http://www.univie.ac.at/webda/webda.html} 

Our M35 member variables range in \Vmag{}-band magnitude from \VoneM{th}~mag to \VtwoM{th}~mag. 
A total of \Mdeblend{} M35 variables are primary sources (top row), while the remaining \Mblend{} variables are ambiguous blends (bottom panel). 
The sole M35 EB is ambiguously blended with a field source (the field source has $P_{\rm m}=0$).
The vast majority of M35 cluster member variables are classified as rotating variables (\MRot{} sources) and pulsating variables (\MPulse{} sources). 

NGC~2158 variables range in \Vmag{}-band magnitude from \VoneN{th}~mag to \VtwoN{th}~mag; a narrower magnitude range than M35, which is unsurprising given the difference in age, turn-off mass, and distance. 
This CMD is much broader than its M35 counterpart, likely caused by an overestimate of membership probability given the cluster's low proper motion. 
A total of \Ndeblend{} NGC~2158 variable members in our variable catalog are identified primary sources (top row), while the remaining \Nblend{} variables are ambiguous blends (bottom panel). 
We observed \EBN{} EBs with membership probabilities ranging between \leastPN{$\%$} and \mostPN{$\%$}. 
One is an ambiguous blend. 
As in the case of M35, the vast majority of our probable NGC~2158 members are rotating variables (\NRot{}) and pulsating variables (\NPulse{}).

\subsection{Period-Magnitude Distribution}
\label{sec:PeriodMag}

\begin{figure}
\centering
\includegraphics[width=0.4\textwidth]{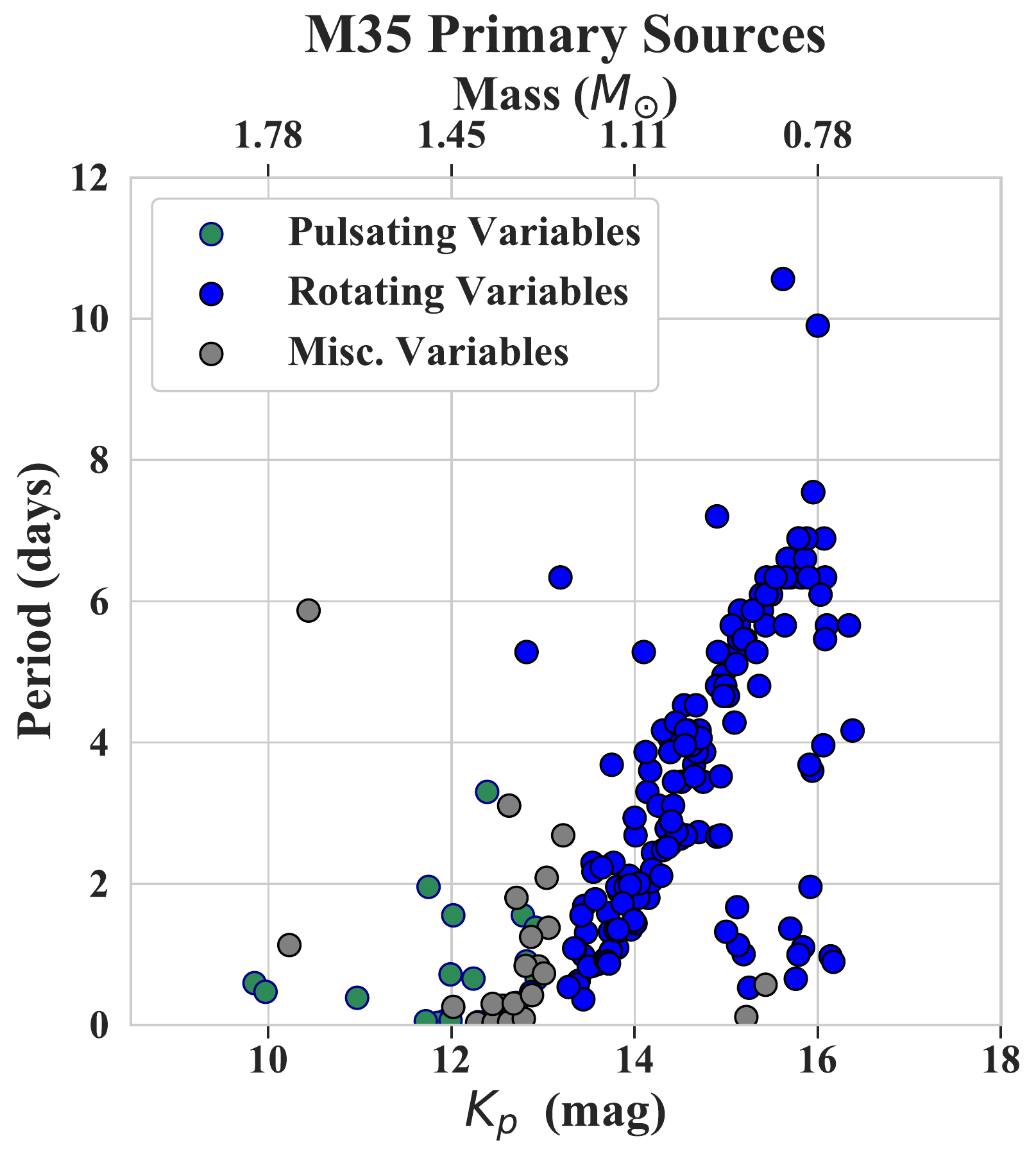}
\caption{The well-correlated rotation period--magnitude relation among our M35 variables. 
We illustrate high-probability members with $P_{m}>0.9$.
This is a chief rotation feature of intermediate aged stars. 
The upper x-axis displays the corresponding masses, produced using isochrone fits to the cluster parameters (age and metallicity) and the absolute \Vmag{} magnitude.
}
\label{fig:MPMag}
\end{figure}

The period-magnitude distribution among our M35 primary variables is shown for members with $P_{m}>0.9$ in Figure~\ref{fig:MPMag} (ambiguous blends are excluded from the figure).
Along the top x-axis, we list the corresponding estimated stellar masses.
The stellar masses in this figure were calculated by fitting an isochrone to the stellar cluster parameters.
These parameters included age, metallicity, and the absolute \Vmag{} source magnitude, which is dependent upon distance and reddening estimates.
The cluster parameters used are listed in Section~\ref{subsec:M35} (M35).

The source intensity is measured in the broad Kepler bandpass, known as the \textit{Kepler magnitude} (\Kepmag{}).
The M35 primary variables range in \Kepmag{} from \MPrimMin{}~mag~--~\MPrimMax{}~mag.
The sources are color-coded to indicate variable class as follows: rotating variables (\textit{blue}), pulsating variables (\textit{green}), and indeterminate variables (\textit{gray}).
The pulsating variables represent the brightest cluster class. 
All M35 pulsating variables have \Kepmag{}~$\leq$~\MPrimPulseMax{}~mag and $P\leq$~\MPrimPulsePMax{} days. 
The M35 rotating variables occupy nearly the full \Kepmag{} magnitude range with the majority of the faintest M35 catalog sources being of this class. 
The rotating variables are also spread across a wide range of source periods, ranging from \MPrimRotPMin{} days to \MPrimRotPMax{} days.
An obvious rotation period-magnitude relation is observed among our M35 variables, which is a chief rotation feature of intermediate aged stars. 
While most of the rotating sources fall near the rotation period-magnitude relation, some are found far from this locus.
Those falling below the sequence may be more rapidly rotating cluster members, which are commonly seen in young clusters like M35. 
Rapidly rotating cluster sources may be tidally spun up, or simply the rapidly rotating tail of a distribution that has yet to converge onto the sequence. 
Sources located above the rotation period-magnitude relation are likely explained by one of the following causes: (a) the target has an incorrect period estimate, (b) the target is blended with an unresolved source, (c) the target is misclassified, or (d) the target is a field source with a proper motion very similar to that of M35.

Unfortunately, there are too few sources to make such a plot for NGC~2158 primary members, as many are blended or have lower membership probabilities. 
Using the membership threshold of \Pm{}, we found that a relationship between the rotation period and stellar mass is not evident for this system.
This is expected given that the detected NGC~2158 variables are brighter than the MSTO and are therefore not representative of the cluster main sequence.\footnote{The NGC~2158 MSTO mass is 1.3~\msun{} \citep{Carraro2002}.}

\section{Comparing Our Results With Other Surveys} 
\label{sec:variabilitysearch}

We compared our variable catalog to the five catalogs that are described in Sections~\ref{subsec:VSXcomp}--\ref{subsec:Libralato}.
To identify matches within each of the catalogs, we used \texttt{Astropy} to perform a search for the nearest neighbors, comparing the periods and associated harmonics for each of the potential matches. 
In the following sections, we describe the catalog contents and the number of comparable sources. 
Comparable sources are those that are not associated with one or more of the following conditions: (a) the source is located within empty regions of the K2C0 super stamp (no match with a separation radius of $r_{\mathrm{s}}<5 \arcsec{}$), (b) the source did not have a counterpart in the UCAC4 catalog and was therefore not in our raw LC sample, (c) the source is outside our period bounds, and (d) the source was outside our \Vmag{} magnitude bounds.

\subsection{VSX Catalog Classification Comparison}
\label{subsec:VSXcomp}

We compared our catalog variables with the variables listed in the \textit{AAVSO International Variable Star Index} \citep[hereafter VSX]{Watson2006,Watson2017}. The \textit{VSX} catalog contains 528,037 variables spread throughout the northern and southern hemispheres.
There was very little overlap between our catalog variables and the VSX catalog sources.
Only 26 VSX variables were located within the $1.5\times 1.5$~degree field centered on M35.
Of these 26 VSX variables, there were eight comparable sources. 
We identified three unique matches when compared with our catalog.
All the matched sources were separated by less than a milli-arcsecond. 
They also all possessed nearly identical periods when provided (less than $0.2\%$ difference) and were classified as the same variable type.
The VSX matches consisted of HN Gem, the hybrid source HD~252154 (which displays LC characteristics reminiscent of both  \dScuti{} and \Gdor{} variables), the $\delta$~Cepheid variable V0371 Gem, and the \dScuti{} source V0392.
The VSX identifiers for all the matched sources are provided in the \textit{Column 70~(Match5)} of our variable catalog. 

\subsection{Meibom M35 Variability Comparison}
\label{subsec:Meibom}

\cite{Meibom2009} (hereafter M09) performed a detailed analysis of the period-color relationship of rotating stars in M35 using data from the WIYN 0.9-meter telescope. 
In addition to conducting a 5-month photometric survey, the authors reviewed data from a decade-long radial-velocity survey, which resulted in the production of a variable catalog listing rotation periods, cluster membership, and binarity for 441 sources within a $40{\arcmin} \times 40{\arcmin} $ field centered on M35.
They concluded that 310 of these rotators were probable cluster members (\Pm{}). 

\begin{figure}
\centering
\includegraphics[width=0.37\textwidth]{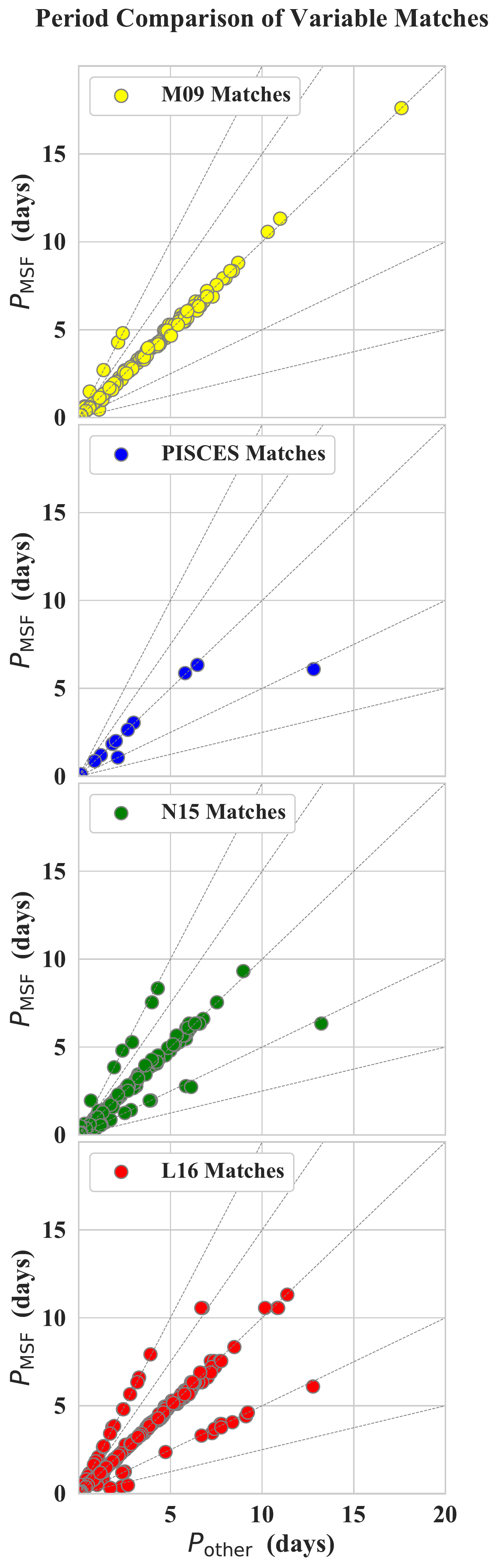}
\caption{Comparison of reported periods for matches found in prior variable catalogs.
The \textit{dashed lines} denote the relation \PMSF{}$=\alpha \times P_{\rm other}$, where $\alpha=[0.25,0.5,1,1.5,2]$ and $P_{\rm other}$ is the period reported in the respective search catalog.}
\label{fig:FullPeriodComparison}
\end{figure}

Of these 441 M09 variables, there were 255 comparable sources.
We identified 189 unique matches when compared with our catalog.
All the matched sources were separated by less than 10 milli-arcseconds.
The unmatched M09 variables tended to have longer periods.
The median period for unmatched sources was $\bar{P}=5.0$~days, while the median period for matched sources was $\bar{P}=3.3$~days.
The M09 variable matches are displayed as \textit{yellow circles} in the first panel of Figure~\ref{fig:FullPeriodComparison}.
The \textit{dotted lines} in the figure depict the relation
$P_{\mathrm{other}}=\alpha \times P_{\mathrm{MSF}}$, where $\alpha=[0.25,0.5,1,1.5,2]$, $P_{\mathrm{other}}$ is the rotation period listed in for the match in the M09 catalog.
Some of the unmatched M09 sources were omitted from our catalog as duplicate secondary sources or due to a lack of compelling periodicity upon inspection. 
The M09 identifiers for matches are provided in \textit{Column 68~(Match3)} of our K2C0 variable catalog.  
Using our membership prescription, as outlined in Section~\ref{sec:membership}, we found that 54\% of uniquely matched sources have an M35 membership probability of \Pm{}. 

\subsection{PISCES NGC~2158 Variability Comparison}
\label{subsec:PISCES}

The first variability search conducted on NGC~2158 was led by \cite{Mochejska2004} as part of the \textit{Planets in Stellar Clusters Extensive Search} (PISCES). 
This photometric survey spanned 20 nights (2003 January 3 to March 11) and observations were taken with the 1.2-m telescope at Fred Lawrence Whipple Observatory.
This group employed an image subtraction technique, achieving ${\sim}5$~millimag precision. 
While the total time-span of the observations is close to our observation window, the PISCES dataset is discontinuous, with a daily ${\sim}16$-hour break in the data.  
A total of 57 variable stars  were found, all exhibiting low-amplitude variability, including 34 EBs and five \dScuti{} variables. 
A second PISCES campaign was performed in 2006, consisting of 260~hours of photometry taken over the course of 59~nights \citep{Mochejska2006}. 
In this second analysis, 40 new variable stars were found, bringing the total number of detected variables to 97 (two sources have since been omitted from this catalog).
The cumulative observation window for both campaigns spanned 13 months. 

The second campaign resulted in the detection of TR1, a hot Jupiter candidate with an expected radius of $1.66~R_{\rm J}$ and a period of 2.3629~days. 
Higher-accuracy LCs were required to better constrain the radius and period.
Using a high-resolution catalog and a neighbor subtraction technique, \cite{Libralato2016} was able to make improved measurements of the depth of the eclipses for this source. 
At the January 2018 \textit{Dwarf Stars and Clusters with K2} conference, Nardiello (co-author of \cite{Libralato2016}) presented results indicating that TR1 is likely an EB.  
TR1 is below our magnitude threshold (\Kepmag{}~$~{\simeq}~18.35$~mag), so it is not present in our catalog. 

Of these 95 PISCES variables, there were 22 comparable sources.
We identified 13 unique matches when compared with our catalog.
All the matched sources were separated by less than a milli-arcsecond.
The variable matches are shown as \textit{blue points} in the second panel of Figure~\ref{fig:FullPeriodComparison}.
Not all matched variables were classified in the same manner.
This classification discrepancy is unsurprising, as the PISCES team noted that more than half of the designated EBs may be ellipsoidal variables.
The PISCES identifiers for matches are provided in the \textit{Column 69~(Match4)} of the K2C0 variable catalog. 
We found that eight matches are probable cluster members (one of which is associated with M35).

\subsection{Nardiello M35 and NGC~2158 Comparison}
\label{subsec:Nardiello}

\cite{Nardiello2015} (hereafter N15) surveyed open clusters M35 and NGC~2158 as part of the \textit{The Asiago Pathfinder for HARPS-N} program. 
They employed ground-based, high-precision (${\sim}5$~millimag), fast-cadence (${\sim}3$ min), multi-band photometric data.
To mitigate blending, the group implemented a point spread function (PSF) neighbor subtraction technique.
They found a total of 519 variables, 273 of these variables were new discoveries. 
No candidate exoplanetary transiting sources were found in this study. 

Of the 519 N15 catalog variables, there were 283 comparable sources. 
We identified 215 unique matches when compared with our catalog.
All the matched sources were separated by less than a milli-arcsecond.
As in the case of the M09 catalog, the unmatched sources tended to have longer periods. The median period for unmatched N15 sources was $\bar{P}=8.5$~days, while for matched sources it was $\bar{P}=2.3$~days. 
The matches are shown in the third panel of Figure~\ref{fig:FullPeriodComparison} (\textit{green points}). 
The N15 identifiers for matches are provided in the \textit{Column 67~(Match2)} of the K2C0 variable catalog. 
We did not always designate the same variable class for our matches.
We omitted a detailed comparison of our catalog with the K2C0 variable search conducted by \cite{LaCourse2015}, which did not implement image subtraction or PSF neighbor subtraction methods.
These results did not offer any information relevant to our catalog that had been omitted by N15.

\subsection{K2-Specific M35 and NGC~2158 Comparison}
\label{subsec:Libralato}

The most direct comparison of our results can be made with the variable catalog generated by \cite{Libralato2016} (hereafter L16).
This group conducted a deep search on the K2C0 super stamp using a PSF neighbor subtraction technique, as described in \cite{Nardiello2015}.  
They employed the high-angular-resolution \textit{Asiago Input Catalog} (AIC), compiled from observations made by the \textit{Asiago Schmidt Telescope}. 
They generated a catalog of 2,848 variable stars.
Nearly 2,000 sources are listed as \textit{candidate variables} and 978 are flagged as ``high probable blends" and/or variables that were difficult to classify.
Variability periods were not provided for 380 catalog sources, which also prevented us from making a direct comparison.

Of the 2,133 L16 catalog variables, there were 730 comparable sources.
This small subset of comparable sources was largely due to their ability to probe to fainter magnitudes.
The AIC reaches magnitudes of \Kepmag{}~${\sim}24$~mag, while our minimum brightness was \Kepmag{}~${\sim}17$~mag.
Another cause was the lack of published periodicity estimates within the L16 catalog.
Moreover, although we employed the same data set, we only use data obtained when the instrument was in fine pointing mode, which comprised roughly half of the campaign window. 
Therefore, our period bounds are narrower.

\begin{figure}
\centering
\includegraphics[width=0.46\textwidth]{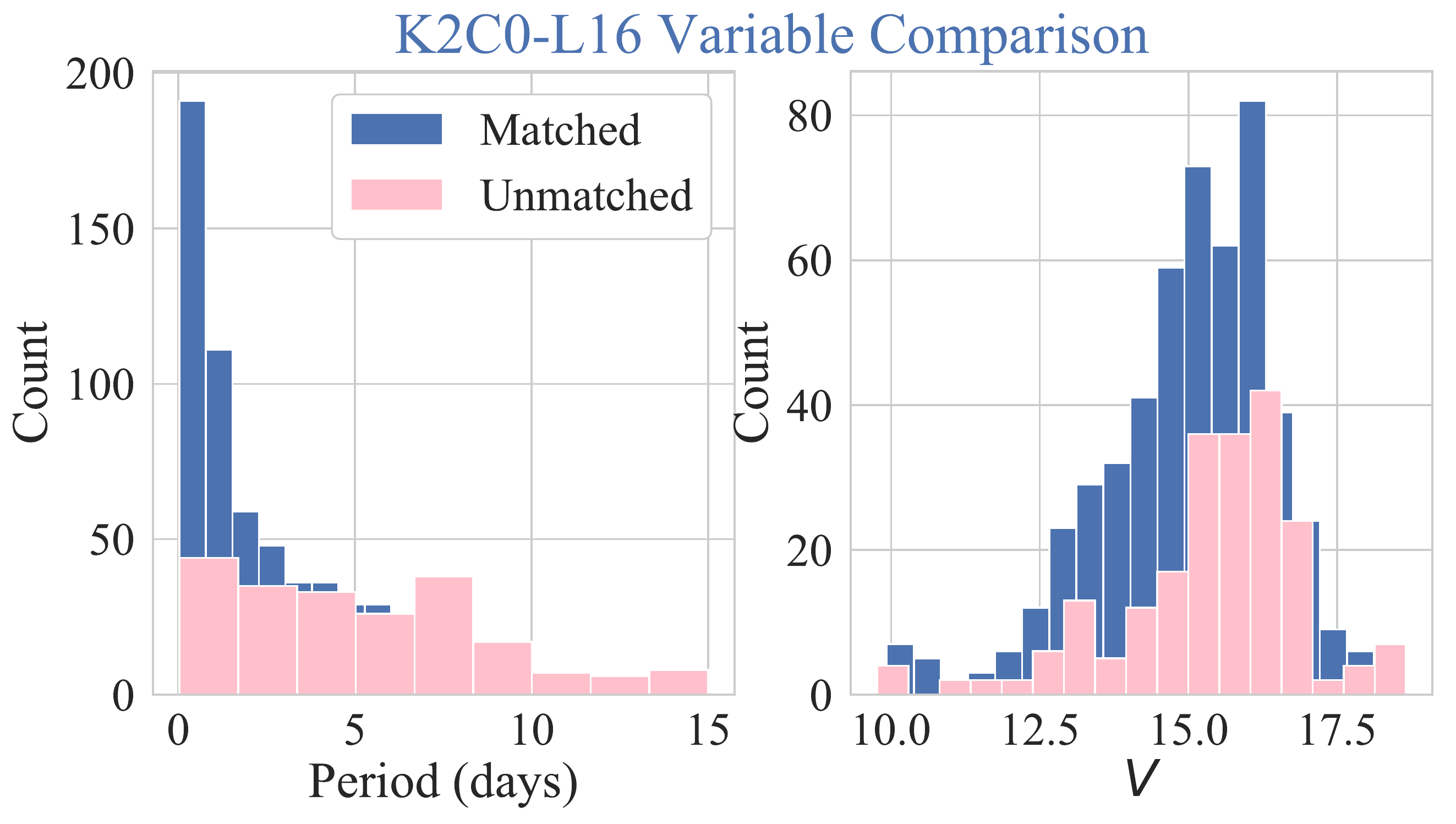}
\caption{Distributions of the period variability and \Vmag{} magnitude for matched (\textit{blue}) and unmatched L16 variables (\textit{pink}). 
Unmatched sources tended to possess longer variability periods with $\bar{P}=4.6$~days for unmatched variables and $\bar{P}=1.5$~days for matched variables. 
Matched sources tended to be slightly brighter ($\bar{V}=15.1$~mag) than unmatched variables ($\bar{V}=15.6$~mag).
This same tendency was seen when comparing our variables to the M09 and N15 catalogs.}
\label{fig:LibComp2}
\end{figure}

Of the 730 comparable sources, we identified 545 unique matches when compared with our catalog.
All the matched sources were separated by less than a milli-arcsecond.
Most of the unmatched sources were omitted from our catalog as duplicate secondaries. 
In fact, $33\%$ of the unmatched L16 variables were listed as being either probable blends or difficult to classify. 
The matches are shown in the bottom panel of Figure~\ref{fig:FullPeriodComparison} (\textit{red points}).
The matched sources tended to be slightly brighter. 
For the unmatched L16 sources, the median was \Vmag{}~$=15.6$~mag, while this value was \Vmag{}~$=15.1$~mag for matched sources.  
Unmatched sources tended to display longer periods of variability.
The median period for unmatched L16 variables was $\bar{P}=4.6$~days, while the median period of matched variables was $\bar{P}=1.5$~day. 
We illustrate the period and magnitude distributions for matched versus unmatched sources in Figure~\ref{fig:LibComp2}.

\section{New Variables}
\label{sec:newvar}

We revealed \newvar{} new variables.
These sources were not identified in any of the variable searches described in Section~\ref{sec:variabilitysearch}. 
Our new variables include two field candidate transiting exoplanets, ten EBs, 55 variables of indeterminate type, 88 \dScuti{} variables, 137 \Gdor{} variables, and 141 rotational variables. 
The exoplanet candidates and EBs are described in further detail in Section~\ref{sec:exo} and Section~\ref{sec:EBs}, respectively.
The fourteen new rotational variables identified as probable NGC~2158 members are particularly valuable given that rotation period measurements are scarce among older stars. 

\begin{figure}
\centering
\includegraphics[width=0.46\textwidth]{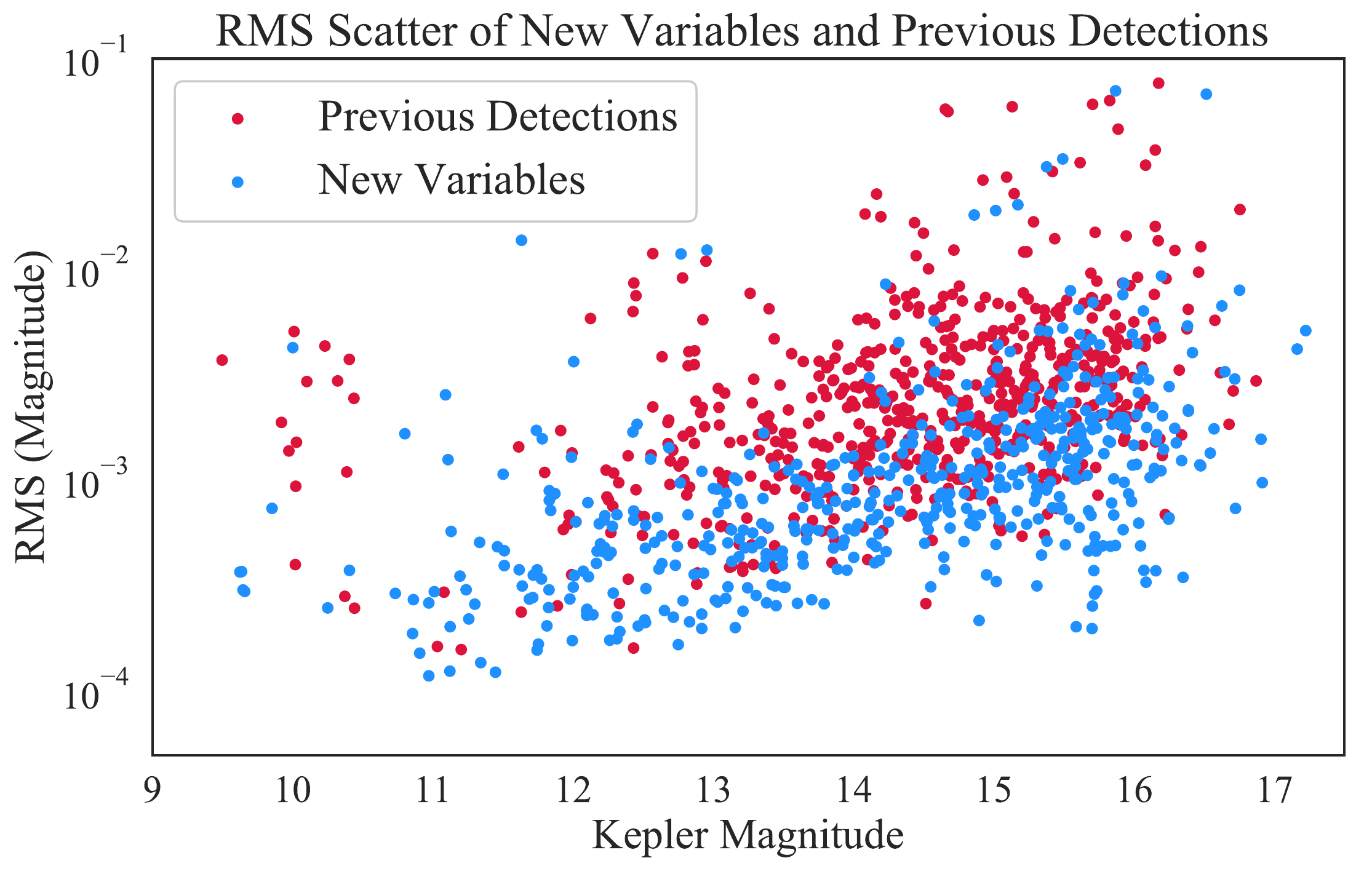}
\caption{The TFA-corrected rms LC scatter for previously detected variables (\textit{red points}) and newly detected variables (\textit{blue points}). 
We define the rms scatter as the $3.5\sigma$ clipped 68.27th percentile of the distribution about the median value of the LC magnitude.
Using bin sizes of one~mag, the median rms for previous detections is a factor of 1.8 to 2.5 larger than that of the new detections. This factor is least for stars of 14~mag--15~mag and greatest for stars of 10~mag--11~mag (\Kepmag{}).}
\label{fig:rms}
\end{figure}

The newly identified variables tended to have smaller amplitude variations than the variables found in prior searches. 
Previously identified variables had a mean oscillation amplitude of ${\sim{}}$13~millimag, while newly identified variables displayed a mean oscillation amplitude of ${\sim{}}$4~millimag. 
Figure~\ref{fig:rms} depicts the TFA-corrected LC rms scatter for previously detected sources (\textit{red points}) to that of the newly detected variables (\textit{blue points}). 
The rms scatter is defined here as the $3.5\sigma$ clipped 68.27th percentile of the distribution about the median value of the LC magnitude.
Comparing the TFA-corrected rms in bin sizes of one~mag, we observed that the median rms of previous detections is larger by a factor of 1.8~--~2.5 than that of our new detections. This factor is smallest for stars with \Kepmag{} magnitude of 14~mag~--~15~mag and greatest for stars with \Kepmag{} of 10~mag~--~11~mag.
The period and \Kepmag{} magnitude distributions of our new variables were similar to that of the unmatched variables. In fact, the mean values of both distributions display $<3\%$ difference.

Our new variables are generally lower in signal-to-noise (S/N).
We found a median LS S/N ratio of 118 for new variables, as compared to 552 for previously identified variables. 
S/N was computed using \texttt{Astrobase} and is described in Section~\ref{ls}.
About 40\% of the newly detected variables are ambiguous blends for which we cannot distinguish a primary source.
In contrast, ambiguous blends represented 27\% of the previously identified variable population.
Using information from Gaia DR2, we find that our new variables tended to be almost twice as distant and $2.5\times$ more luminous.

\section{Exoplanet Candidates}
\label{sec:exo}

\begin{figure*}[]
\centering
\includegraphics[width=0.6\textwidth]{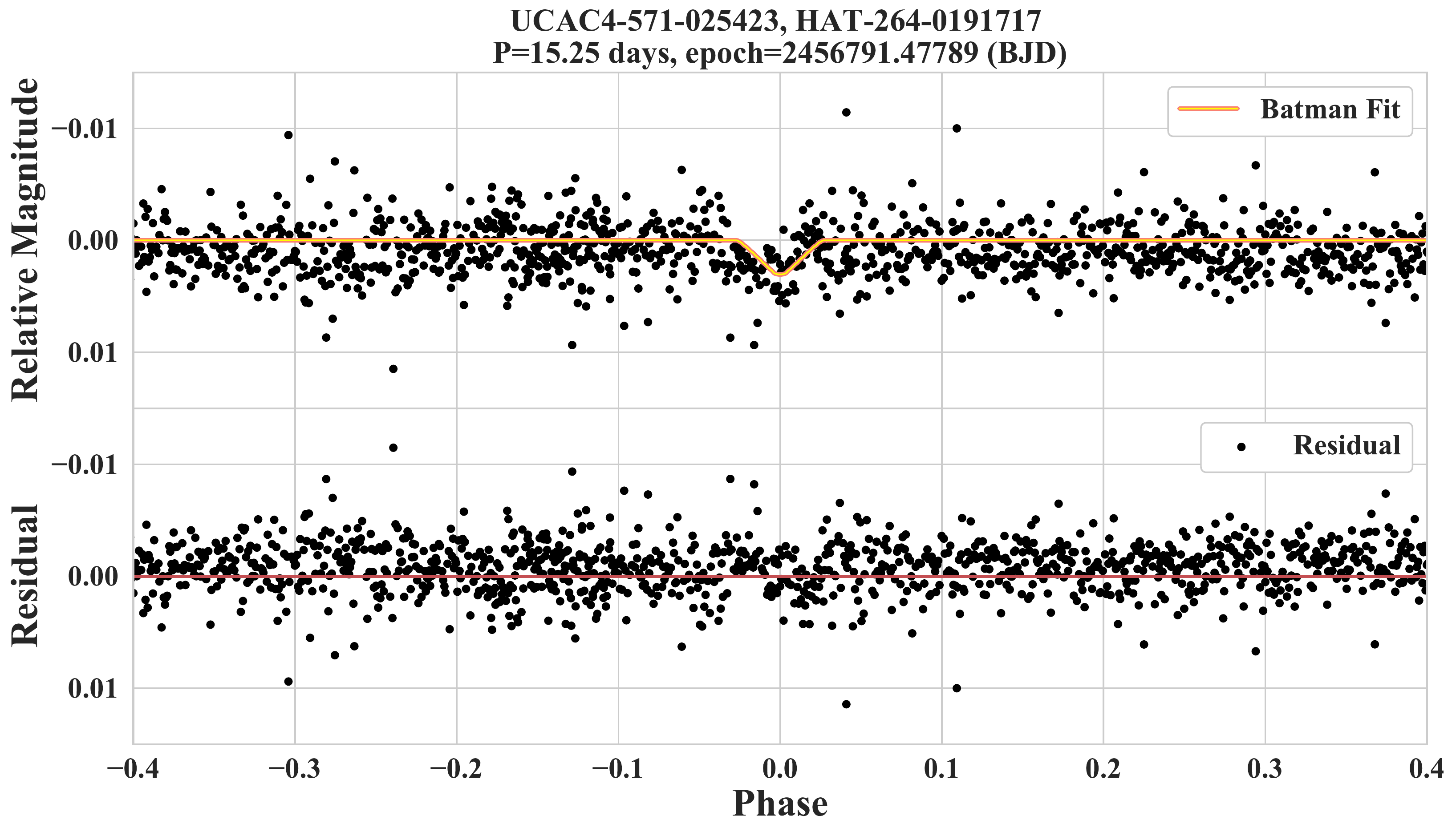}
\caption{\textit{Top:} Phase-folded light curve for UCAC4-571-025423. This is a G9IV type field star exhibiting a \tdBunits{} transit, suggesting a planet of $R_{\rm p}=$~\RpBunits{} with a period of 15.3~days.
The system parameters are listed in Table~\ref{Table:planetparams4}. Only two transits were used in the creation of this light curve. 
This source may be an unresolved binary or a false alarm.
The BATMAN fit is shown in \textit{orange}. 
\textit{Bottom:} The residuals after subtracting the best fit model from the data. This source is a new detection. \\}
\label{fig:FieldTransits2}
\end{figure*}

Our two transiting exoplanet candidates are new variables. 
They are both probable field sources. 
Both have been identified as primary sources (not blended). 
Information is provided for each of them in Table \ref{Table:planetparams1} and Table \ref{Table:planetparams4}, respectively.
Their phase-folded LCs are shown in Figures~\ref{fig:FieldTransits1} and \ref{fig:FieldTransits2}. 
In addition to the Gaia-derived parameters, which are listed in the central panel of the table, we list system parameters obtained from our BLS analysis using \texttt{VARTOOLS} and \texttt{Astrobase}.
We also list parameters obtained from transit modeling with \texttt{BATMAN} (Bad-Ass Transit Model cAlculatioN) \citep{Kreidberg2015}, a Python package that analytically calculates the LC using the methods described in \cite{Mandel2002}.

\begin{table}[t!]
\caption{UCAC4-573-025379}
\centering
\begin{tabularx}{0.35\textwidth}{lr}
\toprule
HATID & HAT-264-0137730 \\
RA (degrees) & 92.428917	\\
DEC (degrees) & 24.528807 \\
\Kepmag{} (mag) & \KpA{} \\
\midrule
Gaia Source ID & 3426297073623137024\\
$T_{\rm eff}$ (K) & \TeffA{} \\
$L_{\star}$ (\Lsun{}) & $0.257~\pm~0.019$ \\
$R_{\star}$ (\Rsun{}) & \RsA{} \\
$A_{G}$ & $0.68~\pm~0.14$ \\ 
E(BP-RP) & $0.34~\pm~0.08$ \\
$\varpi$ (mas) & \varpiA{} \\
Distance (pc) & \distA{} \\
pmra (mas/r) &  $-5.608~\pm~0.084$ \\
pmdec (mas/r) & $-6.437~\pm~0.073$ \\
\midrule
Period (days) & \PeriodA{} \\
Epoch (BJD) & 2456775.234 \\
$t_{\rm depth}$ (millimag) & \tdA{} \\
$t_{\rm dur}$ (hours) & \tdurA{} \\
$R_{\rm  p }$ (\RJ{}) & $1.1~\pm~0.1$  \\
$R_{\rm p}/R_{\star}$ & $0.055~\pm~0.05$ \\
$a/R_{\star}$ & $8^{+4}_{-1}$  \\
Inclination & \incAunits{} \\
\bottomrule
\label{Table:planetparams1}
\end{tabularx}
\end{table} 

\begin{table}[t!]
\caption{UCAC4-571-025423}
\centering
\begin{tabularx}{0.35\textwidth}{lr}
\toprule
HATID & HAT-264-0191717 \\
RA (degrees) & 92.448914\\
DEC (degrees) & 24.185612 \\
\Kepmag{} (mag) &  \KpB{} \\
\midrule
Gaia Source ID & 3425513568507306112\\
$T_{\mathrm{eff}}$ (K) & \TeffB{} \\
$L_{\star}$ (\Lsun{}) & $1.4~\pm~0.4$\\
$R_{\star}$ (\Rsun{}) & \RsB{} \\
$A_{G}$ & $1.7~\pm~0.3$ \\ 
E(BP-RP) & $0.79~\pm~0.15$ \\
$\varpi$ (mas) & \varpiB{} \\
Distance (kpc) & \distB{} \\
pmra (mas/r) & $1.2~\pm~0.1$ \\
pmdec (mas/r) & $-2.5~\pm~0.1$ \\
\midrule
Period (days) &  \PeriodB{} \\
Epoch (BJD) & 2456791.47789\\
$t_{\mathrm{depth}}$ (millimag) & \tdB{} \\
$t_{\mathrm{dur}}$ (hours) & \tdurB{} \\
$R_{\mathrm{p}}$ (\RJ{}) &  \RpB{} \\
$R_{\mathrm{p}}/R_{\star}$ & $0.05~\pm~0.01$ \\
$a/R_{\star}$ & $10~\pm~2$ \\
Inclination & \incBunits{} \\
\bottomrule
\label{Table:planetparams4}
\end{tabularx}
\end{table} 

\subsection{UCAC4-573-025379}
\label{subsec:025379}

\begin{figure*}[t!]
\centering
\includegraphics[width=0.6\textwidth]{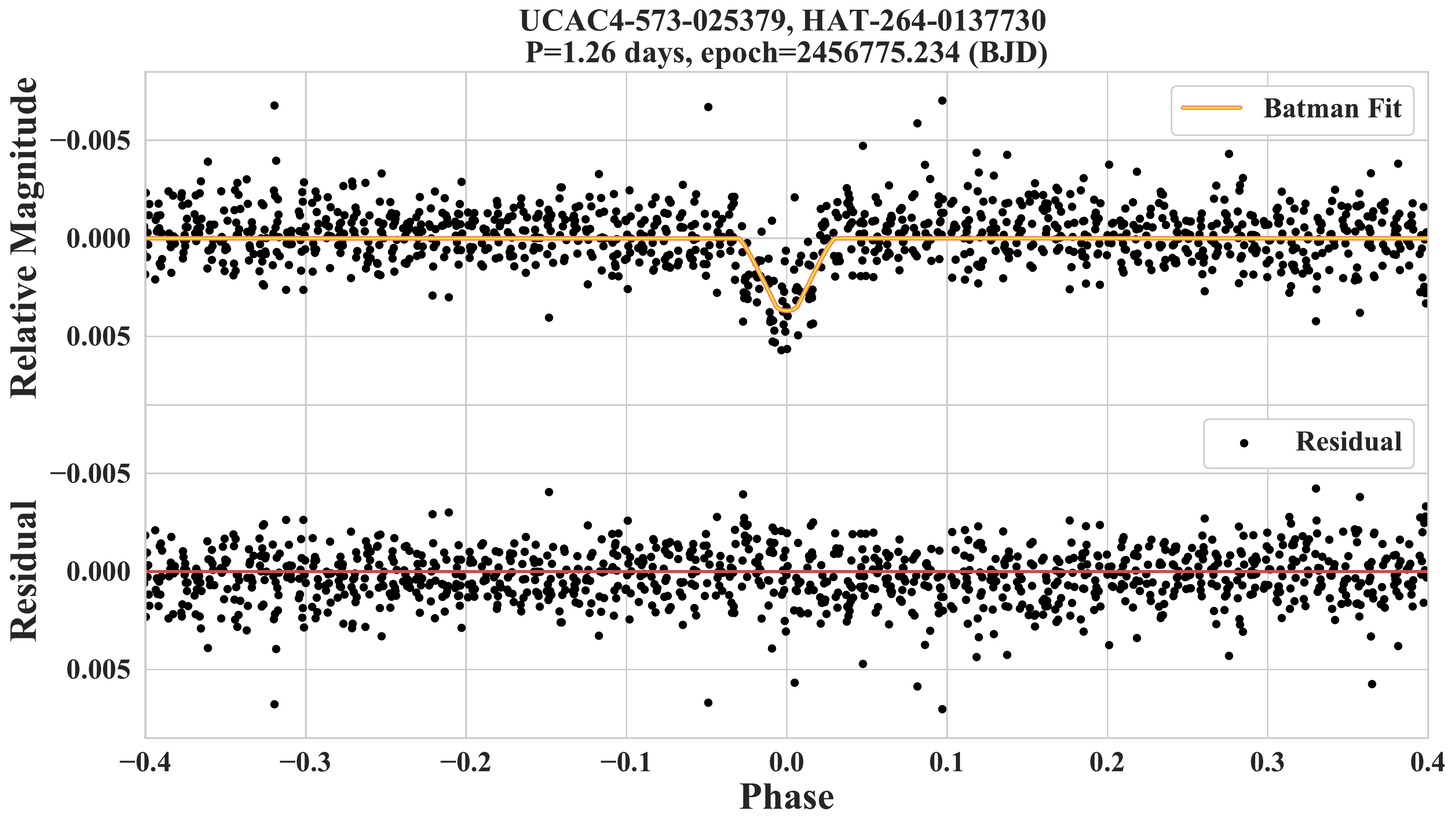}
\caption{\textit{Top:} Phase-folded light curve for UCAC4-573-025379.
This is a K type field dwarf exhibiting a \tdAunits{} transit indicating a planet of $R_{\rm p}=$~\RpAunits{} with a period of 1.3~days. 
The system parameters are listed in Table~\ref{Table:planetparams1}. 
The BATMAN fit is shown in \textit{orange}. 
\textit{Bottom:} The residuals after subtracting the best fit model from the data. This source is a new detection.}
\label{fig:FieldTransits1}
\end{figure*}

Our BLS periodogram search revealed a transiting exoplanet candidate with a period of \PeriodAunits{} for the variable UCAC4-573-025379.
The phase-folded LC for this source is composed of 1,390 cadences and is shown in the top panel of Figure~\ref{fig:FieldTransits1}. 
The residuals after subtracting the best fit model from the data are shown in the bottom panel.  
The \textit{red line} denotes the zero-point for residuals. 
The measured transit depth was \tdAunits{} and the transit duration was measured as \tdurAunits{}. 
The associated star-planet parameters are listed in Table~\ref{Table:planetparams1}. 
The host star has a median brightness of \KpAunits{}. 
Gaia DR2 measurements suggest that the star is a foreground K dwarf with a radius of \RsAunits{}.
Our transit modeling suggests an orbiting planet of \RpAunits{} at an inclination of \incAunits{}.
The star-planet parameters for this system are listed in Table~\ref{Table:planetparams1}. 

UCAC4-573-025379 was matched with the Gaia source 3426297073623137024.
The median $G$~magnitude listed for this source match is 15.56~mag, which is in agreement with our $G$~magnitude calculated using \textit{B~-~V} Johnson-Cousins algorithm.
Gaia provides an effective temperature measurement of \TeffAunits{} and a parallax measurement of \varpiAunits{}~mas, indicating a distance of \distAunits{}. 
The source's parallax and proper motion measurements indicate that is not likely to be a cluster member. 
This source was excluded from the membership catalog generated by \cite{Cantat2018}.

\subsection{UCAC4-571-025423}
\label{subsec:025423}

Our BLS periodogram search revealed a transiting exoplanet candidate with a period of \PeriodBunits{} for the variable UCAC4-571-025423.
The phase-folded LC  for this source is composed of 1,388 cadences and is shown in Figure~\ref{fig:FieldTransits2}.
Also shown in this figure are the residuals for our fit.
The measured transit depth was \tdBunits{} and transit duration was \tdurBunits{}. 
Only two transits were used in the generation of this fit. 
Therefore, while this transiting exoplanet candidate cannot be ruled out, it is not extremely compelling and could turn out to be an unresolved binary or a false alarm.
The host star has a brightness of \KpBunits{}. 
Gaia DR2 measurements suggest that the star is of the type G9IV with a radius of \RsBunits{}. 
Our transit modeling indicates a planet of radius \RpBunits{} at an inclination of \incBunits{}.
The position of this star on an HR diagram of stellar radius versus $T_{\rm eff}$ indicates that some of the stellar properties are probably in error, as this position is not covered by a stellar isochrone younger than the age of the universe. 
The star-planet parameters for this system are listed in Table~\ref{Table:planetparams4}. 

We found a Gaia match for UCAC4-571-025351 (source ID 3425513568507306112).
The mean $G$~magnitude listed for the Gaia match is 16.12~mag, which is in agreement with our calculated $G$~magnitude.
The Gaia parallax measurement contained large error bounds, measuring as $\varpi=$\varpiB{}~mas, indicating a distance of \distB{}~kpc. 
The Gaia parallax and proper motion measurements support the notion that this is likely a field star.
The King model fit indicates that the source is unlikely to be an NGC~2158 cluster member ($P_{\mathrm{m}}<1\%$). 
Gaia provides a wealth of stellar parameters, including the effective temperature, measured as \TeffBunits{}, stellar luminosity ($L_{\star}=1.4~\pm~0.4$~\Lsun{}), the line-of-sight extinction in the $G$ band ($A_{G}=1.67~\pm~0.3$), and a line-of-sight reddening of E(BP-RP)=~$0.82~\pm~0.15$.

\section{Eclipsing Binaries}
\label{sec:EBs}
Of our \totEB{} catalog EBs, ten are new detections and five are probable cluster members.
The sole M35 cluster member EB is a new detection with a cluster membership probability of $P_{\rm m}=0.8$.
Of the \EBN{} NGC~2158 member EBs, one was a new detection and three had been previously identified.
Spectroscopic follow-up has not been conducted on any of the EBs in our variable catalog. 
We found four likely contact binaries, none of which were new detections and all are likely field members.
Our catalog contact binaries include: UCAC4-571-025495, UCAC4-571-024709, UCAC4-573-025524, and UCAC4-571-025426. 
We found one semi-detached EB, UCAC4-573-025574, which is a likely field source.  
One of the field EBs is highly eccentric.
Five EB variables show signs of out-of-transit variation (OOTV), including UCAC4-571-024971, UCAC4-572-024695, UCAC4-573-025041, UCAC4-571-024688, UCAC4-572-024929.
UCAC4-571-024971 also shows the characteristics of a reflection variable.
Eight of our EB variables show a secondary eclipse with a phase offset from 0.5 --- one of which, UCAC4-571-025548, is highly eccentric. 
The other EBs showcasing a phase offset include the newly detected field source UCAC4-573-025427, which is an eccentric EB, and the previously observed sources UCAC4-571-025474, UCAC4-572-025126, UCAC4-571-024775, UCAC4-571-024823, UCAC4-571-024819, and UCAC4-571-024606. 
UCAC4-571-024606 is a probable member of NGC~2158.

\begin{figure}
\centering
\includegraphics[width=0.47\textwidth]{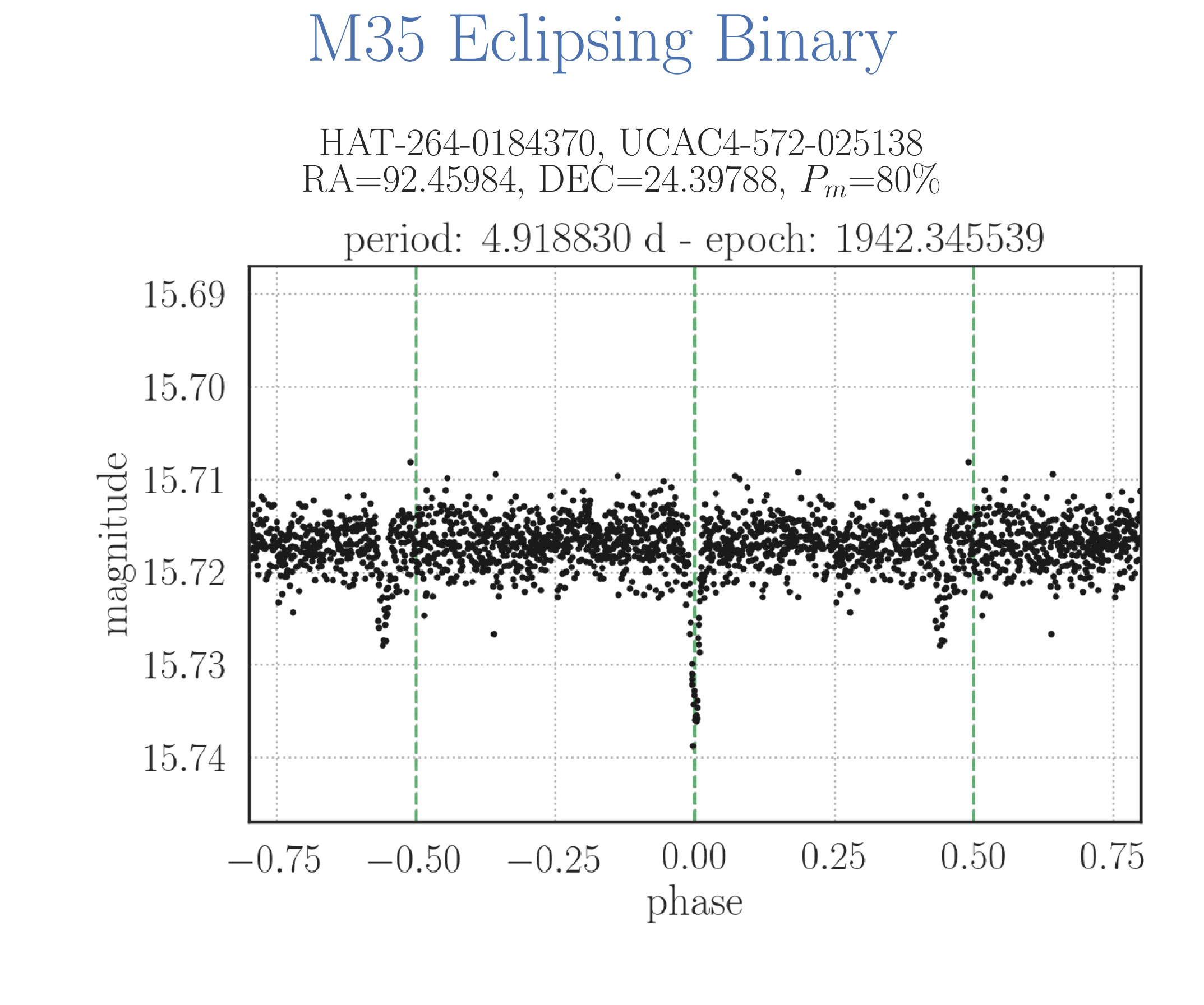}
\caption{Phase-folded LC of our sole M35 EB source (UCAC4-572-025138) with $P=4.9$~days. 
Epoch is provided in BJD-2454833. 
Plotted along the y-axis is the \Kepmag{} magnitude.  
This source is an ambiguous blend with two field EBs.
An offset from 0.5 in phase is seen in the secondary eclipse. 
This M35 variable was not reported in prior analyses. 
Cluster membership is supported by Gaia DR2 proper motion and parallax measurements. 
}
\label{fig:M35EBs}
\end{figure}

UCAC4-572-025138, the sole M35 EB source with $P=4.9$~days, is shown in Figure~\ref{fig:M35EBs}. This was a new detection with \Kepmag{}~$=15.7$~mag.
Unfortunately, the source is ambiguously blended with two field EBs.
The secondary eclipses for this EB are offset by a phase of 0.5. 
Cluster membership is supported by a proper motion analysis and the recent parallax provided by Gaia DR2 ($\varpi=1.241\pm0.090$~mas).

\begin{figure*}
\centering
\includegraphics[width=0.95\textwidth]{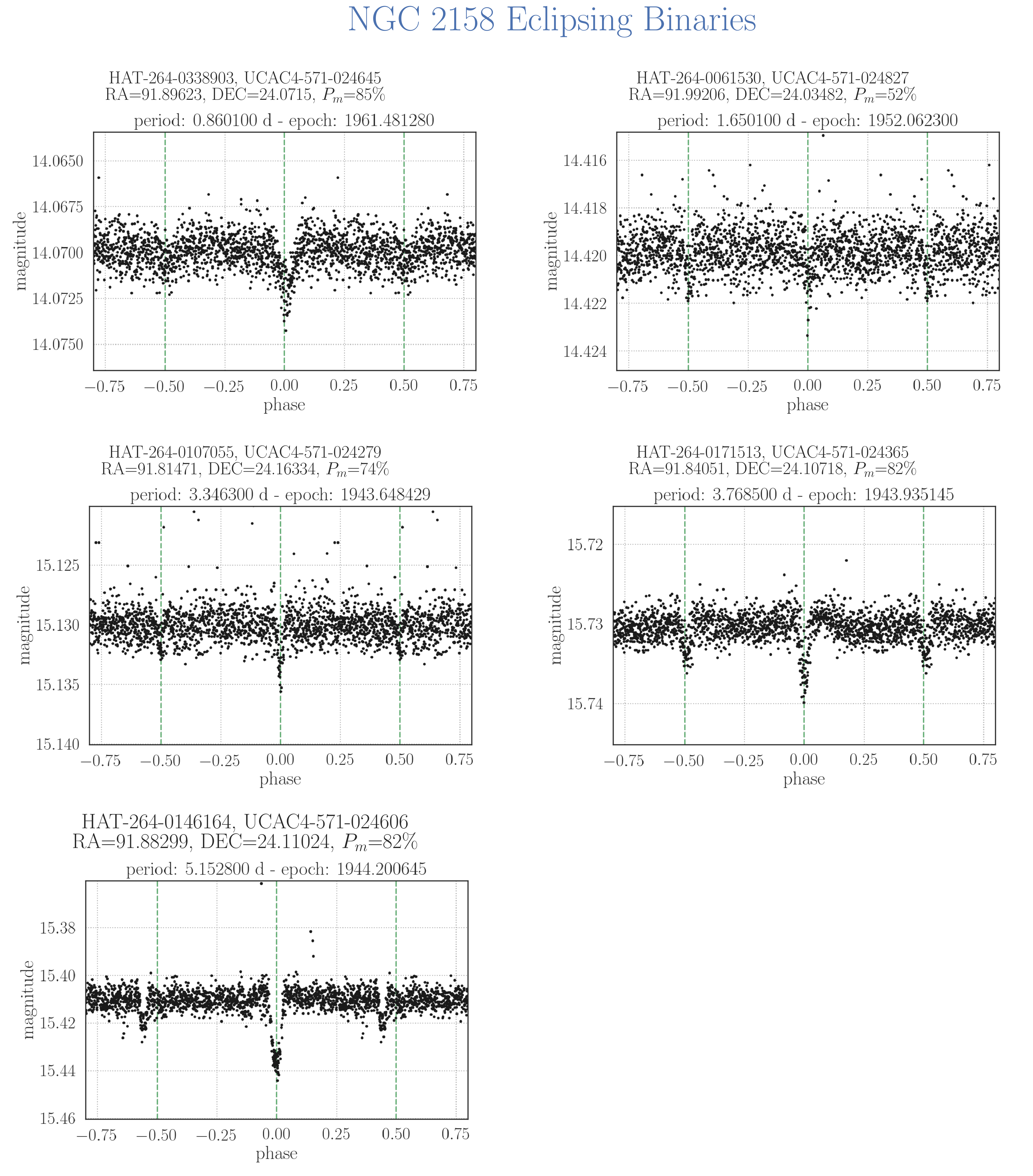}
\caption{Phase-folded LCs for NGC~2158 associated EBs. Epochs are provided in BJD-2454833.  
The y-axis displays the \Kepmag{} magnitude. The EB UCAC4-571-024279 (central left panel) is a new detection.}
\vspace*{4mm}  
\label{fig:NEBs}
\end{figure*}

Our probable NGC~2158 member EBs are shown in Figure~\ref{fig:NEBs}.
They are illustrated in order of ascending period. 
These sources range in \Kepmag{} from 14~mag~--~15.7~mag.
All the NGC~2158 EBs but UCAC4-571-024279 were previously observed by L16.
The Gaia DR2 measured parallax for UCAC4-571-024645 suggests that this source may actually be more distant than the cluster. The distance estimate has large error bounds, however, ranging between 6~kpc~--~10~kpc. Stellar density estimates indicate that this source is an early K dwarf or late G dwarf. 
UCAC4-571-024279 is a new detection and the parallax for this source supports NGC~2158 cluster membership.	
UCAC4-571-024365 has a parallax measurement that supports cluster membership. This EB is ambiguously blended.
A subtle OOTV can be seen in the phase-folded LC.
UCAC4-571-024606 displays a secondary eclipse offset from a phase of 0.5. The parallax of this source supports cluster membership. 
UCAC4-571-024827 has a low membership probability of $P_{m}=0.52$ and may be a field source. 

The phase-folded LCs for our population of field EBs are illustrated throughout Figure~\ref{fig:Field_EBs1}~--~Figure~\ref{fig:Field_EBs5}. 
We show only the brightest member of an ambiguously blended pair/group.
Our field EBs are predominately F and G type variables. 
The brightest field EB is UCAC4-571-025426, a subgiant of class G3, with $L_{\star}=10\pm 2$~\Lsun{} and $T_{\rm eff}=5200~\pm~200$~K. 
This is a semi-detached EB that was observed in prior analyses. 
The second brightest EB, UCAC4-572-025069, is a G class subgiant that shows some ellipsoidal variation. 
This source has $L_{\star}=6.5\pm 0.5$~\Lsun{} and $T_{\rm eff}=4950~\pm~50$~K. 
UCAC4-571-024330 is another EB with OOT ellipsoidal variation. 
The wiggles in this phase-folded LC imply spot modulation.

\section{K2C0 $\delta$ Cepheid Variable}
\label{subsec:Cepheid}

Throughout the duration of the Kepler mission, hundreds of Cepheid variables were observed \citep{Molnar2018}.
Shown in Figure~\ref{fig:Cepheid} is the phase-folded LC of the previously discovered $\delta$~Cepheid variable, V0371 Gem, which is listed in the VSX catalog.
This bright source (\Kepmag{}~=~9.906~mag) is classified as K0 in spectral type.
V0371 Gem is a probable field star. 
The source was also observed by L16 and N15. 
Our periodogram search reveals a period of 2.141186~days for this object. The VSX catalog lists a very similar value of $P=2.1371$~days. 

Filtering out the period and associated harmonics for $P=2.141186$~days, we also observed an overtone of $P=1.288193$~days, which has not been previously reported.
The overtone is shown in the lower panel of Figure~\ref{fig:Cepheid}.
Overtones have been noted in $\delta$ Cepheid variables in the past. 
Double-mode Cepheids are a useful resource for testing stellar evolutionary and pulsation models \citep[e.g.][]{Buchler2007,Buchler2008, Smolec2010}.
V0371 Gem was matched in the Gaia DR2 dataset with the source 3425495186047291136. The Gaia measured parallax indicates a distance of $4~\pm~0.5$~kpc, similar to NGC~2158, however, both the 2D projected distance of the source from the cluster core and the proper motion measurements rule out cluster membership. 
For reference, the source's proper motion is calculated to be $(\mu_{\alpha} \cos \delta,\mu_{\delta})=(-0.54,-1.40)$~mas/yr. 
Gaia analysis estimates a surface temperature of $T_{\rm eff}{\sim}5000$~K, a radius of $R_{\star}{\sim}40$~\Rsun{}, and a luminosity of $L_{\star}=818~\pm~166$~\Lsun{}.
No new Cepheids were found in our variable search, which is unsurprising given that the high luminosity of these sources often permits detection even when blending concerns are not mitigated.

\section{K2C0 Rotational and Pulsational Variables}
\label{sec:RotVarPulse}

We illustrate the phase-folded LCs for three representative \dScuti{} variables, three representative \Gdor{} variables, and three representative rotational variables from our K2C0 variable catalog. 
These represent highlighted examples, as many more are present in our catalog.
The first two columns illustrate candidates observed in prior detections, while the final column illustrates a newly detected variable. 

Figure~\ref{fig:DScuti} displays our three representative \dScuti{} variables. 
In the left panel is UCAC4-571-024831 with a period of $0.1741$~days and an amplitude of 8~millimag.
A Gaia DR2 match for this source measures $T_{\rm eff}{\sim}5800$~K and $L_{\star}{\sim}7.7$~\Lsun{}.
In the center panel is UCAC4-572-024661 with a period of $0.099$~days and an amplitude of 0.009~mag. 
A Gaia DR2 match for this source measures $T_{\rm eff}{\sim}5400$~K and $L_{\star}{\sim}22$~\Lsun{}. 
In the right panel is UCAC4-573-025256 with a period of $0.0584$~days and an amplitude of 7~millimag.
This variable is a new detection. 
A Gaia DR2 match for this source measures $T_{\rm eff}{\sim}7700$~K and $L_{\star}{\sim}11$~\Lsun{}.
The first two panels are field sources, while the third panel has an M35 affiliated membership probability of $P_{m}=0.6$.

Figure~\ref{fig:GDor} displays our three representative \Gdor{} variables. 
In the left panel is UCAC4-571-024231 with a period of $0.3791$~days and an amplitude of 8~millimag.
The Gaia DR2 match for this source measures $T_{\rm eff}{\sim}5300$~K and $L_{\star}{\sim}4$~\Lsun{}.
In the center panel is UCAC4-571-024996 with a period of $0.6363$~days and an amplitude of 7~millimag.
The Gaia DR2 match for this source measures $T_{\rm eff}{\sim}5200$~K and $L_{\star}{\sim}2$~\Lsun{}.
In the right panel is UCAC4-571-025700 with a period of $0.7011$~days and an amplitude of 0.01~mag. 
This variable is a new detection. 
The Gaia DR2 match for this source measures $T_{\rm eff}{\sim}5800$~K and $L_{\star}{\sim}4$~\Lsun{}.
All variables are likely field sources.

Figure~\ref{fig:RotVar} displays our three representative rotational variables. 
In the left panel is UCAC4-573-025241 with a period of $0.42366$~days and an amplitude of 0.06~mag.
The Gaia match for this source measures $T_{\rm eff}{\sim}4300$~K and $L_{\star}{\sim}0.2$~\Lsun{}.
UCAC4-572-025122 is shown in the center panel with a period of $2.08484$~days and an amplitude of 0.01~mag. 
This rotational variable is a member of the M35 cluster, as confirmed by Gaia DR2 proper motion measurements. 
The Gaia DR2 match for this source measures $T_{\rm eff}{\sim}5370$~K and $L_{\star}{\sim}2$~\Lsun{}.
Shown in the right panel is UCAC4-567-023197, a newly detected rotational variable with a period of $2.990$~days and an amplitude of 0.01~mag. 
The Gaia DR2 match for this source measures $T_{\rm eff}{\sim}5900$~K and $L_{\star}{\sim}2$~\Lsun{}.
This rotational variable is a probable field star.

\section{Summary and Conclusions}
\label{sec:summary}

We present the results of a variability search conducted on stars in the K2C0 super stamp. 
This analysis was performed on our publicly released image-subtracted LCs for 3,960 sources.
The source magnitudes range between \Vmin{}~mag~$<$~\Vmag{}~$<$\Vmax{}~mag.
We searched for periodic signatures between 0.03~days to 31~days. 
We detected and classified a total of \totcat{} periodic variables. 
Cluster membership was ascertained for our catalog sources.
Our image subtraction analysis and variability search reveals periodic variables that remain undetected in crowded fields using other reduction methods. Therefore, we stress the need to employ specialized techniques such as image subtraction to fully exploit crowded stellar fields. 

Within our catalog, \newvar{} of the variables are new detections. 
While no new $\delta$~Cepheid variables were found, our analysis revealed an overtone associated with the previously detected source V0371 Gem. 
Our catalog contains a total of \tottransit{} newly detected candidate transiting exoplanets, \totEB{} EBs, and \totvar{} stellar variables (primarily rotating and pulsating sources). 
Of our \totEB{} EBs, seven sources are new detections, including one probable M35 member and one probable NGC~2158 member. 
Among our variable catalog sources, we found \totM{} members of M35 and \totN{} members of NGC~2158. 
A truncated version of our digital table is shown in Table~\ref{table:cattab} of the Appendix.  
The flexible edition of our variable catalog is digitally hosted at \url{https://k2.hatsurveys.org/archive/}.

\acknowledgements
\paragraph{Acknowledgments}
We thank A.~Mao, J.~Auman, J.~Hoffman, C.~Malecki, R.~Mathieu, and J.~Wallace for the insightful conversations related to this work.
This material is based upon work supported by the National Science Foundation Graduate Research Fellowship Program under Grant No.~DGE-1656466.
Any opinions, findings, and conclusions or recommendations expressed in this material are those of the authors and do not necessarily reflect the views of the National Science Foundation.
The data in our analysis were provided by the K2 mission, funded by the NASA Science Mission directorate. Photometric data were downloaded from the Barbara A.~Mikulski Archive for Space Telescopes (MAST). 
This publication makes use of data products from the Two Micron All Sky Survey, which is a joint project of the University of Massachusetts and the Infrared Processing and Analysis Center/California Institute of Technology, funded by the National Aeronautics and Space Administration and the National Science Foundation.
This research has made use of NASA's Astrophysics Data System Bibliographic Services. For our cluster information, we have made extensive use of the WEBDA database, operated at the Department of Theoretical Physics and Astrophysics of the Masaryk University. 
This paper has made use of data from the European Space Agency (ESA)
mission {\it Gaia} (\url{https://www.cosmos.esa.int/gaia}), processed by
the {\it Gaia} Data Processing and Analysis Consortium (DPAC,
\url{https://www.cosmos.esa.int/web/gaia/dpac/consortium}). Funding
for the DPAC has been provided by national institutions, in particular
the institutions participating in the {\it Gaia} Multilateral Agreement.
This research has made use of NASA's Astrophysics Data System
Bibliographic Services.
Lastly, we would like to acknowledge the VizieR catalogue access tool, CDS, Strasbourg, France. The original description of the VizieR service was published in A\&AS 143, 23. 

\facilities{Kepler (K2), 
    Gaia, 
    MAST, 
    ADS, 
    Exoplanet Archive
}
\software{Astrobase\footnote{\url{https://pypi.org/project/astrobase}} \citep{Bhatti2018}, 
    Astropy\footnote{\url{http://www.astropy.org}} \citep{astropy2018}, 
    BATMAN \citep{Kreidberg2015},
    FITSH\footnote{\url{https://fitsh.net}} \citep{pal},
    Matplotlib \citep{matplotlib}, 
    Numpy \citep{numpy}
    Pandas \citep{Pandas}, 
    Scikit-learn \citep{scikit-learn},
    VARTOOLS\footnote{\url{https://www.astro.princeton.edu/~jhartman/vartools.html}} \citep{vartools}
} 


\pagebreak

\appendix

\setlength{\tabcolsep}{0.81em}
\begin{table}[H]
\label{table:cattab}
\caption{A truncated version of the periodic variable catalog obtained from sources in the K2 Campaign-0 super stamp. A description of the column contents can be found below.}
\begin{tabularx}{\textwidth}{lrrrrrrrr}
\toprule
\multicolumn{1}{l}{\textbf{}} & 
\multicolumn{1}{c}{\textbf{1}} & 
\multicolumn{1}{c}{\textbf{2}} & 
\multicolumn{1}{c}{\textbf{3}} & 
\multicolumn{1}{c}{\textbf{4}} & 
\multicolumn{1}{c}{\textbf{5}} & 
\multicolumn{1}{c}{\textbf{6}} & 
\multicolumn{1}{c}{\textbf{7}} & 
\multicolumn{1}{c}{\textbf{8}} \\
\multicolumn{1}{l}{\textbf{No.}} & 
\multicolumn{1}{c}{\textbf{HATID}} & 
\multicolumn{1}{c}{\textbf{UCACID}} & 
\multicolumn{1}{c}{\textbf{RA}} & 
\multicolumn{1}{c}{\textbf{DEC}} & 
\multicolumn{1}{c}{\textbf{Kepmag}} & 
\multicolumn{1}{c}{\textbf{P}} & 
\multicolumn{1}{c}{\textbf{dmag}} & 
\multicolumn{1}{c}{\textbf{Class}} \\
\midrule
0 &  HAT-264-0001505 &  UCAC4-571-025584 &  92.58066 &  24.02087 & 9.91 &  2.1412 &  0.2808 &  Pulsating \\
1 &  HAT-264-0246791 &  UCAC4-573-025534 &  92.52146 &  24.59545 & 15.84 &  0.0356 &  0.0012 &  Misc \\
2 &  HAT-264-0060077 &  UCAC4-571-024903 &  92.06060 &  24.05245 & 13.92 &  0.1301 &  0.0109 &  Misc \\
3 &  HAT-264-0133407 &  UCAC4-571-025201 &  92.28923 &  24.17921 & 15.20 &  0.0899 &  0.0058 &  Misc \\
4 &  HAT-264-0038220 &  UCAC4-572-024259 &  91.93485 &  24.39266 & 13.86 &  0.2156 &  0.0007 & Misc \\
\end{tabularx}

\setlength{\tabcolsep}{0.66em}
\begin{tabularx}{\textwidth}{lrrrrrrrrrr}
\toprule
\multicolumn{1}{l}{\textbf{}} & 
\multicolumn{1}{c}{\textbf{9}} & 
\multicolumn{1}{c}{\textbf{10}} & 
\multicolumn{1}{c}{\textbf{11}} & 
\multicolumn{1}{c}{\textbf{12}} & 
\multicolumn{1}{c}{\textbf{13}} & 
\multicolumn{1}{c}{\textbf{14}} & 
\multicolumn{1}{c}{\textbf{15}} & 
\multicolumn{1}{c}{\textbf{16}} &
\multicolumn{1}{c}{\textbf{17}} &
\multicolumn{1}{c}{\textbf{18}} \\
\multicolumn{1}{l}{\textbf{No.}} & 
\multicolumn{1}{c}{\textbf{Class\_string}} & 
\multicolumn{1}{c}{\textbf{Blend}} & 
\multicolumn{1}{c}{\textbf{LSP1}} & 
\multicolumn{1}{c}{\textbf{LSFAP1}} & 
\multicolumn{1}{c}{\textbf{LSSNR1}} & 
\multicolumn{1}{c}{\textbf{LSP2}} & 
\multicolumn{1}{c}{\textbf{LSFAP2}} & 
\multicolumn{1}{c}{\textbf{LSSNR2}} &
\multicolumn{1}{c}{\textbf{LSP3}} &
\multicolumn{1}{c}{\textbf{LSFAP3}} \\
\midrule
0 & Cepheid & 0 &  2.1268 & 9.7e-256 &  1358.1 &  2.3649 & 2.3e-46 & 173.8 &  1.9441 & 1.1e-18 \\
1& Misc & 0 &  0.1195 & 3.9e-12 & 60.1 &  0.1288 & 1.0e-08 & 45.1 &  3.2670 &  5.3e-02 \\
2& Misc & 0 &  0.1301 & 6.6e-310 &  3045.6 &  0.1008 & 1.0e-138 & 997.1 &  0.1309 & 2.5e-22 \\
3& Misc & 0 &  0.1420 &  9.3e-01 & 12.8 &  0.3375 &  1.0e+00 &  8.2 &  0.1197 &  1.0e+00 \\
4& Misc & 0 &  0.2156 & 3.6e-11 & 36.4 &  0.9239 & 2.3e-07 & 25.4 &  0.6106 & 4.3e-05 \\
\end{tabularx}

\setlength{\tabcolsep}{0.71em}
\begin{tabularx}{\textwidth}{lrrrrrrrrrr}
\toprule
\multicolumn{1}{l}{\textbf{}} & 
\multicolumn{1}{c}{\textbf{19}} & 
\multicolumn{1}{c}{\textbf{20}} & 
\multicolumn{1}{c}{\textbf{21}} & 
\multicolumn{1}{c}{\textbf{22}} & 
\multicolumn{1}{c}{\textbf{23}} & 
\multicolumn{1}{c}{\textbf{24}} & 
\multicolumn{1}{c}{\textbf{25}} & 
\multicolumn{1}{c}{\textbf{26}} &
\multicolumn{1}{c}{\textbf{27}} \\
\multicolumn{1}{l}{\textbf{No.}} & 
\multicolumn{1}{c}{\textbf{LSSNR3}} & 
\multicolumn{1}{c}{\textbf{PDM1}} & 
\multicolumn{1}{c}{\textbf{PDM2}} & 
\multicolumn{1}{c}{\textbf{PDM3}} & 
\multicolumn{1}{c}{\textbf{BLSP1}} & 
\multicolumn{1}{c}{\textbf{BLS1TDur}} & 
\multicolumn{1}{c}{\textbf{BLS1TDepth}} & 
\multicolumn{1}{c}{\textbf{BLSP2}} &
\multicolumn{1}{c}{\textbf{BLS2TDur}} \\
\midrule
0 &     0.0 &  2.1412 &  4.2824 &  0.0405 & NaN & NaN & NaN & NaN & NaN \\
1 &     0.0 &  0.0523 &  0.1916 &  0.3053 & NaN & NaN & NaN & NaN & NaN \\
2 &     0.0 &  0.1301 &  0.2602 &  0.3903 & NaN & NaN & NaN & NaN & NaN \\
3 &     0.0 &  0.0899 &  0.2695 &  0.5389 & NaN & NaN & NaN & NaN & NaN \\
4 &     0.0 &  3.0471 &  7.9224 &  0.1306 & NaN & NaN & NaN & NaN & NaN \\
\end{tabularx}

\setlength{\tabcolsep}{0.56em}
\begin{tabularx}{\textwidth}{lrrrrrrrrrrrr}
\toprule
\multicolumn{1}{l}{\textbf{}} & 
\multicolumn{1}{c}{\textbf{28}} & 
\multicolumn{1}{c}{\textbf{29}} & 
\multicolumn{1}{c}{\textbf{30}} & 
\multicolumn{1}{c}{\textbf{31}} & 
\multicolumn{1}{c}{\textbf{32}} & 
\multicolumn{1}{c}{\textbf{33}} & 
\multicolumn{1}{c}{\textbf{34}} & 
\multicolumn{1}{c}{\textbf{35}} &
\multicolumn{1}{c}{\textbf{36}} &
\multicolumn{1}{c}{\textbf{37}} &
\multicolumn{1}{c}{\textbf{38}} \\
\multicolumn{1}{l}{\textbf{No.}} & 
\multicolumn{1}{c}{\textbf{BLS2TDur}} & 
\multicolumn{1}{c}{\textbf{BLSP3}} & 
\multicolumn{1}{c}{\textbf{BLS3TDur}} & 
\multicolumn{1}{c}{\textbf{BLS3TDepth}} & 
\multicolumn{1}{c}{\textbf{Jmag}} & 
\multicolumn{1}{c}{\textbf{errJ}} & 
\multicolumn{1}{c}{\textbf{Hmag}} & 
\multicolumn{1}{c}{\textbf{errH}} & 
\multicolumn{1}{c}{\textbf{Kmag}} &
\multicolumn{1}{c}{\textbf{errK}} &
\multicolumn{1}{c}{\textbf{Bmag}} \\
\midrule
0 & NaN & NaN & NaN & NaN &   8.91 &  0.02 &   8.60 &  0.02 &   8.47 &  0.02 &  11.61 \\
1 & NaN & NaN & NaN & NaN &  15.17 &  0.04 &  14.84 &  0.05 &  14.85 &  0.08 &  15.80 \\
2 & NaN & NaN & NaN & NaN &  13.31 &  0.02 &  13.06 &  0.03 &  13.04 &  0.03 &  15.36 \\
3 & NaN & NaN & NaN & NaN &  14.06 &  0.03 &  13.74 &  0.03 &  13.57 &  0.03 &  16.70 \\
4 & NaN & NaN & NaN & NaN &  12.02 &  0.02 &  11.31 &  0.02 &  11.13 &  0.02 &  16.22 \\
\end{tabularx}

\setlength{\tabcolsep}{0.76em}
\begin{tabularx}{\textwidth}{lrrrrrrrrrrrrr}
\toprule
\multicolumn{1}{l}{\textbf{}} & 
\multicolumn{1}{c}{\textbf{39}} & 
\multicolumn{1}{c}{\textbf{40}} & 
\multicolumn{1}{c}{\textbf{41}} & 
\multicolumn{1}{c}{\textbf{42}} & 
\multicolumn{1}{c}{\textbf{43}} & 
\multicolumn{1}{c}{\textbf{44}} & 
\multicolumn{1}{c}{\textbf{45}} & 
\multicolumn{1}{c}{\textbf{46}} &
\multicolumn{1}{c}{\textbf{47}} &
\multicolumn{1}{c}{\textbf{48}} &
\multicolumn{1}{c}{\textbf{49}} \\
\multicolumn{1}{l}{\textbf{No.}} & 
\multicolumn{1}{c}{\textbf{Vmag}} & 
\multicolumn{1}{c}{\textbf{Imag}} & 
\multicolumn{1}{c}{\textbf{GaiaID}} & 
\multicolumn{1}{c}{\textbf{d1}} & 
\multicolumn{1}{c}{\textbf{d2}} & 
\multicolumn{1}{c}{\textbf{Teff}} & 
\multicolumn{1}{c}{\textbf{T1}} & 
\multicolumn{1}{c}{\textbf{T2}} & 
\multicolumn{1}{c}{\textbf{Rad}} &
\multicolumn{1}{c}{\textbf{R1}} &
\multicolumn{1}{c}{\textbf{R2}} \\
\midrule
0 &  10.74 &   9.50 &  3425495186047291392 &  3495.3 &   4551.7 &  4897.6 &  4853.4 &  5007.5 &  39.7 &  38.0 &  40.5 \\
1 &  14.81 &  15.56 &  3426302910482311168 &  2355.7 &   4260.8 &  5202.9 &  4731.7 &  5406.0 &   NaN &   NaN &   NaN \\
2 &  14.67 &  13.66 &  3426255562760045056 &  3114.3 &   4304.8 &  5269.0 &  4957.5 &  5454.0 &   NaN &   NaN &   NaN \\
3 &  16.04 &  14.72 &  3426263362420856320 &  3993.6 &   7541.5 &  4960.5 &  4879.0 &  5007.0 &   NaN &   NaN &   NaN \\
4 &  14.69 &  13.11 &  3426279992534733312 &  2697.6 &   3179.7 &  4092.8 &  3993.0 &  4249.0 &   9.0 &   8.4 &   9.4 \\
\end{tabularx}

\setlength{\tabcolsep}{0.94em}
\begin{tabularx}{\textwidth}{lrrrrrrrrr}
\toprule
\multicolumn{1}{l}{\textbf{}} & 
\multicolumn{1}{c}{\textbf{50}} &
\multicolumn{1}{c}{\textbf{51}} &
\multicolumn{1}{c}{\textbf{52}} & 
\multicolumn{1}{c}{\textbf{53}} & 
\multicolumn{1}{c}{\textbf{54}} & 
\multicolumn{1}{c}{\textbf{55}} & 
\multicolumn{1}{c}{\textbf{56}} & 
\multicolumn{1}{c}{\textbf{57}} &
\multicolumn{1}{c}{\textbf{58}} \\
\multicolumn{1}{l}{\textbf{No.}} & 
\multicolumn{1}{c}{\textbf{Lum}} &
\multicolumn{1}{c}{\textbf{L1}} &
\multicolumn{1}{c}{\textbf{L2}} & 
\multicolumn{1}{c}{\textbf{EBV}} & 
\multicolumn{1}{c}{\textbf{PMRA}} & 
\multicolumn{1}{c}{\textbf{PMDEC}} & 
\multicolumn{1}{c}{\textbf{PM35CG18}} & 
\multicolumn{1}{c}{\textbf{PM35CM09}} & 
\multicolumn{1}{c}{\textbf{PM35K13}} \\
\midrule
0 &  818.7 &  652.1 &  985.2 &  0.597 &  0.544 & -1.398 &       NaN &       0.0 &      9.8 \\
1 &    NaN &    NaN &    NaN &  0.564 &  0.599 & -0.629 &       NaN &       0.0 &     74.1 \\
2 &    NaN &    NaN &    NaN &  0.619 &  0.598 & -2.093 &       NaN &      12.0 &     29.6 \\
3 &    NaN &    NaN &    NaN &  0.499 &  0.364 & -2.107 &       NaN &      14.0 &     78.8 \\
4 &   20.5 &   16.3 &   24.6 &  0.535 & -1.827 & -5.453 &       NaN &       0.0 &     36.7 \\
\bottomrule
\end{tabularx}
\end{table}

\setlength{\tabcolsep}{0.85em}
\begin{table}[h!]
\begin{tabularx}{\textwidth}{lrrrrrrrr}
\toprule
\multicolumn{1}{l}{\textbf{}} & 
\multicolumn{1}{c}{\textbf{59}} & 
\multicolumn{1}{c}{\textbf{60}} & 
\multicolumn{1}{c}{\textbf{61}} & 
\multicolumn{1}{c}{\textbf{62}} & 
\multicolumn{1}{c}{\textbf{63}} &
\multicolumn{1}{c}{\textbf{64}} & 
\multicolumn{1}{c}{\textbf{65}} \\
\multicolumn{1}{l}{\textbf{No.}} & 
\multicolumn{1}{c}{\textbf{PM35K13b}} & 
\multicolumn{1}{c}{\textbf{PN2158CG18}} & 
\multicolumn{1}{c}{\textbf{PN2158D06}} & 
\multicolumn{1}{c}{\textbf{PN2158K13}} & 
\multicolumn{1}{c}{\textbf{PN2158K13b}} & 
\multicolumn{1}{c}{\textbf{PM35}} &
\multicolumn{1}{c}{\textbf{PN2158}} \\
\midrule
0 &       0.0 &         NaN &        NaN &        NaN &         NaN &    0.0 &     0.0 \\
1 &       0.0 &         NaN &        NaN &        NaN &         NaN &    0.0 &     0.0 \\
2 &       0.3 &         NaN &        NaN &       23.1 &         0.0 &   12.0 &     8.6 \\
3 &       0.0 &         NaN &        NaN &       89.7 &        99.2 &   14.0 &     0.0 \\
4 &       0.0 &         NaN &        NaN &       56.8 &       100.0 &    0.0 &     4.9 \\
\end{tabularx}

\setlength{\tabcolsep}{1.35em}
\begin{tabularx}{\textwidth}{lrrrrrrrrr}
\toprule
\multicolumn{1}{l}{\textbf{}} & 
\multicolumn{1}{c}{\textbf{66}} & 
\multicolumn{1}{c}{\textbf{67}} & 
\multicolumn{1}{c}{\textbf{68}} &
\multicolumn{1}{c}{\textbf{69}} &
\multicolumn{1}{c}{\textbf{70}} &
\multicolumn{1}{c}{\textbf{71}} \\
\multicolumn{1}{l}{\textbf{No.}} & 
\multicolumn{1}{c}{\textbf{Match1}} & 
\multicolumn{1}{c}{\textbf{Match2}} & 
\multicolumn{1}{c}{\textbf{Match3}} & 
\multicolumn{1}{c}{\textbf{Match4}} &
\multicolumn{1}{c}{\textbf{Match5}} &
\multicolumn{1}{c}{\textbf{Comment}} \\
\midrule
0 &   4710 &   442 &     NaN &    NaN &  V0371 Gem &  VSX match with Delta Cepheid... \\
1 &      NaN &     NaN &     NaN &    NaN &        NaN & Possibly multi-periodic \\
2 &   6386 &    54 &     9 &    NaN &        NaN & -- \\
3 &  13490 &   247 &     NaN &    NaN &        NaN & -- \\
4 &      NaN &     NaN &     NaN &    NaN &        NaN &  Object is spinning too quickly... \\
\bottomrule
\end{tabularx}
\end{table}

\begin{enumerate}
\item \textit{HATID} -- Source identifier as provided by the \textit{Hungarian-made Automated Telescope Network}, also known as the \textit{kepid} in the K2CO LC dataset. If the HATID was unavailable, the source was identified solely using the UCAC source ID. 
\item \textit{UCACID} -- Original source ID in UCAC catalog.
\item \textit{RA} -- Right ascension coordinate in degrees. 
\item \textit{DEC} -- Declination coordinate in degrees.
\item \textit{Kepmag} -- Kepler magnitude, as computed using 2MASS catalog values for sdssg and sdssr. Calculation follows the hierarchical scheme outlined in the Barbara A.~Mikulski Archive for Space Telescopes (MAST).
\item \textit{P} -- The best determined period of the source (in days) from the periodogram analysis.
\item \textit{dmag} -- Amplitude of the magnitude (maximum peak value - minimum peak value) calculated after fitting an order-3 Fourier series to the differential magnitude time series.
\item \textit{Class} -- Variable classification identifier. Possible classification labels include \textit{Transit}, \textit{EB}, \textit{Pulsating}, \textit{Rotating}, and \textit{Misc}.
\item \textit{Class\_string} -- Variable subclassification identifier. This identifier can provide additional information for rotating and pulsating variables. Possible subclassification labels include \textit{Transit}, \textit{EB}, \textit{Cepheid}, \textit{DScuti} (\dScuti{} pulsator), \textit{GDor} (\Gdor{} pulsator), \textit{SPB} (slowly pulsating B-star), \textit{RotVar} (rotational variable of indeterminate type), and \textit{Misc} (indeterminate variable). 
\item \textit{Blend} -- Boolean to indicate whether or not the source is an ambiguous blend. The value 0 indicates that the source is an identified primary variable, while the value 1 indicates that the source is an ambiguous blend. 
\item \textit{LSP1} -- 1st peak period as determined by the Lomb-Scargle periodogram (\texttt{VARTOOLS}).
\item \textit{LSFAP1} -- VARTOOLS-derived formal false alarm probability of the 1st peak period (\texttt{VARTOOLS}).
\item \textit{LSSNR1}  -- Signal-to-noise ratio for the 1st peak period (\texttt{VARTOOLS}).
\item \textit{LSP2} -- 2nd peak period as determined by the Lomb-Scargle periodogram (\texttt{VARTOOLS}).
\item \textit{LSFAP2} -- VARTOOLS-derived formal false alarm probability of the 2nd peak period (\texttt{VARTOOLS}).
\item \textit{LSSNR2}  -- Signal-to-noise ratio for the 2nd peak period (\texttt{VARTOOLS}).
\item \textit{LSP3} -- 3rd peak period as determined by the Lomb-Scargle periodogram (\texttt{VARTOOLS}).
\item \textit{LSFAP3} -- VARTOOLS-derived formal false alarm probability of the 3rd peak period (\texttt{VARTOOLS}).
\item \textit{LSSNR3}  -- Signal-to-noise ratio for the 3rd peak period (\texttt{VARTOOLS}). 
\item \textit{PDM1} -- 1st peak period as determined by the PDM periodogram (\texttt{Astrobase}).
\item \textit{PDM2} -- 2nd peak period as determined by the PDM periodogram (\texttt{Astrobase}).
\item \textit{PDM3} -- 3rd peak period as determined by the PDM periodogram (\texttt{Astrobase}).
\item \textit{BLSP1} -- 1st BLS peak period in days (\texttt{Astrobase}). Values are only listed for candidate EB and transit sources. 
\item \textit{BLS1tdepth} -- 1st BLS peak period transit depth in mag (\texttt{Astrobase}). 
\item \textit{BLS1tdur} -- 1st BLS peak period transit duration in days 
(\texttt{Astrobase}).
\item \textit{BLSP2} -- 2nd BLS peak period in days (\texttt{Astrobase}). Values are only listed for candidate EB and transit sources. 
\item \textit{BLS2tdepth} -- 2nd BLS peak period transit depth in mag (\texttt{Astrobase}).
\item \textit{BLS2tdur} -- 2nd BLS peak period transit duration in days (\texttt{Astrobase}).
\item \textit{BLSP3} -- 3rd BLS peak period in days (\texttt{Astrobase}).
\item \textit{BLS3tdepth} -- 3rd BLS peak period transit depth in mag (\texttt{Astrobase}). Values are only listed for candidate EB and transit sources. 
\item \textit{BLS3tdur} -- 3rd BLS peak period transit duration in days (\texttt{Astrobase}).
\item \textit{Jmag} -- \Jmag{} band magnitude (2MASS).
\item \textit{errJ} -- Error on the value of the \Jmag{} band magnitude (2MASS).
\item \textit{Hmag} -- \Hmag{} band magnitude (2MASS).
\item \textit{errH} -- Error on the value of the \Hmag{} band magnitude (2MASS).
\item \textit{Kmag} -- \Kmag{} band magnitude (2MASS).
\item \textit{errK} -- Error on the value of the \Kmag{} band magnitude (2MASS).
\item \textit{Bmag} -- \Bmag{} band magnitude (2MASS).
\item \textit{Vmag} -- \Vmag{} band magnitude (2MASS).
\item \textit{Imag} -- \Imag{} band magnitude (2MASS).
\item \textit{E(B-V)} -- E(B-V) extinction as provided by \cite{Schlafly2011}.
\item \textit{GaiaID} -- Designated Gaia ID for source match. 
\item \textit{d1} -- Minimum source distance in parsecs derived from the Gaia DR2 absolute stellar parallax at the reference epoch and the standard error of parallax. 
\item \textit{d2} -- Maximum source distance in parsecs derived from the Gaia DR2 absolute stellar parallax at the reference epoch and the standard error of parallax.
\item \textit{Teff} -- Gaia DR2 estimate of stellar effective temperature (in Kelvin) from Apsis-Priam.
\item \textit{T1} -- Gaia DR2 uncertainty (lower) on \textit{Teff} estimate from Apsis-Priam. This is the 16th percentile of the probability distribution function over \textit{Teff}.
\item \textit{T2} -- Gaia DR2 uncertainty (upper) on \textit{Teff} estimate from Apsis-Priam. This is the 84th percentile of the probability distribution function over \textit{Teff}.
\item \textit{Rad} -- Gaia DR2 stellar radius estimate from Apsis-Priam. Listed in units of solar radii. 
\item \textit{R1} -- Gaia DR2 uncertainty (lower) on radius estimate from Apsis-FLAME. This is the 16th percentile of the probability distribution function over radius. Listed in units of solar radii. 
\item \textit{R2} -- Gaia DR2 uncertainty (upper) on radius estimate from Apsis-FLAME. This is the 84th percentile of the probability distribution function over radius. Listed in units of solar radii. 
\item \textit{Lum} -- Gaia DR2 estimate of luminosity from Apsis-FLAME. Listed in units of solar luminosity. 
\item \textit{L1} -- Gaia DR2 uncertainty (lower) on luminosity estimate from Apsis-FLAME. This is the 16th percentile of the probability distribution function over luminosity. Listed in units of solar luminosity. 
\item \textit{L2} -- Gaia DR2 uncertainty (upper) on luminosity estimate from Apsis-FLAME. This is the 84th percentile of the probability distribution function over luminosity. Listed in units of solar luminosity. 
\item \textit{PMRA} -- Proper motion in RA (milli-arcseconds/yr) from Gaia DR2.
\item \textit{PMDEC} -- Proper motion in DEC (milli-arcseconds/yr) from Gaia DR2.
\item \textit{PM35CG18} -- M35 membership probability  from  Gaia \citep{Cantat2018}.
\item \textit{PM35CM09} -- M35 membership probability, determined with proper motion measurements \citep{Bouy2015}.
\item \textit{PM35K13} -- M35 membership probability, determined with proper motion measurements \citep{Kharchenko2013}.
\item \textit{PM35K13b} -- M35 membership probability from  CMD \citep{Kharchenko2013}.
\item \textit{PN2158CG18} -- Membership probability for affiliation with NGC 2158 from  Gaia \citep{Cantat2018}. Oddly, some of these probabilities were reported as $100\%$.
\item \textit{PN2158D06} -- NGC~2158 membership probability, determined with proper motion measurements \citep{Dias2006}. 
\item \textit{PN2158K13} -- NGC~2158 membership probability, determined with proper motion measurements \citep{Kharchenko2013}.
\item \textit{PN2158K13b} -- NGC~2158 membership probability from  CMD \citep{Kharchenko2013}.
\item \textit{PM35} -- the overall designated M35 membership probability used for the source, employing the method described in Section~\ref{sec:membership}.
\item \textit{PN2158} -- the overall designated NGC~2158 membership probability used for the source, employing the method described in Section~\ref{sec:membership}, which includes a King model fit to weight cluster membership according to the radial distance from the cluster core.
\item \textit{Match1} -- Corresponding source catalog match from the variable catalog presented in \cite{Libralato2016}.
\item \textit{Match2} -- Corresponding source catalog match from the variable catalog presented in \cite{Nardiello2015}.
\item \textit{Match3} -- Corresponding source catalog match from the variable catalog presented in \cite{Meibom2009}.
\item \textit{Match4} -- Corresponding source catalog match from the variable catalog presented in \cite{Mochejska2004}.
\item \textit{Match5} -- Corresponding source catalog match from the VSX Catalog.
\item \textit{Comment} -- Source commentary.
\end{enumerate}

\pagebreak

\begin{figure*}[b]
\centering
\includegraphics[width=0.75\textwidth]{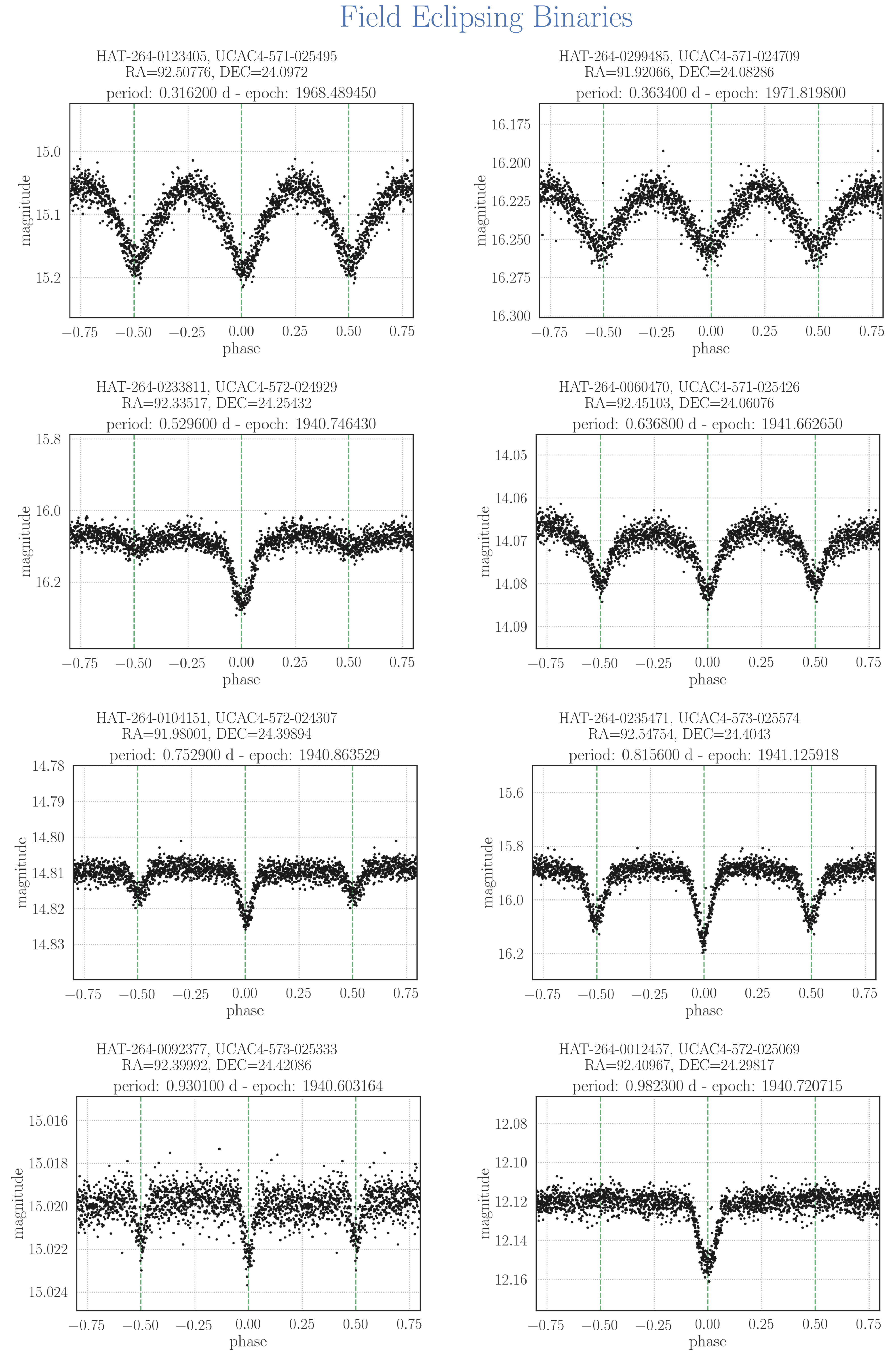}
\caption{Phase-folded LCs for the EB variables associated with the field, displayed in ascending order of period. Epochs are provided in BJD-2454833. 
Plotted along the y-axis is the \Kepmag{} magnitude of the source.}
\label{fig:Field_EBs1}
\end{figure*}

\begin{figure*}
\centering
\includegraphics[width=0.75\textwidth]{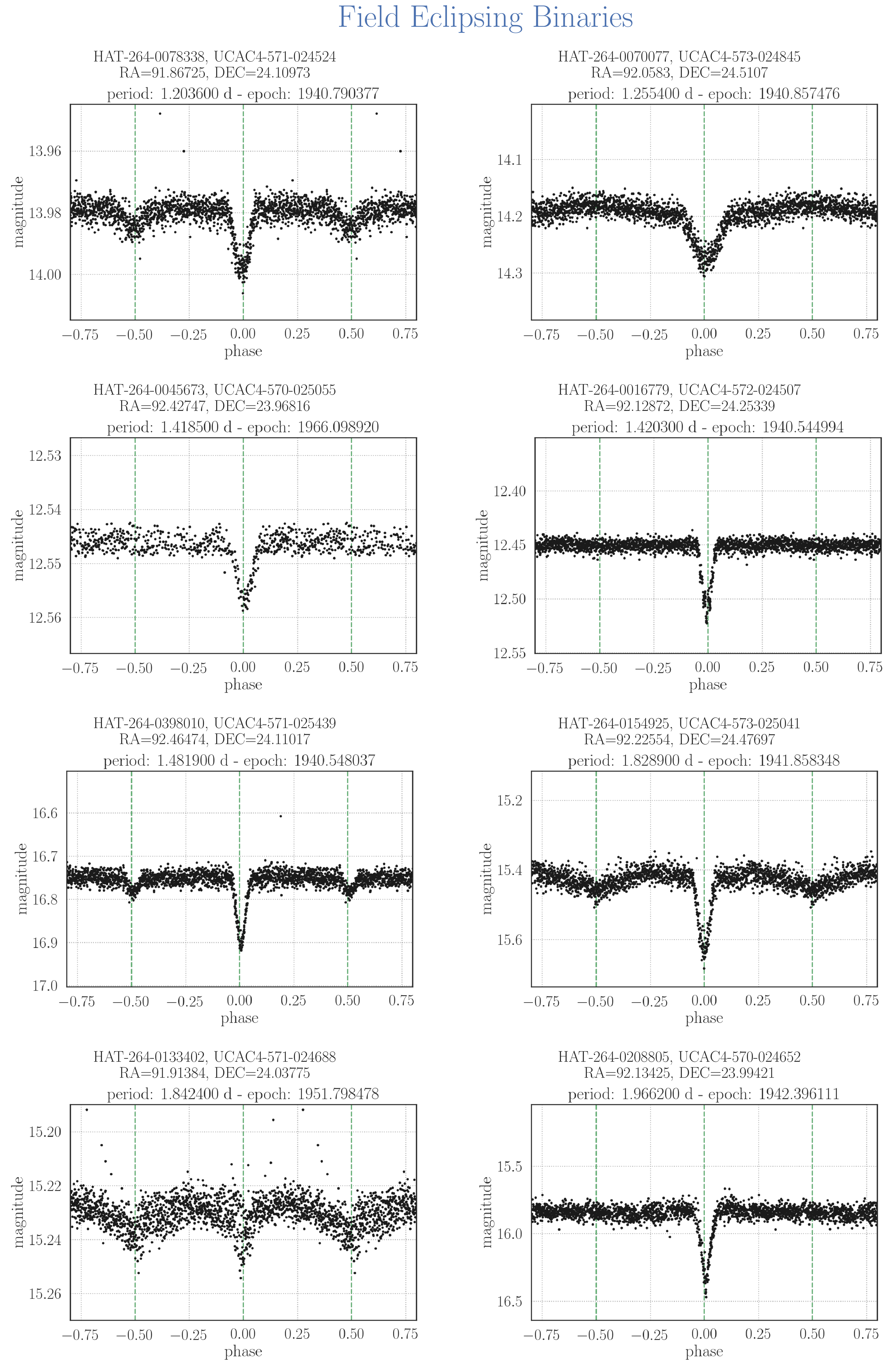}
\caption{Phase-folded LCs for the EB catalog variables associated with the field, displayed in ascending order of period. Epochs are provided in BJD-2454833. 
Plotted along the y-axis is the \Kepmag{} magnitude of the source.}
\label{fig:Field_EBs2}
\end{figure*}

\begin{figure*}
\centering
\includegraphics[width=0.8\textwidth]{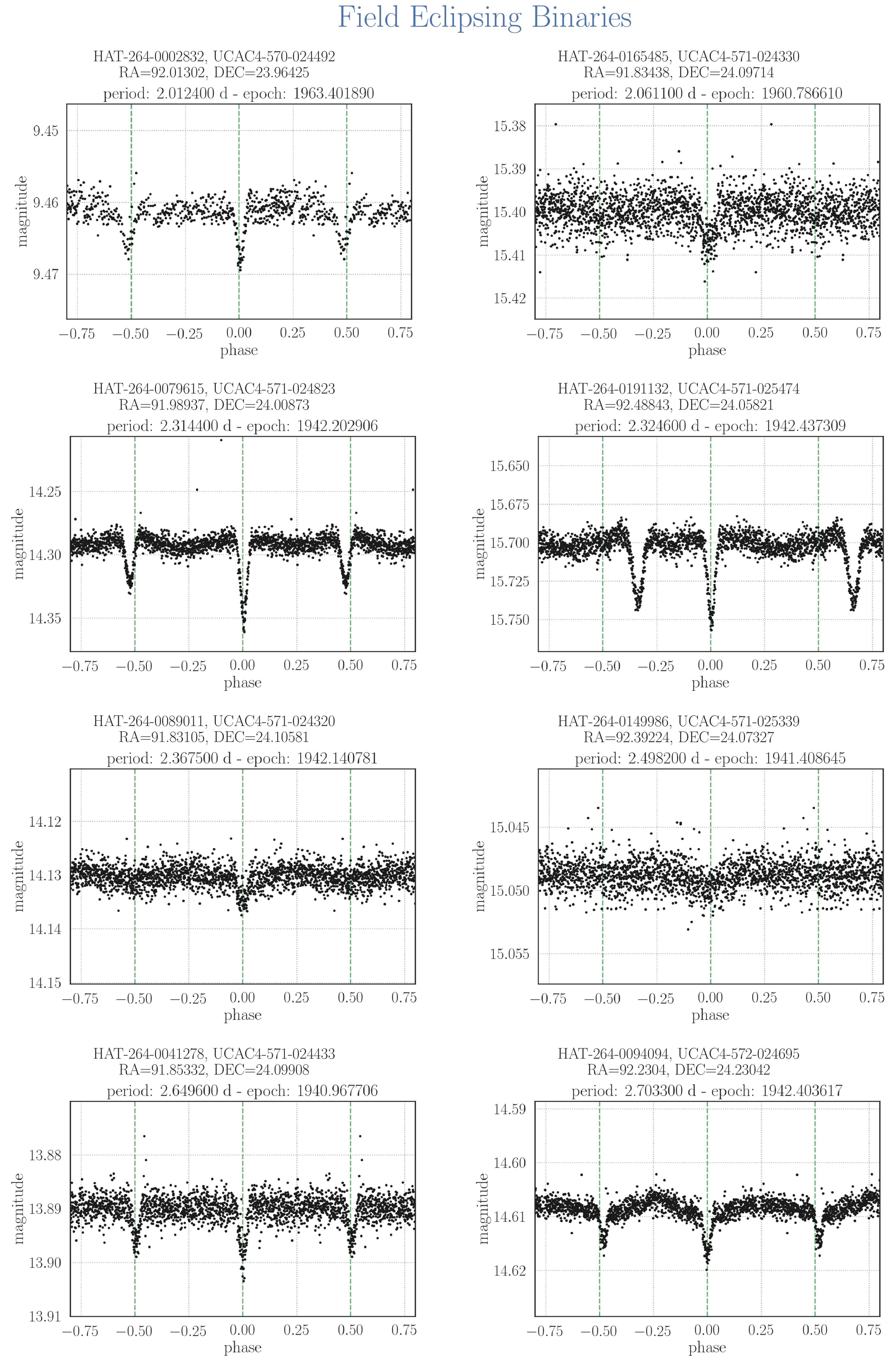}
\caption{Phase-folded LCs for the  EB catalog variables associated with the field, displayed in ascending order of period. 
Epochs are provided in BJD-2454833. 
Plotted along the y-axis is the \Kepmag{} magnitude of the source.}
\label{fig:Field_EBs3}
\end{figure*}

\begin{figure*}
\centering
\includegraphics[width=0.8\textwidth]{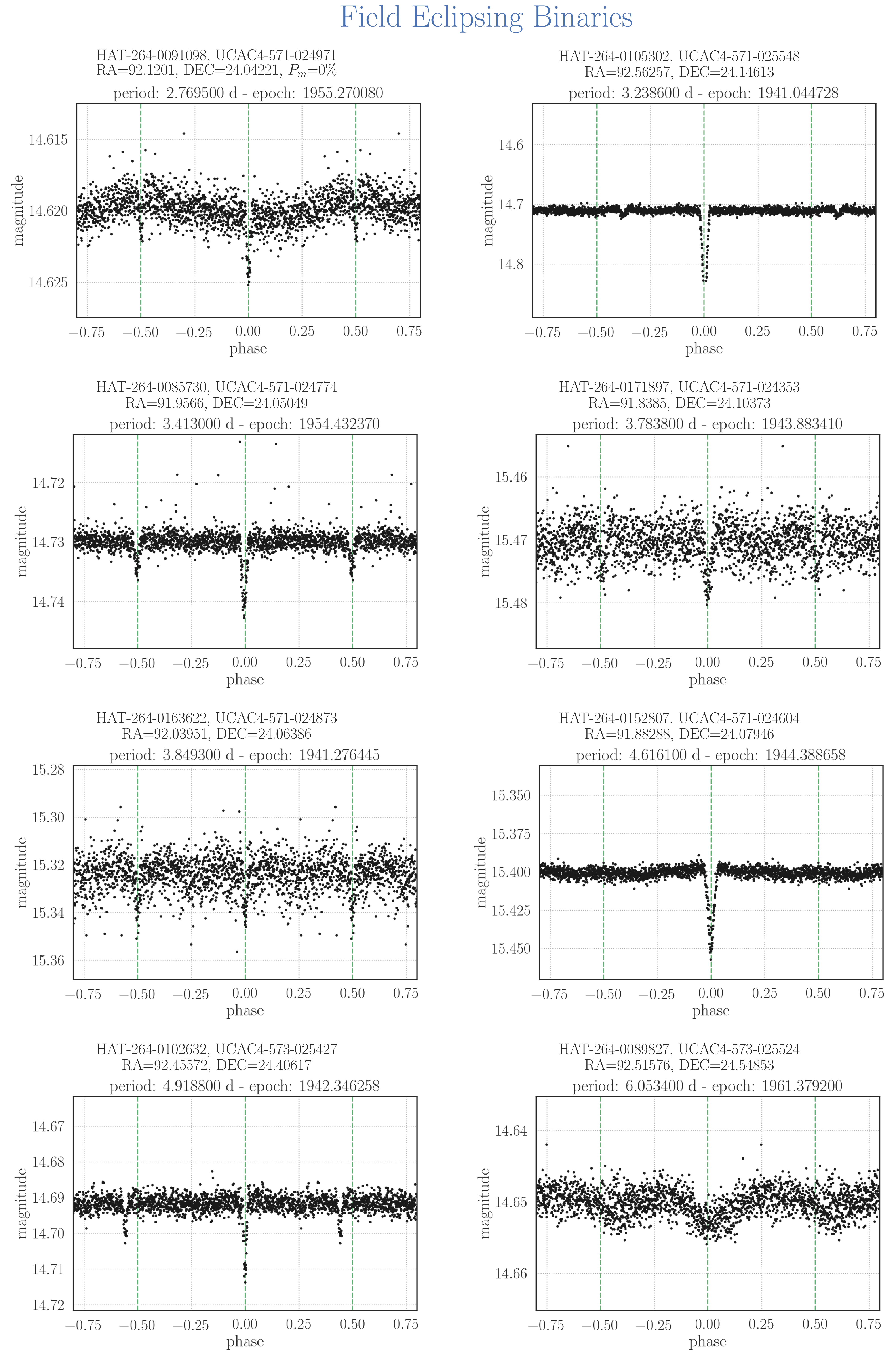}
\caption{Phase-folded LCs for the  EB catalog variables associated with the field, displayed in ascending order of period. 
Epochs are provided in BJD-2454833. 
Plotted along the y-axis is the \Kepmag{} magnitude of the source.}
\label{fig:Field_EBs4}
\end{figure*}

\begin{figure*}
\centering
\includegraphics[width=0.95\textwidth]{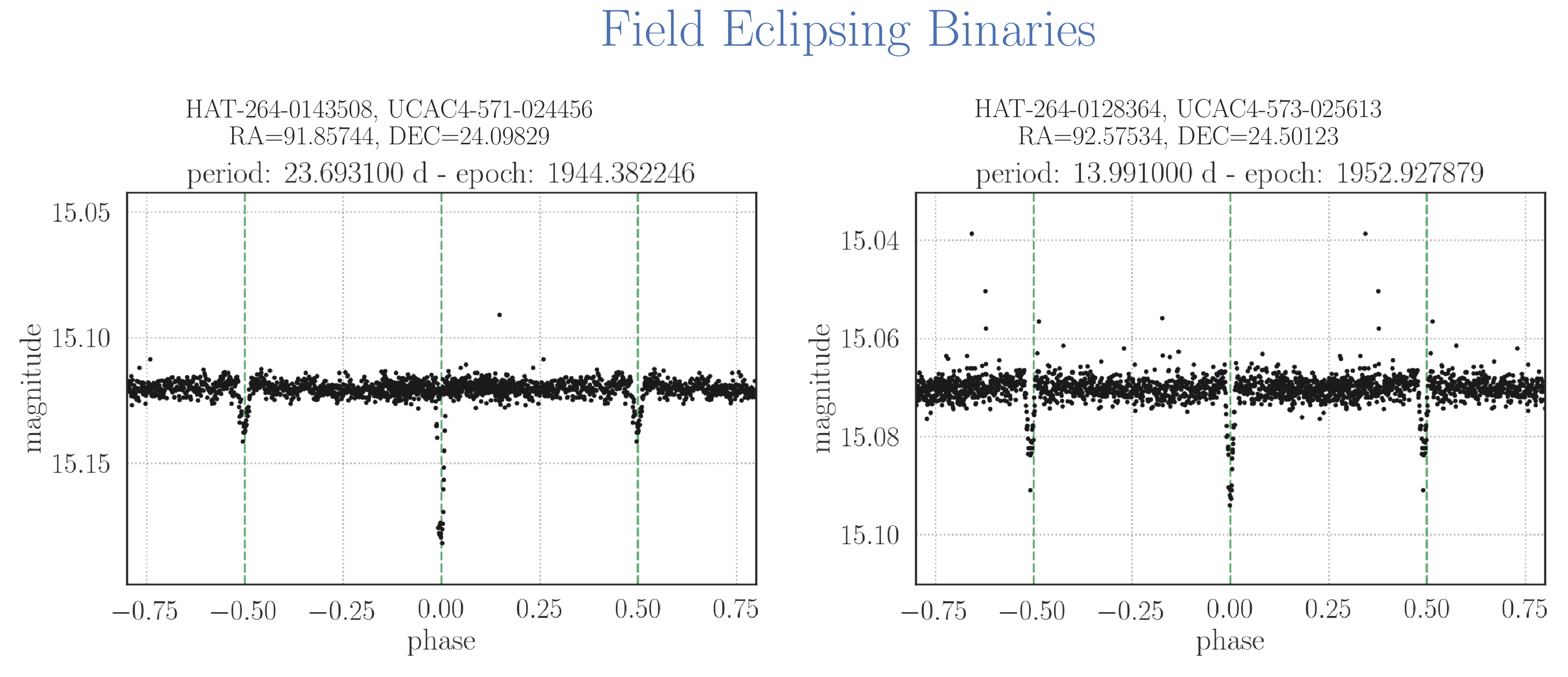}
\caption{Phase-folded LCs for the  EB catalog variables associated with the field, displayed in ascending order of period. 
Epochs are provided in BJD-2454833. 
Plotted along the y-axis is the \Kepmag{} magnitude of the source.}
\label{fig:Field_EBs5}
\end{figure*}

\begin{figure}
\centering
\includegraphics[width=0.95\textwidth]{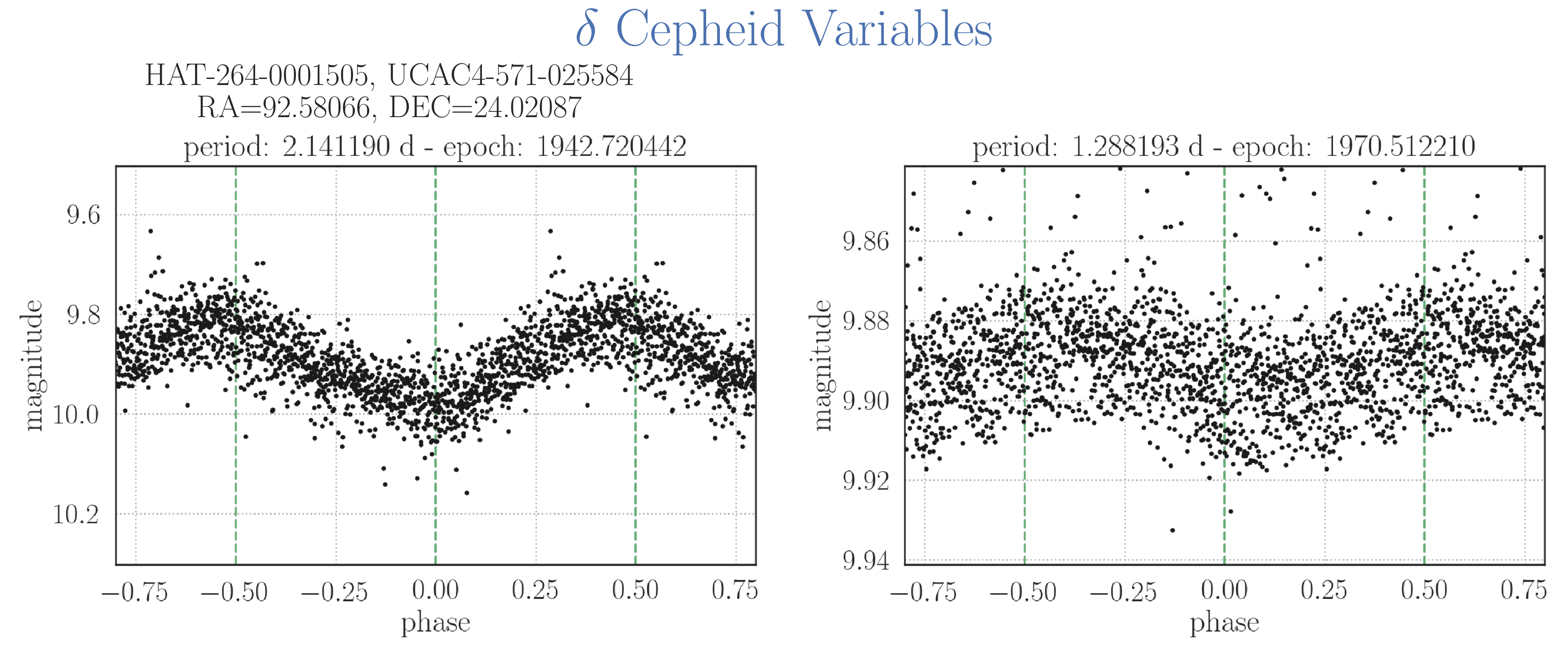}
\caption{Phase-folded LC for V0371 Gem (UCAC4-571-025584), our catalog $\delta$~Cepheid. 
The \Kepmag{} magnitude is shown along the y-axis.
\textit{Left Panel:} the strongest variability period at 2.141186~days.
\textit{Right Panel:} the $P=1.288193$~day signal detected after filtering out the dominant period and associated harmonics.
This source is not associated with either open cluster. This is confirmed by Gaia DR2 proper motion and parallax measurements. 
The period noted in the VSX catalog is 2.1371~days. The epoch is provided in BJD-2454833. }
\label{fig:Cepheid}
\end{figure}

\begin{figure}
\centering
\includegraphics[width=\textwidth]{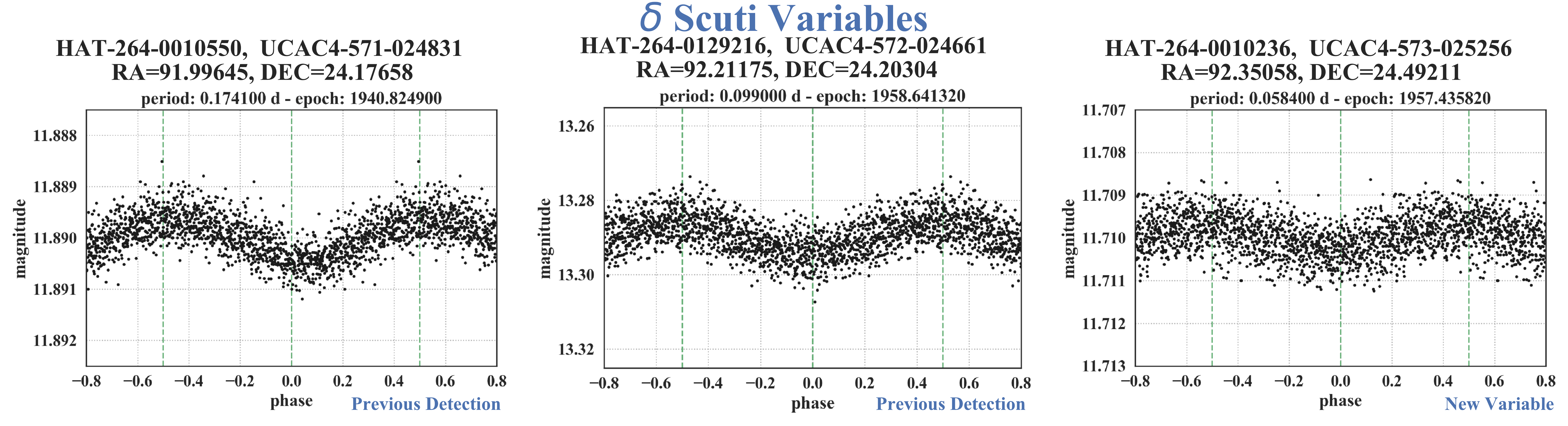}
\caption{LCs of three representative \dScuti{} variables in our catalog. 
Plotted along the y-axis is the \Kepmag{} magnitude.
The first two columns display previously detected sources and the third column displays a new detection. 
\textit{Left:} UCAC4-571-024831 with a period of $0.1741$~days and an amplitude of 8~millimag.
\textit{Center:} UCAC4-572-024661 with a period of $0.099$~days and an amplitude 0.009~mag. 
\textit{Right:} UCAC4-573-025256 with a period of $0.0584$~days and an amplitude of 7~millimag.
UCAC4-573-025256 has an M35 membership probability of 60\%, while the others are field sources.}
\label{fig:DScuti}
\end{figure}

\begin{figure}
\centering
\includegraphics[width=\textwidth]{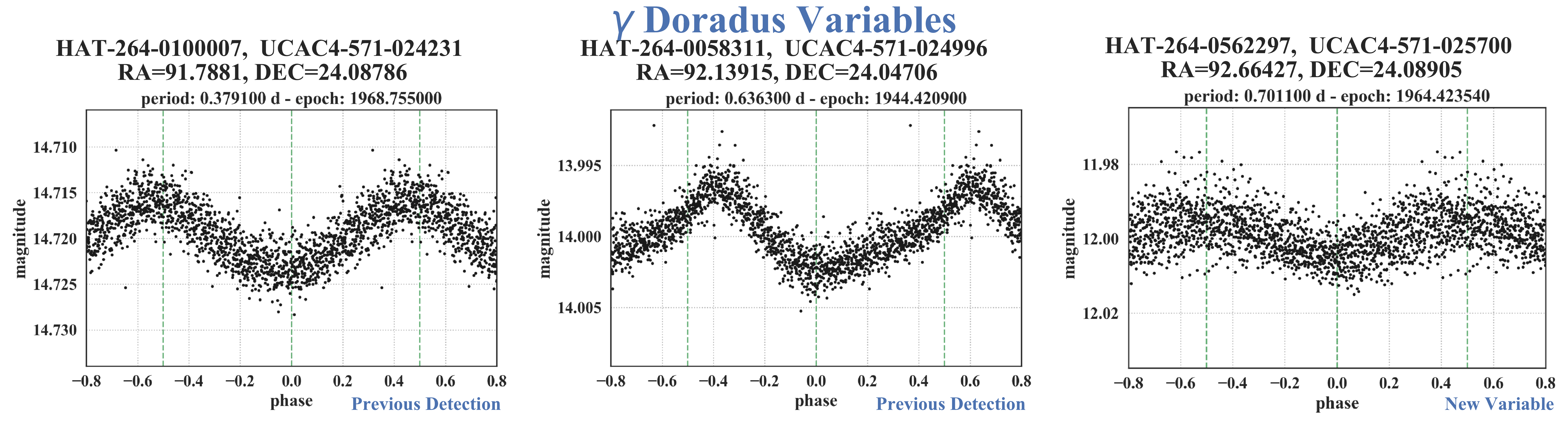}
\caption{LCs of three representative \Gdor{} variables in our catalog. 
Plotted along the y-axis is the \Kepmag{} magnitude.
The first two columns display previously detected sources and the third column displays a new detection. 
\textit{Left:} UCAC4-571-024231 with a period of $0.3791$~days and an amplitude of 8~millimag.
\textit{Center:} UCAC4-571-024996 with a period of $0.6363$~days and an amplitude of 7~millimag. 
\textit{Right:} UCAC4-571-025700 with a period of $0.7011$~days and an amplitude of 0.01~mag. 
All sources are confirmed by Gaia to be field stars.}
\label{fig:GDor}
\end{figure}

\begin{figure}
\centering
\includegraphics[width=\textwidth]{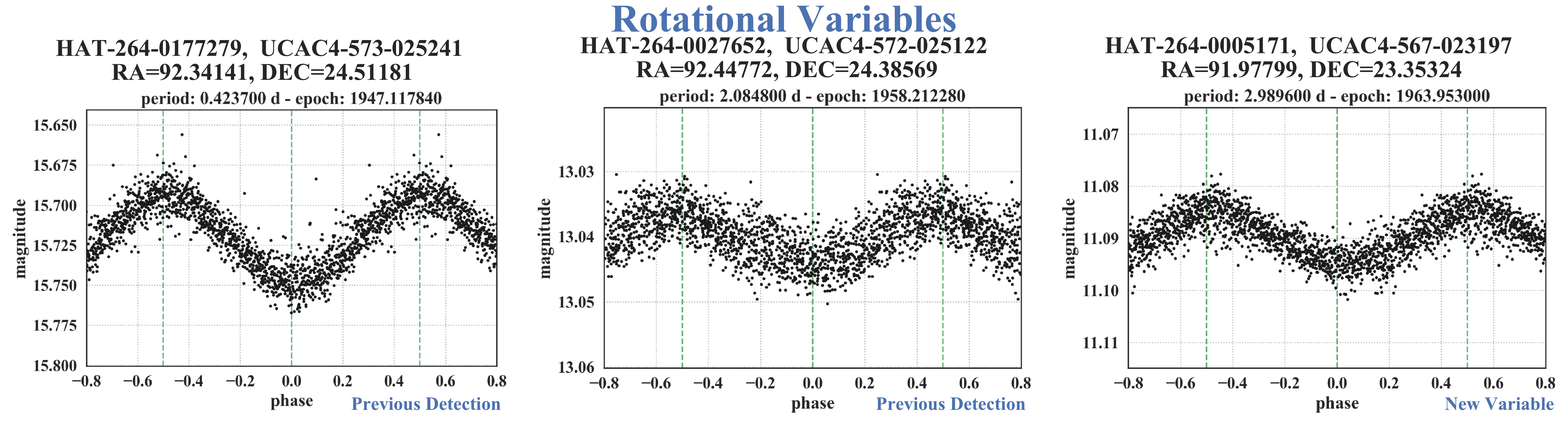}
\caption{LCs of three representative rotational variables in our catalog. 
Plotted along the y-axis is the \Kepmag{} magnitude.
The first two columns display previously detected sources and the third column displays a new detection. 
\textit{Left:} UCAC4-573-025241 with a period of $0.4237$~days and an amplitude of 0.06~mag.
The Gaia DR2 match for this source measures $T_{\rm eff}{\sim}4400$~K and $L_{\star}{\sim}0.2$~\Lsun{}.
\textit{Center:} UCAC4-572-025122 with a period of $2.08484$~days and amplitude of 0.01~mag. 
The Gaia DR2 match for this source measures $T_{\rm eff}{\sim}5370$~K and $L_{\star}{\sim}2$~\Lsun{}.
\textit{Right:} UCAC4-567-023197 with a period of $2.9896$~days and an amplitude of 0.01~mag. 
The Gaia DR2 match for this source measures $T_{\rm eff}{\sim}5900$~K and $L_{\star}{\sim}2$~\Lsun{}.
The proper motion and parallax of UCAC4-572-025122 indicate M35 membership, while the other two are field sources.
}
\label{fig:RotVar}
\end{figure}

\end{document}